\documentclass[12pt]{article}

\usepackage{newtxtext,newtxmath}

\usepackage{graphicx}

\usepackage[letterpaper,margin=1in]{geometry}

\renewenvironment{abstract}
	{\quotation}
	{\endquotation}

\date{}

\makeatletter
\renewcommand{\fnum@figure}{\textbf{Figure \thefigure}}
\renewcommand{\fnum@table}{\textbf{Table \thetable}}
\makeatother

\usepackage{scicite}

\usepackage{url}

\def\scititle{
	Directed assembly of tetrahedral patchy particles
}
\title{\bfseries \boldmath \scititle}

\author{
	Xin Yin$^{1}$,
	Ekaterina Kostyurina$^{1}$,
    Bert Nickel$^{1}$,
	Tim Liedl$^{1\ast}$,
    Gregor Posnjak$^{1\ast}$\and
	\small$^{1}$Faculty of Physics and CeNS, Ludwig-Maximilian-University Munich, 80539 Munich, Bavaria, Germany.\and
	\small$^\ast$Corresponding author. Email: tim.liedl@lmu.de; gregor.posnjak@lmu.de
}

\begin{document} 

\maketitle

\begin{abstract} \bfseries \boldmath

Colloidal particles with prescribed valency such as the tetrahedral patchy particles have long been seen as a viable route to technologically relevant open lattice structures on the scale of hundreds of nanometers. However, conceptual  limitations and resulting competing local bonding configurations often lead to mixed lattice phases. Here, we present a DNA-origami enabled approach to controlling the attachment of tetrapod building blocks in predictable ways. By varying the relative strength of two designed binding configurations we are able to direct the assembly of tetrapod particles into diamond cubic, twinned diamonds, stacking-disordered mixtures, hexagonal diamonds, and sII clathrates. Under specific conditions, the diamond structures are interpenetrated by additional networks, resulting in triple cubic and triple hexagonal diamond structures. The 440 nm large unit cell of the clathrates shifts structural reflections into the visible range, giving these rationally designed, self-assembled crystals structural color.

\end{abstract}

A single type of a building block can assemble in multiple ways to form different structures. For example, carbon atoms can form different allotropes such as graphite, graphene, carbon nanotubes, and diamond. Although the building blocks are identical, the way they connect dictates the mechanical, chemical, and conductive properties of the assembled materials. 

Colloids provide model systems for studying the fundamental principles of crystallization and can facilitate the assembly and growth of diverse types of crystalline structures\cite{macfarlane_nanoparticle_2011, zhang_general_2013}. At the same time, the material  properties of colloidal lattices whose characteristic length scales exceed those of atomistic and molecular crystals by more than two orders of magnitude, give rise to engineerable phenomena such as structural color and photonic band gaps\cite{joannopoulos_photonic_1997}. However, open lattice structures, desirable for photonic band gap materials and exemplified by the diamond structure, are particularly demanding for self-assembly, as structures with a high packing ratio are typically entropically favored. \cite{zhang_general_2013} 
Assembly of identical particles typically leads to closely packed structures such as hexagonal close packed, FCC, SC, or BCC. Other types of structures can also be assembled by varying the type of the interaction potential and mixing particles of different sizes~\cite{hueckel_ionic_2020}. In an especially fruitful approach researchers combined metallic particles of different shapes with DNA as the binding material to realize a large diversity of crystalline structures~\cite{macfarlane_nanoparticle_2011, Jones_programmable_2015,zhou_space-tiled_2024}, including clathrate structures~\cite{lin_clathrate_2017}, and even quasicrystals~\cite{zhou_colloidal_2024}.

A promising way of forming open structures such as the diamond cubic through self-assembly was proposed in 2005~\cite{zhang_self-assembly_2005}, when spherical particles with four binding patches placed in a tetrahedral configuration were suggested as large-scale analogues of carbon atoms~\cite{nelson_toward_2002}.  Here, the arrangement of interacting patches mimics the directionality of bonds at smaller scales. Theoretical studies found that such tetrahedral patchy particles have a diverse phase diagram, including amorphous glassy phases, diamond cubic (DC), and hexagonal diamond (HD) structures\cite{romano_phase_2010,romano_crystallization_2011}. In numerical studies, the DC and HD structures of these particles commonly have very similar energies, leading to coexistence of both phases, stacking-disordered mixtures of these two phases\cite{baran_interplay_2024,tarasewicz_self-assembly_2025}, and even clathrates, which appear only under specific conditions\cite{noya_assembly_2019,baran_interplay_2024}. This is contrasted by naturally occurring crystals of this type, for example, carbon-based diamond predominantly occurs in pure DC structure. Carbon-based HD only forms together with cubic diamond under extreme conditions \cite{frondel_lonsdaleite_1967}, however, recent advances have enabled the synthesis of diamond with different degrees of twinning and stacking disorder \cite{huang_tracking_2023,tong_structural_2024,ying_enhancing_2025}, and even pure HD structure \cite{yang_synthesis_2025,lai_bulk_2026}. In a different example, water molecules commonly crystallize in almost pure HD structure (the so-called Ice I$_h$), with some degree of stacking disorder. However, there are over 20 distinct crystalline water ice phases which are stable at different temperatures and pressures \cite{lee_multiple_2026}. Recently, researchers were able to stabilize the diamond cubic structure in water ice (Ice I$_c$) \cite{komatsu_ice_2020,huang_tracking_2023}, and even ice with sII clathrate structure (Ice XVI) \cite{falenty_formation_2014}. 

Clathrates are structures in which polyhedral cages of different types are arranged in crystalline lattices. They are found in many naturally occurring materials such as clathrate hydrates\cite{lu_complex_2007,hassanpouryouzband_gas_2020}, where guest atoms or molecules populate the large polyhedral cages and stabilize the structures with additional interactions. 
Under specific conditions, silicon\cite{gryko_low-density_2000}, and germanium compounds\cite{guloy_guest-free_2006} can form guest-free clathrate structures. In the case of water ice, the clathrates are formed at low pressures with neon serving as the guest, but become metastable after extraction of the gas~\cite{falenty_formation_2014}. In contrast, in engineered colloidal systems, clathrate architectures would be achievable through other mechanisms, without the need for a stabilizing guest molecule \cite{romano_patterning_2012,lin_clathrate_2017,noya_assembly_2019}.

Multiple approaches of controlling the stability and nucleation of the different phases in tetrahedral patchy particles have been proposed and explored numerically. Variation of the angular "width" of the patch and the range of the interaction potential produced mixed phases of DC and HD \cite{romano_crystallization_2011}, however, with an additional external potential the sII clathrate or diamond cubic structures can be stabilized \cite{baran_interplay_2024}. With polychromatic bonds, each of the binding patches is assigned specific affinity for only some of the other "colors" of patches. In simulations, this approach, combined with advanced optimisation algorithms, was able to circumvent kinetic traps that could lead to non-targeted assembly~\cite{rovigatti_simple_2022}. 

Experimental realizations of crystalline assemblies from patchy particles have been limited due to the difficulty of their manufacturing\cite{wang_colloids_2012,kim_patchy_2025}. In recent years, DNA origami has emerged as a customizable platform for patchy colloids, offering programmable particle shapes and precise control over their binding. DNA origami particles of various shapes and valences have been assembled in lattices of rhombohedral, simple cubic, body centered cubic, simple hexagonal, face centered cubic, diamond cubic, and other symmetries \cite{zhang_3D_2018, tian_ordered_2020, wang_dna_2021, wang_ph-induced_2022, michelson_three-dimensional_2022, dai_programming_2023}. Notably, the programmability of DNA origami has been harnessed to control the binding assignments of multiple monomers \cite{kahn_encoding_2025}, enabling careful adjustment of binding relations for prevention of kinetic traps to isolate specific open structures, such as the pyrochlore structure~\cite{russo_sat-assembly_2022,liu_inverse_2024}. 

\begin{figure}
	\centering
	\includegraphics[width=0.9\textwidth]{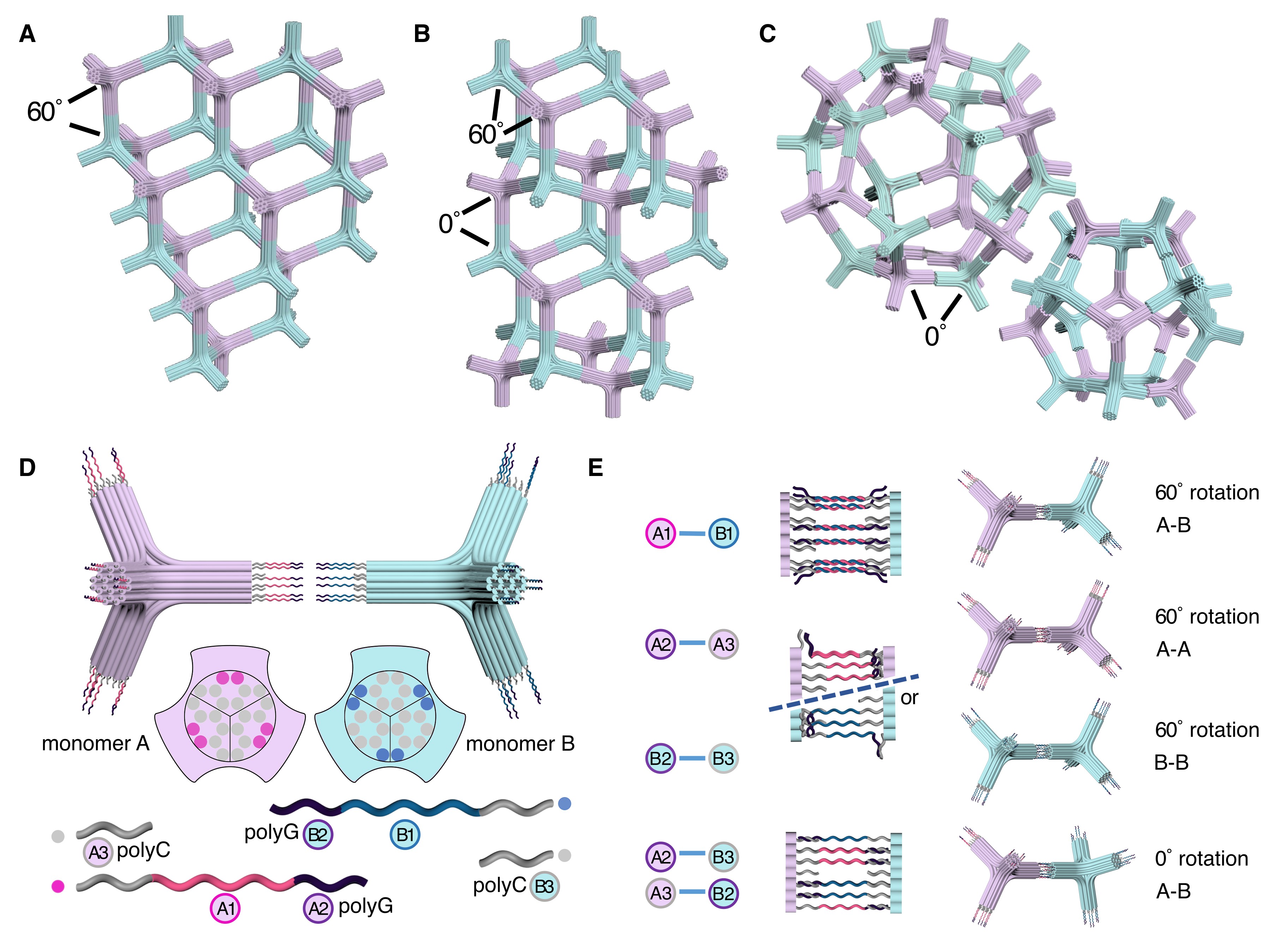} 
	\caption{\textbf{Tetrapod assemblies with tunable torsional bonds}
		(\textbf{A}) The rod-connected diamond cubic structure, where all the bonds are in the staggered configuration.  (\textbf{B}) The hexagonal diamond structure, where each tetrapod has one eclipsed and three staggered bonds.  (\textbf{C}) The building blocks of the sII clathrate structure: (left) hexakaidecahedral cage with 4 hexagonal and 12 pentagonal surfaces, where the hexagons are arranged with tetrahedral symmetry, (right) dodecahedron with 12 pentagonal surfaces. All the bonds between the tetrapods in the sII clathrate structure are in the eclipsed configuration. (\textbf{D}) Models of two tetrapod monomers, mA and mB, where the pink and turquois cylinders represent double-stranded DNA. Single-stranded sequences extend from the ends of each arm at the positions indicated in the schematic cross sections. Specific regions of these extensions are labeled A1 - A3 and B1 - B3, with A1 containing bases that are complementary to those of B1, A2 and B2 consisting of two or three guanines and A3 and B3 always consisting of 3 cytosines. (\textbf{E}) The four torsional bonds between the monomers in our system: binding between the A1 sequence on mA and B1 on mB, or A2 to A3 binding between two A monomers, or B2 to B3 binding between two B monomers all lead to 60° rotation (staggered configuration). In contrast, A2 to B3, and A3 to B2 binding between mA and mB results in 0° rotation (eclipsed configuration).}
	\label{clathrate_figure1}
\end{figure}

A powerful approach to reliably form specific phases arises from the fact that in ordered assemblies of tetrahedral patchy particles, the monomers are arranged in two possible configurations - either the staggered, or the eclipsed configuration.  For example, in the diamond cubic structure (Figure~\ref{clathrate_figure1}A) the tetrapods form staggered bonds exclusively, where each pair of neighbours is rotated by 60°. In contrast, in hexagonal diamond each tetrapod has three staggered bonds within the $\{001\}$ crystal plane and a single eclipsed bond with 0° rotation in the $\langle 001 \rangle$ direction (Figure~\ref{clathrate_figure1}B), and in the sII clathrate all bonds are in the eclipsed configuration (Figure~\ref{clathrate_figure1}C). If one is able to design the interactions of the monomers so that they lead to specific configurations, then it should be possible to guide the assembly to one of the possible structures\cite{romano_patterning_2012}. Experimentally, this is difficult to realize because a torsional binding potential is needed. In a notable example, tetrahedral clusters of spheres with retracted DNA binding patches were assembled in a close-packed diamond structure~\cite{he_colloidal_2020}, with the shape of the clusters providing the steric restriction to form exclusively staggered configurations. In an alternative approach, we have shown that DNA origami can be used to design exclusively staggered bonds, resulting in pure diamond cubic phases~\cite{posnjak_diamond-lattice_2024}. While torsional bonds are a powerful way of restricting the assembly to specific structures, such tetrahedral patchy particles have thus far captured only a fraction of the predicted behavior of atomic and molecular analogues.

In this work, we design the binding interaction so that each binding patch can form both the eclipsed and the staggered configuration. We then vary the relative strength of the two competing bond configurations by changing the DNA sequences on the binding patches, and thus control the resulting lattice assembly where a certain level of ambiguity allows for annealing and error correction.
We explore the phase space of relative interaction strengths of the competing modes and growth conditions, and find a rich array of possible structures, starting from pure diamond cubic crystals, assembled with exclusively staggered configurations, through diamond cubic structures with multiple twinning planes, continuously progressing through disordered stacking of multiple cubic and hexagonal planes, and even achieving pure hexagonal diamond crystals. Surprisingly, we discovered structures with multiple interpenetrating diamond networks, which we call triple diamond structures. Finally, under conditions where eclipsed bonds are the predetermined mode of binding, we observe pure sII clathrate-type structures, which are diamond cubic networks of $6^4$ $5^{12}$ polyhedral cages, with the intermediate space filled by the smaller $5^{12}$ cages. The unit cell of this clathrate structure contains 136 DNA origami tetrapods and has an edge length of 440 nm leading to structural reflections of visible light.

\subsection*{Competing bond configurations}

To realize a colloidal system with two competing bond configurations, we prepared two types of DNA origami tetrapod monomers, mA and mB, where the terminal surfaces of each tetrapod arm present 24 sites for placing user-defined DNA sequences (Figure~\ref{clathrate_figure1}D; design details are provided in Figure~\ref{mA_cadnano}, Figure~\ref{mB_cadnano}, and materials and methods). 18 of the 24 sites were designed to carry DNA strands terminating in 3 cytosines (CCC), initially with the intention of preventing blunt end stacking of DNA helices and therefore introducing repulsive interactions between the ends of the tetrapod arms~\cite{ijas_label-free_2022}. The remaining six sites bear DNA extensions that mediate binding between the tetrapods. Each binding extension comprises three segments: in the first segment, three thymines act as a spacer (TTT); the second provides a specific binding region with complementary sequences on mA and mB; the third region is a poly-guanine segment (two or three guanines) fostering site-non-specific binding with the poly(C) extensions. The extensions on each arm are arranged in a three-fold symmetry pattern, where the patterns on all patches of monomer mB are rotated by 60° relative to the ones on the monomer mA (inset to Fig.~\ref{clathrate_figure1}D). Figure~\ref{clathrate_figure1}E explains the two possible DNA binding modes: (i) hybridization between complementary specific binding regions (A1 to B1), and (ii) hybridization between the poly(G) segments and the poly(C) extensions (any sequence 2 to any sequence 3). The latter mode can occur both between the two different monomer types (mA-mB) as well as between monomers of the same type (mA-mA, or mB-mB). Figure~\ref{clathrate_figure1}E also depicts the various interactions and the resulting binding configurations. 

We systematically tuned the system by designing a series of binding extensions with varying binding strength, while always providing a poly(G) segment at their ends. Throughout our experiments, we found that these poly(G) segments were crucial for successful crystal growth even in the cases where the binding configuration between monomers is realized through the longer specific sequence (Figure~\ref{6ntcompare}). We hypothesize that the poly(G) segments transiently bind to the abundant poly(C) extensions, which were initially intended to prevent aggregation. In contrast, they may provide a weak attractive interaction, which helps the system to reach its minimal energy state. Each type of binding extension was named after the following convention: length and type of spacer nucleotides + number of nucleotides (nt) in the specific binding region + length of poly(G) segment. For example, an extension of the sequence TTT ATGACT GGG would be named "3T 6nt 3G". The names, sequences, and the calculated hybridization energy of the binding extensions are listed in Table~\ref{SequenceTable}. 
The DNA origami tetrapods were folded and purified as described in the Methods. For crystal growth, the two types of tetrapods were mixed with the selected type of extension strands, and the mixture was annealed at various concentrations of MgCl$_{2}$ ranging between 18 and 70 mM according to the crystallization protocol described in the Methods. To increase their mechanical stability, the crystals were then silicified via wet chemistry in bulk \cite{nguyen_dna-origami-templated_2019}. Finally, the samples were dried and imaged with Scanning Electron Microscopy (SEM), or examined with small-angle x-ray scattering (SAXS) in solution (see materials and methods for experimental details).

\subsection*{Diamond cubic, twinning and stacking disorder, hexagonal diamond, and sII clathrate}

\begin{figure} 
	\centering
	\includegraphics[width=0.9\textwidth]{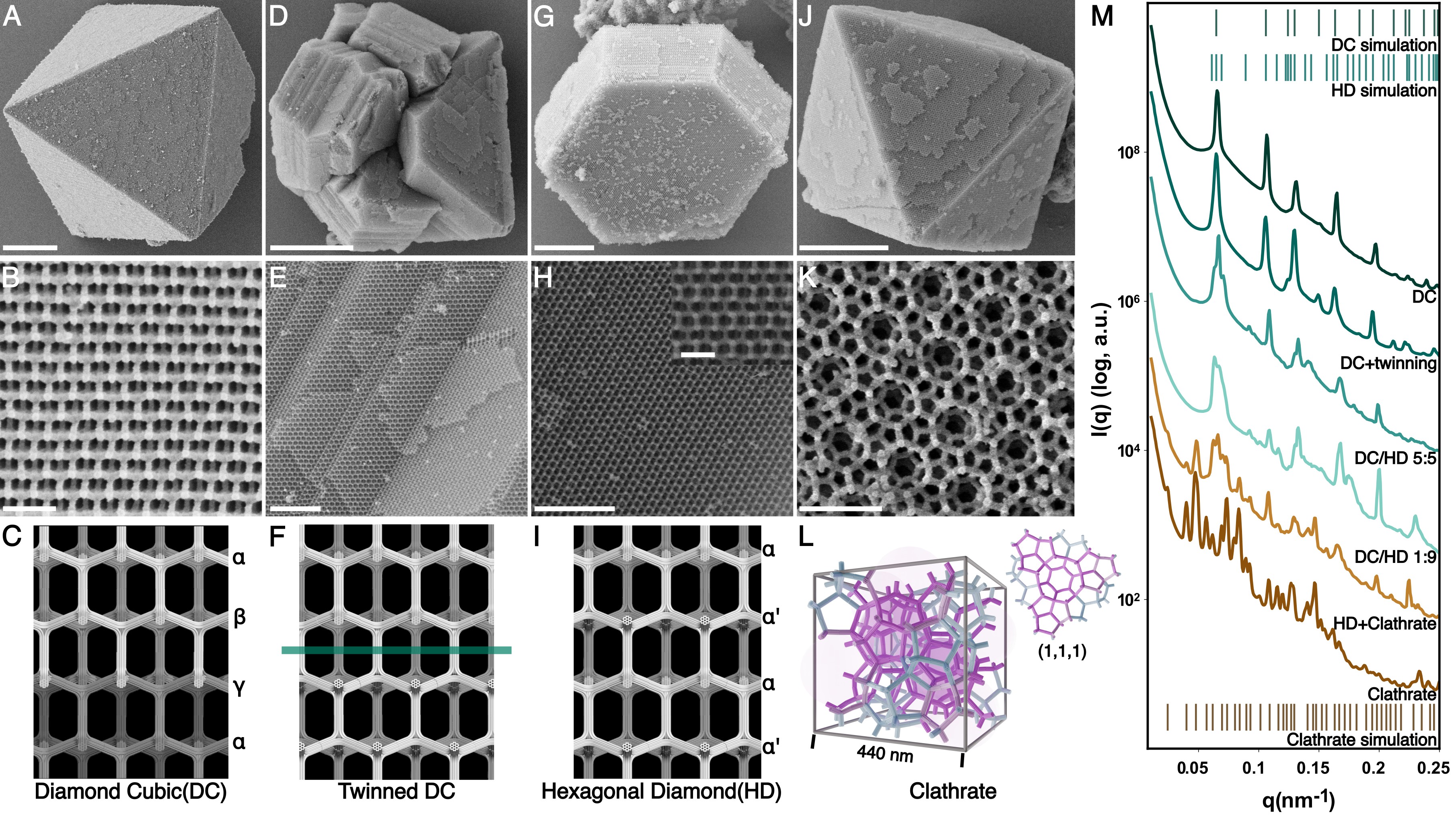} 
	\caption{\textbf{Variety of crystals grown from DNA origami tetrapods }
		(\textbf{A} to \textbf{C}) Scanning electron microscopy (SEM) images of pure diamond cubic (DC) crystals showing the (A) octahedral habit, (B) zoomed-in view of a tilted $\{111\}$ surface, and (c) a 3D model of the structure. (\textbf{D} to \textbf{F}) Diamond cubic crystals with multiple twinning planes, showing (D) habits of irregular shape, (E) zoomed-in view, and (F) a model of twinning planes (green line).  (\textbf{G} to \textbf{I}) Pure hexagonal diamond (HD) crystals with (G) typical hexagonal prism habit,  (\textbf{H}) zoomed-in view of a $\{100\}$ crystal surface, and (I) model of the HD structure. (\textbf{J} to \textbf{L}) SII clathrate crystals with (J) typical octahedral habits, (K) zoomed-in view of a $\{111\}$ crystal surface, and (L) the model of the unit cell, with the inset showing the $\{111\}$ plane. The $6^4 5^{12}$ cages are shown in purple. (\textbf{M}) Small-angle X-ray scattering (SAXS) intensities of the different tetrapod crystal types. The vertical lines indicate calculated Bragg  peaks. Scale bars in \textbf{A}, \textbf{D}, \textbf{J}: 10 $\mu$m, in \textbf{G}: 3 $\mu$m, in \textbf{E}, \textbf{H}: 1 $\mu$m, in \textbf{K}: 500 nm, in \textbf{B}, and inset to \textbf{H}: 200 nm.}
	\label{clathrate_figure2} 
\end{figure}

We first tested the torsional bond system by replicating the diamond cubic structure, which was previously obtained for a single monomer~\cite{posnjak_diamond-lattice_2024} (Figure~\ref{3T3G}). There, the torsional potential between the tetrapods was designed for binding in the staggered configuration, yielding pure diamond cubic structures that exhibit almost exclusively octahedral habits with very rare twinned crystals. We prepared a solution containing equimolar concentrations of monomers mA and mB with the binding sequence "3T 6nt 2G" (Figure~\ref{clathrate_figure2}A-C and Figure~\ref{6nt2G}). Here, the staggered bonds remain dominant, however eclipsed bonds are not entirely excluded, leading to more frequent occurrence of twinned crystals compared to the single-monomer system. Note that twinning occurs in DC structures when a pair of $\{111\}$ planes is connected with exclusively eclipsed bonds, while the previous and following layers are connected with staggered bonds (Figure~\ref{clathrate_figure2}D-F). The effect of rare twinning planes on the SAXS scattering intensities is very subtle, hardly differing from the intensities of the DC structure (Figure~\ref{clathrate_figure2}M).

By adding an additional guanine to the ends of the binding sequences – strengthening the binding mode (ii)  – and by shortening the central sequence – weakening the binding mode (i) – the relative strength and frequency of eclipsed configurations is increased, and consequently the number of observed twinning planes grows (Figure~\ref{clathrate_figure2}D,E, Figure~\ref{6nt3G}, and Figure~\ref{5nt3G}). Interestingly, the strengthening of eclipsed configurations did not lead to disordered structures, but instead resulted in the formation of twinning planes (Figure~\ref{clathrate_figure2}F). Similarly, raising the Mg$^{2+}$ concentration in the growth solution of these samples also lead to the formation of more and more twinning planes. Notably, the morphology of single crystals, which are octahedral for purely diamond cubic structures, becomes distorted with the increasing number of twinning planes. Each twinning plane causes a change in orientation of the $\{111\}$ crystal facet, resulting in a zig-zag appearance of the otherwise flat surfaces (Figure~\ref{clathrate_figure2}E). In crystals with multiple twinning planes, some of the angles in the facet corners can switch from 60° to 120° (Figure \ref{6nt3G}G,I,K and Figure~\ref{5nt3G}E,G,I). This even lead to hexagonal prism habits with jagged side-surfaces at high Mg$^{2+}$concentrations, despite the bulk of the crystals still having the diamond cubic structure (Figure~\ref{clathrate_figure2}D, Figure~\ref{5nt3G}G,I).

Increasing the relative strength of the eclipsed binding configuration further at moderate Mg$^{2+}$ concentrations resulted in the stacking-disordered polytypic diamond structure, where multiple layers of DC structure are mixed with multiple layers of the HD structure in an irregular fashion (Figure~\ref{stacking4nt}). This behavior has been numerically predicted for tetrahedral patchy particles \cite{baran_interplay_2024,tarasewicz_self-assembly_2025}, and observed both in carbon diamond \cite{salzmann_extent_2015} and water ice \cite{huang_tracking_2023}, and even in DNA origami crystals for FCC and simple hexagonal lattices~\cite{dai_programming_2023}. The stacking-disordered crystals have hexagonal shape, with uneven sides resulting from changes in preferred facets of the two phases. Increasing the concentration of Mg$^{2+}$ had the effect of increasing the proportion of HD structure in the stacking-disordered diamond (Fig.~\ref{stacking4nt}, \ref{4nt26_30}, and \ref{4nt40_45}). The stacking disordered samples can be recognised in SAXS data as mixtures of DC and HD signals (Figure~\ref{clathrate_figure2}M, supplementary text). 

At Mg$^{2+}$ concentrations around 40 mM, HD single crystals appeared, for which only few isolated staggered layers could be identified in the SEM images (Fig.~\ref{clathrate_figure2}G-I, Fig.~\ref{Hex}). The HD structure demands an exact stoichiometric ratio and arrangement of bond configurations on each tetrapod (three staggered bonds in the $\{100\}$ plane and a single eclipsed bond in the $[001]$ direction). Possibly, local variations in crystallization conditions shift the balance of bond competition in our system from the relative narrow range that is needed for pure HD structures. As a result, samples with pure single HD crystals also contained stacking-disordered crystals, as well as sII clathrate crystals.


\begin{figure}
	\centering 
	\includegraphics[width=0.9\textwidth]{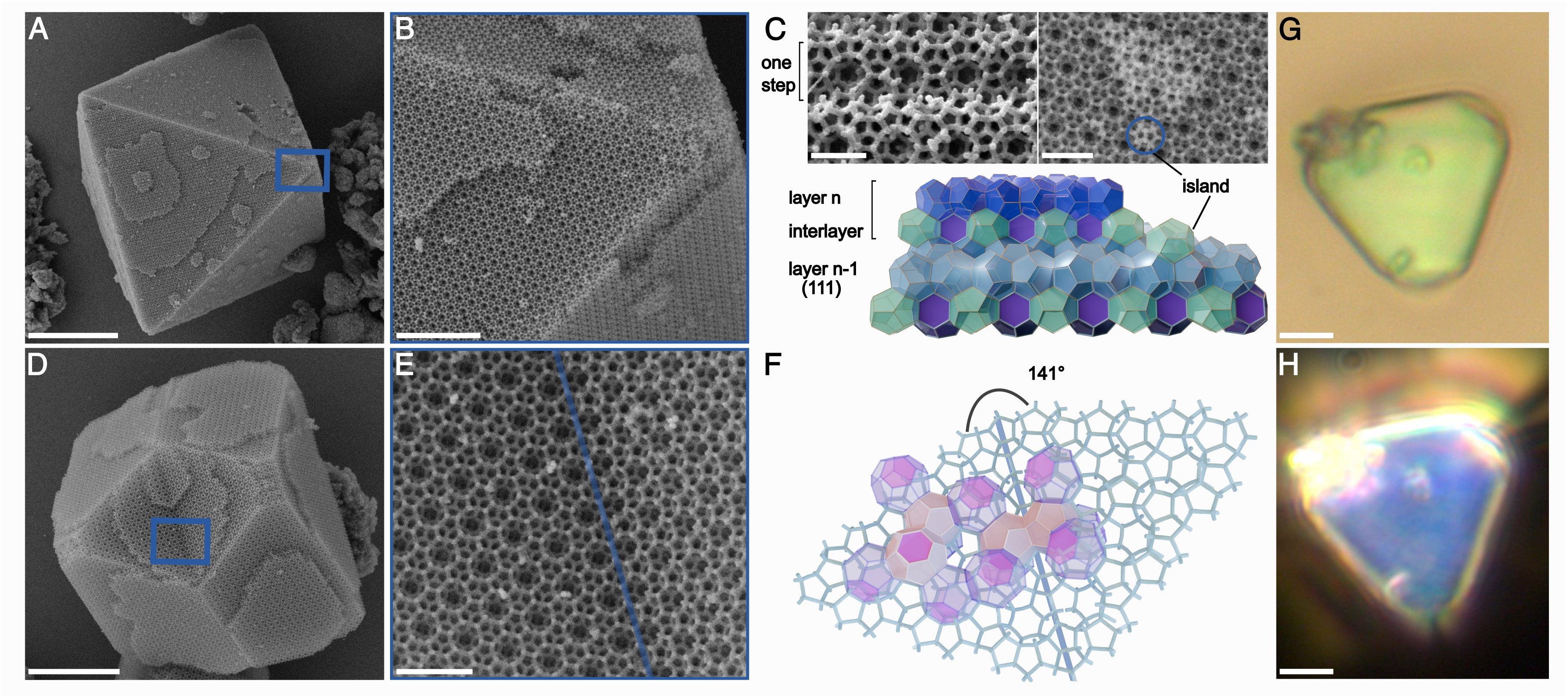} 
		\caption{\textbf{Step edges and twinning in sII clathrate.}
		(\textbf{A, B}) SEM images of a sII clathrate single crystal with octahedral morphology displaying multiple edge steps and islands of pentagonal dodecahedra cages. (\textbf{C}) Top left: SEM image of one edge step; top right: SEM image of one surface step and one island. Bottom: Model of the edge step and island. One step consists of two layers of pentagonal dodecahedra cages and one island appears as one isolated pentagonal dodecahedra cage. (\textbf{D, E}) SEM images of an sII clathrate single crystal with one twinning plane. (\textbf{F}) Model showing a pair of twinned surface (111) planes and the arrangement of hexakaidecahedral cages with highlighted hexagonal contact areas (magenta).  Conceptually, these areas can fuse such that they form a staggered (left, everywhere in the bulk region) or eclipsed (right, only in the twinning plane) configuration. The model emphasizes selected cavities for clarity. (\textbf{G}) Optical bright field microscope image of sII clathrate single crystal showing green structural color. (\textbf{H}) Optical dark field microscope image of sII clathrate single crystal showing blue structural color. Scale bars: \textbf{A}: 10 $\mu$m, \textbf{B}: 2 $\mu$m, \textbf{C}: 200 nm,  \textbf{D}, \textbf{G}, \textbf{H}: 5 $\mu$m, \textbf{E}: 4 $\mu$m, \textbf{F}: 500 nm.}
	\label{clathrate_figure3}
\end{figure}

We were able to isolate sII clathrate structures by strengthening the eclipsed binding mode even further. We achieved this by "turning off" the binding mode (i), so that eclipsed poly(G) to poly(C)-mediated bonds between the two different types of tetrapods dominated. This lead to robust and high-quality sII clathrate growth across a wide range of conditions, and at Mg concentrations below 40 mM samples contained exclusively sII clathrate crystals (Figure~\ref{clathrate_figure2}J-L, Figure~\ref{4nt_7T26}, Figure~\ref{4nt_7T34}).  This structure has a unit cell composed of 136 tetrapods, arranged into 20-monomer pentagonal dodecahedra [5$^{12}$], and 28-monomer hexakaidecahedral cages [6$^4$5$^{12}$], composed of 4 hexagonal and 12 pentagonal sides. The hexagons in the large cages are arranged with tetrahedral symmetry, and pairs of hexakaidecahedra connect by sharing hexagonal sides to form a diamond cubic structure, where the smaller cages fill the space between them (Figure~\ref{clathrate_figure2}K,L). Based on the designed 70 nm center-to-center distance between two neighboring tetrapods, the clathrate unit cell has a size of 440 nm, matching the SAXS measurement of 445 $\pm$ 3 nm(Figure~\ref{saxssimulation}). The designed tetrahedral angle of 109.5° between the legs of the tetrapods closely matches the 108° internal angles in pentagons, however, the hexagons require an angle of 120$^\circ$, deforming the DNA origami tetrapods (Figure~\ref{deformed_tetrapod}). Apparently, the lowest energy state of the clathrate structure here overcompensates this deformation energy. The additional deformation in the hexakaidecahedra likely contributes to the fact that we only observed isolated dodecahedra on well-formed crystal surfaces, but never the hexakaidecahedra (Figures~\ref{clathrate_figure3}C and \ref{clathrate_detail}). Interestingly, in the pentagons a stringent alternation between mA and mB monomers is impossible, thus every pentagon contains at least one eclipsed bond between equal monomers, where, based on the patch design, a staggered configuration should occur. We hypothesize that this mismatched bond is made possible because of either the flexibility of all involved binding strands facilitating the necessary rotation; either the formation of G-quadruplexes due to the high guanine content in the binding extensions and the relatively high concentration of Mg$^{2+}$; or binding of the poly(G) extensions to the poly(C) extensions, which are not located on the outer ring of attachment sites. 
 
Despite the large unit cell, the clathrate crystals form well-defined crystal planes, with only a few steps across relatively large facets (Figure~\ref{clathrate_figure3}A,B, Figure~\ref{clathrate_detail}, Figure~\ref{clathrate}). The steps comprise two layers of pentagonal dodecahedra (Figure~\ref{clathrate_figure3}C), corresponding to an addition of one full hexadecahedral cage. With our well-thermalized samples, individual full dodecahedral cages could form on an otherwise well-defined crystal surface as multi-monomer islands, which are analogues of adatoms. Notably, we also observed twinning in octahedral clathrate crystals (Figure~\ref{clathrate_figure3}D,E). This is a consequence of the tetrahedral arrangement of the four hexagons in the hexakaidecahedral cages, which is analogous to the orientation of tetrapod arms. Here, two neighboring hexakaidecahedral cages can be arranged in either the staggered or in the eclipsed configuration, just as in the diamond structure (Figure~\ref{clathrate_figure3}F). When all the hexakaidecahedra are in the staggered configuration, this leads to their diamond cubic arrangement, e.g. the sII clathrate structure. However, when a plane of eclipsed configurations is nucleated, this manifests as a mirror plane. An interesting distinction between twinning here and in diamond crystals, is that here the twinning occurs because of the arrangement of the hexagons on the hexakaidecahedral cages, and not because of the configuration of the bonds between neighboring monomers - in both cases all the tetrapod bonds are in the eclipsed configuration. Even though both configurations could accommodate the additional space-filling dodecahedral cages, we speculate that the staggered configuration of hexakaidecahedra causes lower overall deformation, leading to stable growth of the sII clathrate structure without stacking disorder.


As the unit cell in sII clathrate crystals contains far more monomers than that of the DC structure – 136 compared to 8  – it is also much larger: 440 nm vs. 170 nm, respectively (Figure~\ref{clathrate_figure2}E). Most prominently, the $\{111\}$ crystal planes of the diamond lattice of the hexakaidecahedral cages in the clathrate structure are $440$ nm $ /  \sqrt{3}=254$ nm apart, with constructive interference expected for light with wavelengths twice this distance. Consequently, we observed colorful structural reflections both in bright field and dark field microscopy images of silicified clathrate DNA origami crystals in air (Figure~\ref{clathrate_figure3}G and H). 

\begin{figure}
	\centering
	\includegraphics[width=0.6\textwidth]{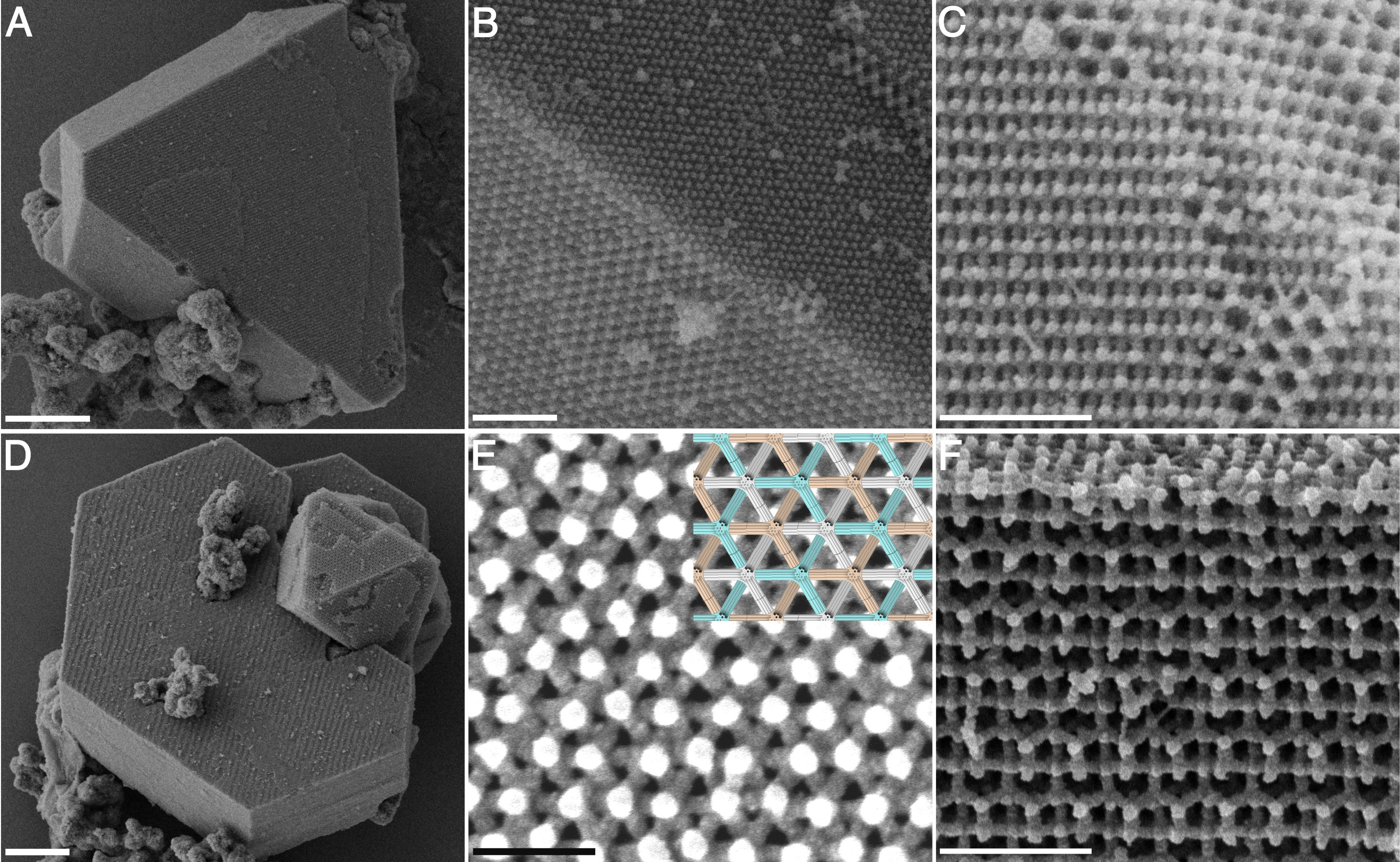}
	\caption{\textbf{Triple diamond.}
		At magnesium concentrations above 40 mM, growth of interpenetrating crystals is facilitated. (\textbf{A, B, C}) SEM images of triple diamond cubic structures. (\textbf{D, E, F}) SEM images of triple hexagonal diamond structures. (\textbf{E})  SEM image of the \{111\} facet of a triple diamond cubic crystal. (\textbf{F})  Corresponding model with the three interpenetrating lattices colored in gray, brown and turquoise. Scale bars: (\textbf{A, D}) 5 $\mu$m; (\textbf{B, C, F}) 500 nm; (\textbf{E}) 200 nm.}
	\label{clathrate_figure4}
\end{figure}

\subsection*{Triple Diamond}
At Mg$^{2+}$ concentrations above 40 mM, we observed the emergence of triple interpenetrating lattices in both cubic and hexagonal diamond structures (Figures~\ref{clathrate_figure4}, \ref{4nt40_45}B, \ref{DB4nt3G}, \ref{4nt50_55}, \ref{4nt60_70}, \ref{4nt_7T45}, \ref{4nt_7T55}, \ref{4nt_7T65}). Because of the reduced electrostatic repulsion between the tetrapods in the presence of divalent cations, additional monomers can occupy interstitial positions in the lattices. Within a single diamond lattice, a second and a third lattices penetrate the structure with 1/3 and 2/3 lateral offsets along the $[1 1 \overline{2}]$ direction for diamond cubic, and along the $[ 1 \overline{1} 0]$ direction for the hexagonal lattice (Figure~\ref{triple_side}), forming a characteristically dense pattern in the $\{111\}$ (DC; Figure~\ref{clathrate_figure4}B) and $(001)$ planes (HD; Figure~\ref{clathrate_figure4}E).  The side surfaces of the triple diamond crystals have a stripped appearance, corresponding to the denser crystal planes (Figure~\ref{clathrate_figure4}C and F). The additional lattices seem to affect the nucleation and stability of the $\{111\}$, and $(001)$ crystal planes for the DC and HD structures, respectively. In the case of DC, this can be seen in Fig.~\ref{clathrate_figure4}A, where the $\{111\}$ plane with the triple lattice changes shape from triangular to triangular with truncated corners, extending the vertices into edges. The other, less dense $\{111\}$ planes are shown in Figure~\ref{clathrate_figure4}B,C. In the case of HD, the morphology of the crystals remained hexagonal, but the aspect ratio (width vs. thickness) increased (Figures~\ref{clathrate_figure4}D, \ref{4nt60_70}, \ref{triple_side} and \ref{triple_inside}), hinting at reduced nucleation rate of $(001)$ crystal planes in this regime and consequential preferred growth in the $\langle 100 \rangle$ directions. Similarly to the single lattices at lower Mg$^{2+}$ concentrations, the triple HD structures were mostly stacking-disorder HD/DC mixtures (Figure~\ref{triple_side}E, Figure~\ref{tripleHD_side}). Figure~\ref{saxssimulation_1} shows SAXS data comparing the single and interpenetrating DC and HD lattices and displays the lattice indices (supplementary text).

\begin{figure}
	\centering
	\includegraphics[width=0.6\textwidth]{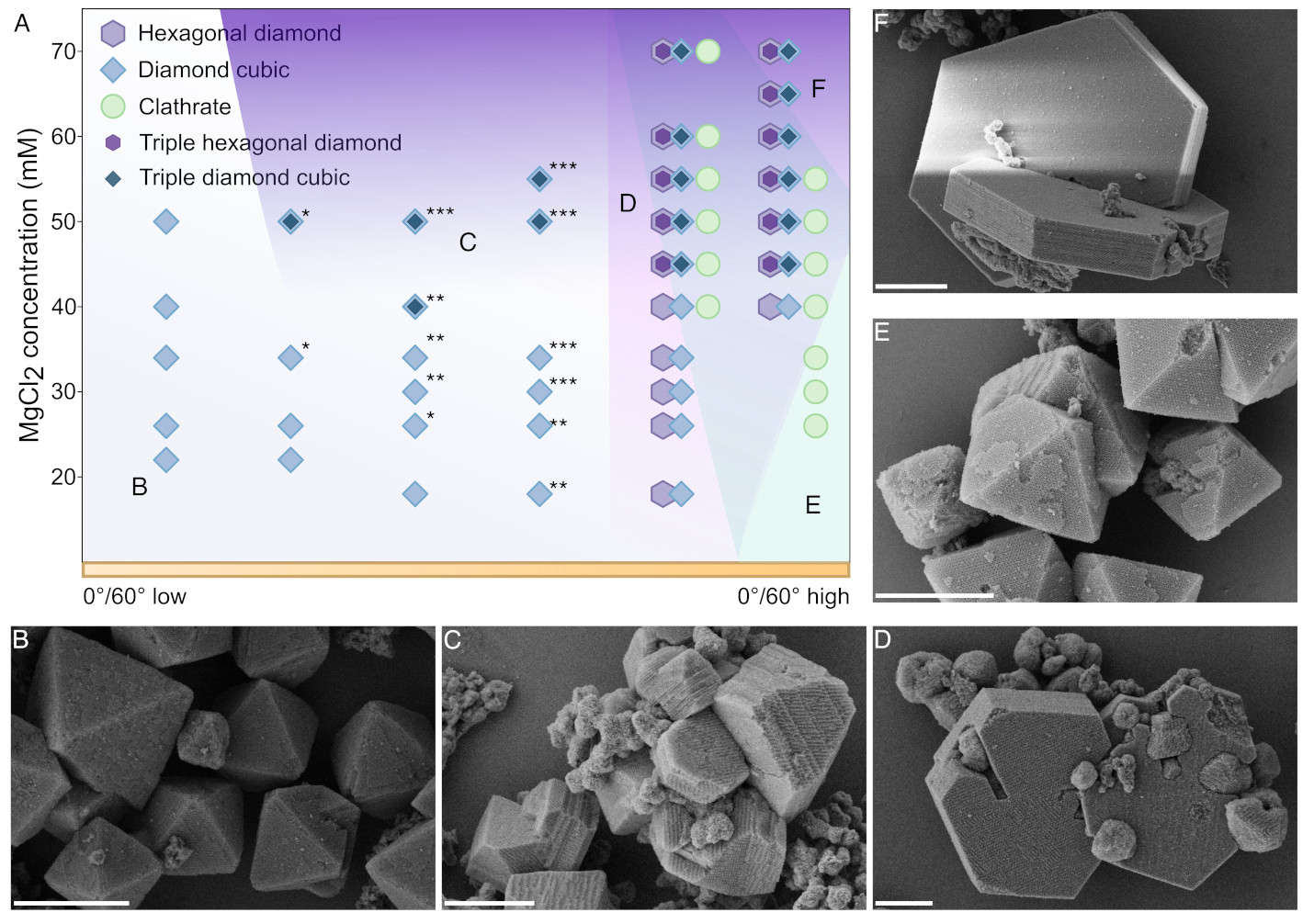} 

	\caption{\textbf{Phase diagram of lattice structures achievable with origami tetrapods.}
		(\textbf{A}) Occurrence of structures, plotted over the configuration bias (eclipsed/staggered, x-axis), and the MgCl$_2$ concentration (y-axis). Each blue diamond, purple hexagon or green circle denotes the observation of diamond cubic, hexagonal diamond, or sII clathrates, respectively, at the indicated conditions. Touching diamond and hexagonal symbols correspond to stacking-disordered mixture of DC and HD. Dark filling of the diamonds and hexagons points to the occurrence of triple lattices. Asterisks denote the amount of twinning in the DC crystals: * for most crystals with single twinning plane, ** for multiple twinning planes in most crystals, and *** for multiple twinning planes in the same direction, leading to crystals with hexagonal morphology. (\textbf{B-F}) SEM images of typical structures for different areas of the phase diagram: (\textbf{B}) Pure diamond cubic crystals,  (\textbf{C}) triple DC with multiple twinning planes, (\textbf{D}) triple HD, mixed with octahedral sII clathrate crystals, (\textbf{E}) pure sII clathrate, and (\textbf{F}) triple HD. Scale bars: 10 $\mu$m.}
	\label{clathrate_figure5}
\end{figure}

\subsection*{Phase Diagram}
We summarize the occurrence of the different crystal structures with respect to the relative configuration strength (x-axis), and Mg$^{2+}$ concentration (y-axis) in the phase diagram shown in Figure~\ref{clathrate_figure5} (a diagram with extended annotations is shown in Figure~\ref{phasediagram_samples}). By changing the relative binding configuration strength, we can tune the structures from pure DC on the left side of the diagram, where the staggered connections are dominant, all the way to pure sII clathrate structures on the right side of the diagram, where eclipsed configurations are preferred. In the intermediate bonding regime and at moderate Mg$^{2+}$ concentrations in the lower central part of the diagram, the otherwise pure DC structures exhibit multiple twinning planes, leading to irregular shapes of the crystals. Further to the right, the balance between staggered and eclipsed configurations enables formation of HD, however, mostly mixed with DC in stacking-disordered structures, with the occurrence of high-purity HD single-crystals. At higher Mg$^{2+}$ concentrations, the DC and HD structures form triple lattices, leading to changed morphologies of the crystals. 
Next to confirming the crystal structures, we were also able to estimate the fractions of the different allotropes by comparing the SAXS scattering intensities of the samples (Figure~\ref{clathrate_figure2}M and Figures~\ref{4nt3G_fraction}, \ref{4nt7_fraction}, \ref{34mM_fraction}, \ref{50mM_fraction}).

\subsection*{Discussion}


We have established a versatile crystallization system in which we direct the assembly of tetrahedral patchy particles through a rich phase diagram of crystalline allotropes. We achieve this with competing binding modes between DNA origami tetrapods. In our approach, all four arms of the tetrapod are equivalent and therefore offer an entropic advantage compared to assemblies with poly-chromatic patches, where multiple unique binding assignments are used.  The equivalent interaction patches also mimic the crystallization behavior at the atomic scale, where non-binding electron orbitals influence the assembly and stability of different structures. This enabled us to produce pure DC and sII clathrate structures, as well as capture the stacking-disordered mixtures of DC and HD structures. While we were able to produce large single crystals of hexagonal diamond, such samples contained admixtures of the other phases. This is not surprising, as the hexagonal diamond structure demands an exact balance and distribution of staggered and eclipsed bonds, and local variation in conditions can easily shift the balance either way. In this regime, the type of the next nucleated layer (either staggered or eclipsed) is not deterministic, but once a layer of a certain type is nucleated, it will grow efficiently, leading to crystals with large lateral dimensions. For the DC and sII clathrate structures the situation is more favorable, as they are formed exclusively from one type of binding configuration in a wide range of conditions, leading to efficient and robust crystal growth, as seen from well-defined crystal facets with very few steps. 

We showcase structural color emerging from the clathrate crystals, illustrating the potential of a rationally designed and self-assembling biomaterial that is transparent across the entire visible spectrum and here exhibits directional reflections. The rod-connected structure of the DNA origami tetrapod-assembled crystals provides a low volume fill fraction that is suitable for further addition of materials with different chemical and physical properties\cite{ermatov_fabrication_2025}, as well as an ideal geometry for achieving wide photonic band gaps. We envision, that with the addition of high-refractive index coatings, the large periodicity of these assemblies will enable photonic band engineering in the visible range.




%
\bibliography{clathrate_sci} 

@article{he_colloidal_2020,
	title = {Colloidal diamond},
	volume = {585},
	copyright = {2020 The Author(s), under exclusive licence to Springer Nature Limited},
	issn = {1476-4687},
	url = {https://www.nature.com/articles/s41586-020-2718-6},
	doi = {10.1038/s41586-020-2718-6},
	language = {en},
	number = {7826},
	urldate = {2022-09-14},
	journal = {Nature},
	author = {He, Mingxin and Gales, Johnathon P. and Ducrot, Étienne and Gong, Zhe and Yi, Gi-Ra and Sacanna, Stefano and Pine, David J.},
	month = sep,
	year = {2020},
	pages = {524--529},
}

@article{ijas_label-free_2022,
	title = {A label-free light-scattering method to resolve assembly and disassembly of {DNA} nanostructures},
	volume = {121},
	issn = {0006-3495},
	url = {https://www.sciencedirect.com/science/article/pii/S0006349522008748},
	doi = {10.1016/j.bpj.2022.10.036},
	language = {en},
	number = {24},
	urldate = {2023-05-04},
	journal = {Biophysical Journal},
	author = {Ijäs, Heini and Liedl, Tim and Linko, Veikko and Posnjak, Gregor},
	month = dec,
	year = {2022},
	pages = {4800--4809},
}

@article{hueckel_ionic_2020,
	title = {Ionic solids from common colloids},
	volume = {580},
	copyright = {2020 The Author(s), under exclusive licence to Springer Nature Limited},
	issn = {1476-4687},
	url = {https://www.nature.com/articles/s41586-020-2205-0},
	doi = {10.1038/s41586-020-2205-0},
	language = {en},
	number = {7804},
	urldate = {2023-07-23},
	journal = {Nature},
	publisher = {Nature Publishing Group},
	author = {Hueckel, Theodore and Hocky, Glen M. and Palacci, Jeremie and Sacanna, Stefano},
	month = apr,
	year = {2020},
	note = {Number: 7804},
	pages = {487--490},
}

@article{stahl_facile_2014,
	title = {Facile and {Scalable} {Preparation} of {Pure} and {Dense} {DNA} {Origami} {Solutions}},
	volume = {53},
	copyright = {© 2014 The Authors. Published by Wiley-VCH Verlag GmbH \& Co. KGaA. This is an open access article under the terms of the Creative Commons Attribution Non-Commercial License, which permits use, distribution and reproduction in any medium, provided the original work is properly cited and is not used for commercial purposes},
	issn = {1521-3773},
	url = {https://onlinelibrary.wiley.com/doi/abs/10.1002/anie.201405991},
	doi = {10.1002/anie.201405991},
	language = {en},
	number = {47},
	urldate = {2023-08-02},
	journal = {Angewandte Chemie International Edition},
	author = {Stahl, Evi and Martin, Thomas G. and Praetorius, Florian and Dietz, Hendrik},
	year = {2014},
	pages = {12735--12740},
}

@article{rovigatti_simple_2022,
	title = {A simple solution to the problem of self-assembling cubic diamond crystals},
	volume = {14},
	url = {https://pubs.rsc.org/en/content/articlelanding/2022/nr/d2nr03533b},
	doi = {10.1039/D2NR03533B},
	language = {en},
	number = {38},
	urldate = {2023-08-15},
	journal = {Nanoscale},
	author = {Rovigatti, Lorenzo and Russo, John and Romano, Flavio and Matthies, Michael and Kroc, Lukáš and Šulc, Petr},
	year = {2022},
	pages = {14268--14275},
}

@article{tian_ordered_2020,
	title = {Ordered three-dimensional nanomaterials using {DNA}-prescribed and valence-controlled material voxels},
	volume = {19},
	copyright = {2020 This is a U.S. government work and not under copyright protection in the U.S.; foreign copyright protection may apply},
	issn = {1476-4660},
	url = {https://www.nature.com/articles/s41563-019-0550-x},
	doi = {10.1038/s41563-019-0550-x},
	language = {en},
	number = {7},
	urldate = {2023-08-15},
	journal = {Nature Materials},
	author = {Tian, Ye and Lhermitte, Julien R. and Bai, Lin and Vo, Thi and Xin, Huolin L. and Li, Huilin and Li, Ruipeng and Fukuto, Masafumi and Yager, Kevin G. and Kahn, Jason S. and Xiong, Yan and Minevich, Brian and Kumar, Sanat K. and Gang, Oleg},
	month = jul,
	year = {2020},
	pages = {789--796},
}

@article{majewski_resilient_2021,
	title = {Resilient three-dimensional ordered architectures assembled from nanoparticles by {DNA}},
	volume = {7},
	url = {https://www.science.org/doi/full/10.1126/sciadv.abf0617},
	doi = {10.1126/sciadv.abf0617},
	number = {12},
	urldate = {2023-08-15},
	journal = {Science Advances},
	author = {Majewski, Pawel W. and Michelson, Aaron and Cordeiro, Marco A. L. and Tian, Cheng and Ma, Chunli and Kisslinger, Kim and Tian, Ye and Liu, Wenyan and Stach, Eric A. and Yager, Kevin G. and Gang, Oleg},
	month = mar,
	year = {2021},
	pages = {eabf0617},
}

@article{dietz_folding_2009,
	title = {Folding {DNA} into {Twisted} and {Curved} {Nanoscale} {Shapes}},
	volume = {325},
	url = {https://www.science.org/doi/full/10.1126/science.1174251},
	doi = {10.1126/science.1174251},
	number = {5941},
	urldate = {2023-08-15},
	journal = {Science},
	author = {Dietz, Hendrik and Douglas, Shawn M. and Shih, William M.},
	month = aug,
	year = {2009},
	pages = {725--730},
}

@article{nguyen_dna-origami-templated_2019,
	title = {{DNA}-{Origami}-{Templated} {Silica} {Growth} by {Sol}–{Gel} {Chemistry}},
	volume = {58},
	issn = {1521-3773},
	url = {https://onlinelibrary.wiley.com/doi/abs/10.1002/anie.201811323},
	doi = {10.1002/anie.201811323},
	language = {en},
	number = {3},
	urldate = {2023-08-15},
	journal = {Angewandte Chemie International Edition},
	author = {Nguyen, Linh and Döblinger, Markus and Liedl, Tim and Heuer-Jungemann, Amelie},
	year = {2019},
	pages = {912--916},
}

@article{douglas_rapid_2009,
	title = {Rapid prototyping of {3D} {DNA}-origami shapes with {caDNAno}},
	volume = {37},
	issn = {0305-1048},
	url = {https://doi.org/10.1093/nar/gkp436},
	doi = {10.1093/nar/gkp436},
	number = {15},
	urldate = {2023-08-15},
	journal = {Nucleic Acids Research},
	author = {Douglas, Shawn M. and Marblestone, Adam H. and Teerapittayanon, Surat and Vazquez, Alejandro and Church, George M. and Shih, William M.},
	month = aug,
	year = {2009},
	pages = {5001--5006},
}

@article{zhang_3d_2018,
	title = {{3D} {DNA} origami crystals},
	volume = {30},
	issn = {0935-9648},
	number = {28},
	journal = {Advanced Materials},
	author = {Zhang, Tao and Hartl, Caroline and Frank, Kilian and Heuer‐Jungemann, Amelie and Fischer, Stefan and Nickels, Philipp C and Nickel, Bert and Liedl, Tim},
	year = {2018},
	pages = {1800273},
}

@article{michelson_three-dimensional_2022,
	title = {Three-dimensional visualization of nanoparticle lattices and multimaterial frameworks},
	volume = {376},
	issn = {0036-8075},
	number = {6589},
	journal = {Science},
	author = {Michelson, Aaron and Minevich, Brian and Emamy, Hamed and Huang, Xiaojing and Chu, Yong S and Yan, Hanfei and Gang, Oleg},
	year = {2022},
	pages = {203--207},
}

@article{zhang_self-assembly_2005,
	title = {Self-{Assembly} of {Patchy} {Particles} into {Diamond} {Structures} through {Molecular} {Mimicry}},
	volume = {21},
	issn = {0743-7463},
	url = {https://doi.org/10.1021/la0513611},
	doi = {10.1021/la0513611},
	number = {25},
	urldate = {2023-08-16},
	journal = {Langmuir},
	author = {{Zhang} and Keys, Aaron S. and Chen, Ting and Glotzer, Sharon C.},
	month = dec,
	year = {2005},
	pages = {11547--11551},
}

@article{romano_crystallization_2011,
	title = {Crystallization of tetrahedral patchy particles in silico},
	volume = {134},
	issn = {0021-9606},
	url = {https://doi.org/10.1063/1.3578182},
	doi = {10.1063/1.3578182},
	number = {17},
	urldate = {2023-08-16},
	journal = {The Journal of Chemical Physics},
	author = {Romano, Flavio and Sanz, Eduardo and Sciortino, Francesco},
	month = may,
	year = {2011},
	pages = {174502},
}

@article{wang_colloids_2012,
	title = {Colloids with valence and specific directional bonding},
	volume = {491},
	copyright = {2012 Springer Nature Limited},
	issn = {1476-4687},
	url = {https://www.nature.com/articles/nature11564},
	doi = {10.1038/nature11564},
	language = {en},
	number = {7422},
	urldate = {2023-08-16},
	journal = {Nature},
	author = {Wang, Yufeng and Wang, Yu and Breed, Dana R. and Manoharan, Vinothan N. and Feng, Lang and Hollingsworth, Andrew D. and Weck, Marcus and Pine, David J.},
	month = nov,
	year = {2012},
	pages = {51--55},
}

@article{macfarlane_nanoparticle_2011,
	title = {Nanoparticle {Superlattice} {Engineering} with {DNA}},
	volume = {334},
	url = {https://www.science.org/doi/10.1126/science.1210493},
	doi = {10.1126/science.1210493},
	number = {6053},
	urldate = {2023-08-18},
	journal = {Science},
	author = {Macfarlane, Robert J. and Lee, Byeongdu and Jones, Matthew R. and Harris, Nadine and Schatz, George C. and Mirkin, Chad A.},
	month = oct,
	year = {2011},
	pages = {204--208},
}

@article{lin_clathrate_2017,
	title = {Clathrate colloidal crystals},
	volume = {355},
	url = {https://www.science.org/doi/10.1126/science.aal3919},
	doi = {10.1126/science.aal3919},
	number = {6328},
	urldate = {2023-08-18},
	journal = {Science},
	publisher = {American Association for the Advancement of Science},
	author = {Lin, Haixin and Lee, Sangmin and Sun, Lin and Spellings, Matthew and Engel, Michael and Glotzer, Sharon C. and Mirkin, Chad A.},
	month = mar,
	year = {2017},
	pages = {931--935},
}

@article{joannopoulos_photonic_1997,
	title = {Photonic crystals: putting a new twist on light},
	volume = {386},
	copyright = {1997 Springer Nature Limited},
	issn = {1476-4687},
	shorttitle = {Photonic crystals},
	url = {https://www.nature.com/articles/386143a0},
	doi = {10.1038/386143a0},
	language = {en},
	number = {6621},
	urldate = {2023-08-24},
	journal = {Nature},
	author = {Joannopoulos, J. D. and Villeneuve, Pierre R. and Fan, Shanhui},
	month = mar,
	year = {1997},
	pages = {143--149},
}

@article{wang_dna_2021,
	title = {{DNA} origami single crystals with {Wulff} shapes},
	volume = {12},
	copyright = {2021 The Author(s)},
	issn = {2041-1723},
	url = {https://www.nature.com/articles/s41467-021-23332-4},
	doi = {10.1038/s41467-021-23332-4},
	language = {en},
	number = {1},
	urldate = {2023-11-01},
	journal = {Nature Communications},
	publisher = {Nature Publishing Group},
	author = {Wang, Yong and Dai, Lizhi and Ding, Zhiyuan and Ji, Min and Liu, Jiliang and Xing, Hang and Liu, Xiaoguo and Ke, Yonggang and Fan, Chunhai and Wang, Peng and Tian, Ye},
	month = may,
	year = {2021},
	note = {Number: 1},
	pages = {3011},
}

@article{posnjak_diamond-lattice_2024,
	title = {Diamond-lattice photonic crystals assembled from {DNA} origami},
	volume = {384},
	url = {https://www.science.org/doi/10.1126/science.adl2733},
	doi = {10.1126/science.adl2733},
	number = {6697},
	urldate = {2024-06-07},
	journal = {Science},
	publisher = {American Association for the Advancement of Science},
	author = {Posnjak, Gregor and Yin, Xin and Butler, Paul and Bienek, Oliver and Dass, Mihir and Lee, Seungwoo and Sharp, Ian D. and Liedl, Tim},
	month = may,
	year = {2024},
	pages = {781--785},
}

@article{liu_inverse_2024,
	title = {Inverse design of a pyrochlore lattice of {DNA} origami through model-driven experiments},
	volume = {384},
	url = {https://www.science.org/doi/abs/10.1126/science.adl5549},
	doi = {10.1126/science.adl5549},
	number = {6697},
	urldate = {2024-07-23},
	journal = {Science},
	publisher = {American Association for the Advancement of Science},
	author = {Liu, Hao and Matthies, Michael and Russo, John and Rovigatti, Lorenzo and Narayanan, Raghu Pradeep and Diep, Thong and McKeen, Daniel and Gang, Oleg and Stephanopoulos, Nicholas and Sciortino, Francesco and Yan, Hao and Romano, Flavio and Šulc, Petr},
	month = may,
	year = {2024},
	pages = {776--781},
}

@article{dai_programming_2023,
	title = {Programming the morphology of {DNA} origami crystals by magnesium ion strength},
	volume = {120},
	url = {https://www.pnas.org/doi/full/10.1073/pnas.2302142120},
	doi = {10.1073/pnas.2302142120},
	number = {28},
	urldate = {2024-09-17},
	journal = {Proceedings of the National Academy of Sciences},
	publisher = {Proceedings of the National Academy of Sciences},
	author = {Dai, Lizhi and Hu, Xiaoxue and Ji, Min and Ma, Ningning and Xing, Hang and Zhu, Jun-Jie and Min, Qianhao and Tian, Ye},
	month = jul,
	year = {2023},
	pages = {e2302142120},
}

@article{zhou_colloidal_2024,
	title = {Colloidal quasicrystals engineered with {DNA}},
	volume = {23},
	copyright = {2023 The Author(s), under exclusive licence to Springer Nature Limited},
	issn = {1476-4660},
	url = {https://www.nature.com/articles/s41563-023-01706-x},
	doi = {10.1038/s41563-023-01706-x},
	language = {en},
	number = {3},
	urldate = {2024-09-20},
	journal = {Nature Materials},
	publisher = {Nature Publishing Group},
	author = {Zhou, Wenjie and Lim, Yein and Lin, Haixin and Lee, Sangmin and Li, Yuanwei and Huang, Ziyin and Du, Jingshan S. and Lee, Byeongdu and Wang, Shunzhi and Sánchez-Iglesias, Ana and Grzelczak, Marek and Liz-Marzán, Luis M. and Glotzer, Sharon C. and Mirkin, Chad A.},
	month = mar,
	year = {2024},
	pages = {424--428},
}

@article{tarasewicz_self-assembly_2025,
	title = {Self-assembly of chromatic patchy particles with tetrahedrally arranged patches},
	volume = {21},
	issn = {1744-6848},
	url = {https://pubs.rsc.org/en/content/articlelanding/2025/sm/d4sm01210k},
	doi = {10.1039/D4SM01210K},
	language = {en},
	urldate = {2025-02-05},
	journal = {Soft Matter},
	publisher = {The Royal Society of Chemistry},
	author = {Tarasewicz, Dariusz and Raczyłło, Edyta and Rżysko, Wojciech and Baran, Lukasz},
	month = jan,
	year = {2025},
	pages = {1203--1211},
}

@article{noya_assembly_2019,
	title = {Assembly of clathrates from tetrahedral patchy colloids with narrow patches},
	volume = {151},
	issn = {0021-9606},
	url = {https://doi.org/10.1063/1.5109382},
	doi = {10.1063/1.5109382},
	number = {9},
	urldate = {2025-03-26},
	journal = {The Journal of Chemical Physics},
	author = {Noya, Eva G. and Zubieta, Itziar and Pine, David J. and Sciortino, Francesco},
	month = sep,
	year = {2019},
	pages = {094502},
}

@article{romano_patterning_2012,
	title = {Patterning symmetry in the rational design of colloidal crystals},
	volume = {3},
	copyright = {2012 The Author(s)},
	issn = {2041-1723},
	url = {https://www.nature.com/articles/ncomms1968},
	doi = {10.1038/ncomms1968},
	language = {en},
	number = {1},
	urldate = {2025-03-26},
	journal = {Nature Communications},
	publisher = {Nature Publishing Group},
	author = {Romano, Flavio and Sciortino, Francesco},
	month = jul,
	year = {2012},
	pages = {975},
}

@article{romano_phase_2010,
	title = {Phase diagram of a tetrahedral patchy particle model for different interaction ranges},
	volume = {132},
	issn = {0021-9606},
	url = {https://doi.org/10.1063/1.3393777},
	doi = {10.1063/1.3393777},
	number = {18},
	urldate = {2025-03-26},
	journal = {The Journal of Chemical Physics},
	author = {Romano, Flavio and Sanz, Eduardo and Sciortino, Francesco},
	month = may,
	year = {2010},
	pages = {184501},
}

@article{baran_interplay_2024,
	title = {Interplay between the {Formation} of {Colloidal} {Clathrate} and {Cubic} {Diamond} {Crystals}},
	volume = {128},
	issn = {1520-6106},
	url = {https://doi.org/10.1021/acs.jpcb.4c02456},
	doi = {10.1021/acs.jpcb.4c02456},
	number = {23},
	urldate = {2025-03-26},
	journal = {The Journal of Physical Chemistry B},
	publisher = {American Chemical Society},
	author = {Baran, Lukasz and Tarasewicz, Dariusz and Rżysko, Wojciech},
	month = jun,
	year = {2024},
	pages = {5792--5801},
}

@article{wang_ph-induced_2022,
	title = {{pH}-{Induced} {Symmetry} {Conversion} of {DNA} {Origami} {Lattices}},
	volume = {61},
	copyright = {© 2022 Wiley-VCH GmbH},
	issn = {1521-3773},
	url = {https://onlinelibrary.wiley.com/doi/abs/10.1002/anie.202208290},
	doi = {10.1002/anie.202208290},
	language = {en},
	number = {40},
	urldate = {2025-05-14},
	journal = {Angewandte Chemie International Edition},
	author = {Wang, Yong and Yan, Xuehui and Zhou, Zhaoyu and Ma, Ningning and Tian, Ye},
	year = {2022},
	note = {\_eprint: https://onlinelibrary.wiley.com/doi/pdf/10.1002/anie.202208290},
	pages = {e202208290},
}

@article{ermatov_fabrication_2025,
	title = {Fabrication of {Functional} {3D} {Nanoarchitectures} via {Atomic} {Layer} {Deposition} on {DNA} {Origami} {Crystals}},
	volume = {147},
	issn = {0002-7863},
	url = {https://doi.org/10.1021/jacs.4c17232},
	doi = {10.1021/jacs.4c17232},
	number = {11},
	urldate = {2025-06-05},
	journal = {Journal of the American Chemical Society},
	publisher = {American Chemical Society},
	author = {Ermatov, Arthur and Kost, Melisande and Yin, Xin and Butler, Paul and Dass, Mihir and Sharp, Ian D. and Liedl, Tim and Bein, Thomas and Posnjak, Gregor},
	month = mar,
	year = {2025},
	pages = {9519--9527},
}

@article{kim_patchy_2025,
	title = {Patchy nanoparticles by atomic stencilling},
	volume = {646},
	copyright = {2025 The Author(s)},
	issn = {1476-4687},
	url = {https://www.nature.com/articles/s41586-025-09605-8},
	doi = {10.1038/s41586-025-09605-8},
	language = {en},
	number = {8085},
	urldate = {2026-03-10},
	journal = {Nature},
	publisher = {Nature Publishing Group},
	author = {Kim, Ahyoung and Kim, Chansong and Waltmann, Tommy and Vo, Thi and Kim, Eun Mi and Kim, Junseok and Shao, Yu-Tsun and Michelson, Aaron and Crockett, John R. and Kalutantirige, Falon C. and Yang, Eric and Yao, Lehan and Hwang, Chu-Yun and Zhang, Yugang and Liu, Yu-Shen and An, Hyosung and Gao, Zirui and Kim, Jiyeon and Mandal, Sohini and Muller, David A. and Fichthorn, Kristen A. and Glotzer, Sharon C. and Chen, Qian},
	month = oct,
	year = {2025},
	pages = {592--600},
}

@article{nelson_toward_2002,
	title = {Toward a {Tetravalent} {Chemistry} of {Colloids}},
	volume = {2},
	issn = {1530-6984},
	url = {https://doi.org/10.1021/nl0202096},
	doi = {10.1021/nl0202096},
	number = {10},
	urldate = {2026-05-19},
	journal = {Nano Letters},
	publisher = {American Chemical Society},
	author = {Nelson, David R.},
	month = oct,
	year = {2002},
	pages = {1125--1129},
}

@article{komatsu_ice_2020,
	title = {Ice {Ic} without stacking disorder by evacuating hydrogen from hydrogen hydrate},
	volume = {11},
	copyright = {2020 The Author(s)},
	issn = {2041-1723},
	url = {https://www.nature.com/articles/s41467-020-14346-5},
	doi = {10.1038/s41467-020-14346-5},
	language = {en},
	number = {1},
	urldate = {2026-06-08},
	journal = {Nature Communications},
	publisher = {Nature Publishing Group},
	author = {Komatsu, Kazuki and Machida, Shinichi and Noritake, Fumiya and Hattori, Takanori and Sano-Furukawa, Asami and Yamane, Ryo and Yamashita, Keishiro and Kagi, Hiroyuki},
	month = feb,
	year = {2020},
	pages = {464},
}

@article{falenty_formation_2014,
	title = {Formation and properties of ice {XVI} obtained by emptying a type {sII} clathrate hydrate},
	volume = {516},
	copyright = {2014 Springer Nature Limited},
	issn = {1476-4687},
	url = {https://www.nature.com/articles/nature14014},
	doi = {10.1038/nature14014},
	language = {en},
	number = {7530},
	urldate = {2026-06-08},
	journal = {Nature},
	publisher = {Nature Publishing Group},
	author = {Falenty, Andrzej and Hansen, Thomas C. and Kuhs, Werner F.},
	month = dec,
	year = {2014},
	pages = {231--233},
}

@article{kahn_encoding_2025,
	title = {Encoding hierarchical {3D} architecture through inverse design of programmable bonds},
	volume = {24},
	copyright = {2025 The Author(s)},
	issn = {1476-4660},
	url = {https://www.nature.com/articles/s41563-025-02263-1},
	doi = {10.1038/s41563-025-02263-1},
	language = {en},
	number = {8},
	urldate = {2026-06-16},
	journal = {Nature Materials},
	publisher = {Nature Publishing Group},
	author = {Kahn, Jason S. and Minevich, Brian and Michelson, Aaron and Emamy, Hamed and Wu, Jiahao and Ji, Huajian and Yun, Alexia and Kisslinger, Kim and Xiang, Shuting and Yu, Nanfang and Kumar, Sanat K. and Gang, Oleg},
	month = aug,
	year = {2025},
	pages = {1273--1282},
}

@article{russo_sat-assembly_2022,
	title = {{SAT}-assembly: a new approach for designing self-assembling systems},
	volume = {34},
	issn = {0953-8984},
	shorttitle = {{SAT}-assembly},
	url = {https://doi.org/10.1088/1361-648X/ac5479},
	doi = {10.1088/1361-648X/ac5479},
	language = {en},
	number = {35},
	urldate = {2026-06-16},
	journal = {Journal of Physics: Condensed Matter},
	publisher = {IOP Publishing},
	author = {Russo, John and Romano, Flavio and Kroc, Lukáš and Sciortino, Francesco and Rovigatti, Lorenzo and Šulc, Petr},
	month = jun,
	year = {2022},
	pages = {354002},
}

@article{blanchet_versatile_2015,
	title = {Versatile sample environments and automation for biological solution {X}-ray scattering experiments at the {P12} beamline ({PETRA} {III}, {DESY})},
	volume = {48},
	issn = {1600-5767},
	url = {//journals.iucr.org/paper?ge5013},
	doi = {10.1107/S160057671500254X},
	language = {en},
	number = {2},
	urldate = {2026-06-18},
	journal = {Journal of Applied Crystallography},
	publisher = {International Union of Crystallography},
	author = {Blanchet, C. E. and Spilotros, A. and Schwemmer, F. and Graewert, M. A. and Kikhney, A. and Jeffries, C. M. and Franke, D. and Mark, D. and Zengerle, R. and Cipriani, F. and Fiedler, S. and Roessle, M. and Svergun, D. I.},
	month = apr,
	year = {2015},
	pages = {431--443},
}

@article{haas_new_2023,
	title = {The new small-angle {X}-ray scattering beamline for materials research at {PETRA} {III}: {SAXSMAT} beamline {P62}},
	volume = {30},
	copyright = {https://creativecommons.org/licenses/by/4.0/},
	issn = {1600-5775},
	shorttitle = {The new small-angle {X}-ray scattering beamline for materials research at {PETRA} {III}},
	url = {https://journals.iucr.org/s/issues/2023/06/00/ay5619/},
	doi = {10.1107/S1600577523008603},
	language = {en},
	number = {6},
	urldate = {2026-06-18},
	journal = {Journal of Synchrotron Radiation},
	publisher = {International Union of Crystallography},
	author = {Haas, S. and Sun, X. and Conceição, A. L. C. and Horbach, J. and Pfeffer, S.},
	month = nov,
	year = {2023},
	pages = {1156--1167},
}

@article{douglas_dna-nanotube-induced_2007,
	title = {{DNA}-nanotube-induced alignment of membrane proteins for {NMR} structure determination},
	volume = {104},
	issn = {0027-8424, 1091-6490},
	url = {https://pnas.org/doi/full/10.1073/pnas.0700930104},
	doi = {10.1073/pnas.0700930104},
	language = {en},
	number = {16},
	urldate = {2026-06-18},
	journal = {Proceedings of the National Academy of Sciences},
	author = {Douglas, Shawn M. and Chou, James J. and Shih, William M.},
	month = apr,
	year = {2007},
	pages = {6644--6648},
}

@article{guloy_guest-free_2006,
	title = {A guest-free germanium clathrate},
	volume = {443},
	copyright = {2006 Springer Nature Limited},
	issn = {1476-4687},
	url = {https://www.nature.com/articles/nature05145},
	doi = {10.1038/nature05145},
	language = {en},
	number = {7109},
	urldate = {2026-06-18},
	journal = {Nature},
	publisher = {Nature Publishing Group},
	author = {Guloy, Arnold M. and Ramlau, Reiner and Tang, Zhongjia and Schnelle, Walter and Baitinger, Michael and Grin, Yuri},
	month = sep,
	year = {2006},
	pages = {320--323},
}

@article{gryko_low-density_2000,
	title = {Low-density framework form of crystalline silicon with a wide optical band gap},
	volume = {62},
	copyright = {http://link.aps.org/licenses/aps-default-license},
	issn = {0163-1829, 1095-3795},
	url = {https://link.aps.org/doi/10.1103/PhysRevB.62.R7707},
	doi = {10.1103/PhysRevB.62.R7707},
	language = {en},
	number = {12},
	urldate = {2026-06-22},
	journal = {Physical Review B},
	author = {Gryko, Jan and McMillan, Paul F. and Marzke, Robert F. and Ramachandran, Ganesh K. and Patton, Derek and Deb, Sudip K. and Sankey, Otto F.},
	month = sep,
	year = {2000},
	pages = {R7707--R7710},
}

@article{momma_vesta_2011,
	title = {{VESTA} 3 for three-dimensional visualization of crystal, volumetric and morphology data},
	volume = {44},
	issn = {0021-8898},
	url = {https://journals.iucr.org/j/issues/2011/06/00/db5098/},
	doi = {10.1107/S0021889811038970},
	language = {en},
	number = {6},
	urldate = {2026-06-23},
	journal = {Journal of Applied Crystallography},
	publisher = {International Union of Crystallography},
	author = {Momma, K. and Izumi, F.},
	month = dec,
	year = {2011},
	pages = {1272--1276},
}

@article{ying_enhancing_2025,
	title = {Enhancing the hardness of diamond through twin refinement and interlocked twins},
	volume = {4},
	copyright = {2025 The Author(s), under exclusive licence to Springer Nature Limited},
	issn = {2731-0582},
	url = {https://www.nature.com/articles/s44160-024-00707-1},
	doi = {10.1038/s44160-024-00707-1},
	language = {en},
	number = {3},
	urldate = {2026-07-04},
	journal = {Nature Synthesis},
	publisher = {Nature Publishing Group},
	author = {Ying, Pan and Li, Baozhong and Ma, Mengdong and Gao, Yufei and Sun, Rongxin and Li, Zihe and Chen, Shuai and Zhang, Bin and Li, Hefei and Liu, Bing and Sun, Lei and Zhao, Song and Tong, Ke and Hu, Wentao and Pan, Yilong and Tang, Guodong and Yu, Dongli and Zhao, Zhisheng and Xu, Bo and Tian, Yongjun},
	month = mar,
	year = {2025},
	pages = {391--398},
}

@article{zhou_space-tiled_2024,
	title = {Space-tiled colloidal crystals from {DNA}-forced shape-complementary polyhedra pairing},
	volume = {383},
	url = {https://www.science.org/doi/10.1126/science.adj1021},
	doi = {10.1126/science.adj1021},
	number = {6680},
	urldate = {2026-07-05},
	journal = {Science},
	publisher = {American Association for the Advancement of Science},
	author = {Zhou, Wenjie and Li, Yuanwei and Je, Kwanghwi and Vo, Thi and Lin, Haixin and Partridge, Benjamin E. and Huang, Ziyin and Glotzer, Sharon C. and Mirkin, Chad A.},
	month = jan,
	year = {2024},
	pages = {312--319},
}

@article{zhang_general_2013,
	title = {A general approach to {DNA}-programmable atom equivalents},
	volume = {12},
	copyright = {2013 Springer Nature Limited},
	issn = {1476-4660},
	url = {https://www.nature.com/articles/nmat3647},
	doi = {10.1038/nmat3647},
	language = {en},
	number = {8},
	urldate = {2026-07-08},
	journal = {Nature Materials},
	publisher = {Nature Publishing Group},
	author = {Zhang, Chuan and Macfarlane, Robert J. and Young, Kaylie L. and Choi, Chung Hang J. and Hao, Liangliang and Auyeung, Evelyn and Liu, Guoliang and Zhou, Xiaozhu and Mirkin, Chad A.},
	month = aug,
	year = {2013},
	pages = {741--746},
}

@article{jones_programmable_2015,
	title = {Programmable materials and the nature of the {DNA} bond},
	volume = {347},
	url = {https://www.science.org/doi/10.1126/science.1260901},
	doi = {10.1126/science.1260901},
	number = {6224},
	urldate = {2026-07-08},
	journal = {Science},
	publisher = {American Association for the Advancement of Science},
	author = {Jones, Matthew R. and Seeman, Nadrian C. and Mirkin, Chad A.},
	month = feb,
	year = {2015},
	pages = {1260901},
}

@article{hassanpouryouzband_gas_2020,
	title = {Gas hydrates in sustainable chemistry},
	volume = {49},
	issn = {0306-0012},
	url = {https://doi.org/10.1039/c8cs00989a},
	doi = {10.1039/c8cs00989a},
	number = {15},
	urldate = {2026-07-08},
	journal = {Chemical Society Reviews},
	author = {Hassanpouryouzband, Aliakbar and Joonaki, Edris and Vasheghani Farahani, Mehrdad and Takeya, Satoshi and Ruppel, Carolyn and Yang, Jinhai and English, Niall J. and Schicks, Judith M. and Edlmann, Katriona and Mehrabian, Hadi and Aman, Zachary M. and Tohidi, Bahman},
	month = aug,
	year = {2020},
	pages = {5225--5309},
}

@article{lu_complex_2007,
	title = {Complex gas hydrate from the {Cascadia} margin},
	volume = {445},
	copyright = {2006 Springer Nature Limited},
	issn = {1476-4687},
	url = {https://www.nature.com/articles/nature05463},
	doi = {10.1038/nature05463},
	language = {en},
	number = {7125},
	urldate = {2026-07-08},
	journal = {Nature},
	publisher = {Nature Publishing Group},
	author = {Lu, Hailong and Seo, Yu-taek and Lee, Jong-won and Moudrakovski, Igor and Ripmeester, John A. and Chapman, N. Ross and Coffin, Richard B. and Gardner, Graeme and Pohlman, John},
	month = jan,
	year = {2007},
	pages = {303--306},
}

@article{frondel_lonsdaleite_1967,
	title = {Lonsdaleite, a {Hexagonal} {Polymorph} of {Diamond}},
	volume = {214},
	copyright = {1967 Springer Nature Limited},
	issn = {1476-4687},
	url = {https://www.nature.com/articles/214587a0},
	doi = {10.1038/214587a0},
	language = {en},
	number = {5088},
	urldate = {2026-07-08},
	journal = {Nature},
	publisher = {Nature Publishing Group},
	author = {Frondel, Clifford and Marvin, Ursula B.},
	month = may,
	year = {1967},
	pages = {587--589},
}

@article{lai_bulk_2026,
	title = {Bulk hexagonal diamond},
	volume = {651},
	copyright = {2026 The Author(s), under exclusive licence to Springer Nature Limited},
	issn = {1476-4687},
	url = {https://www.nature.com/articles/s41586-026-10212-4},
	doi = {10.1038/s41586-026-10212-4},
	language = {en},
	number = {8106},
	urldate = {2026-07-08},
	journal = {Nature},
	publisher = {Nature Publishing Group},
	author = {Lai, Shoulong and Yang, Xigui and Shi, Jiuyang and Liu, Shijie and Guo, Ying and Yan, Longbin and Zang, Jinhao and Zhang, Zhuangfei and Jia, Qiuhan and Sun, Jian and Cheng, Shaobo and Shan, Chongxin},
	month = mar,
	year = {2026},
	pages = {621--625},
}

@article{yang_synthesis_2025,
	title = {Synthesis of bulk hexagonal diamond},
	volume = {644},
	copyright = {2025 The Author(s), under exclusive licence to Springer Nature Limited},
	issn = {1476-4687},
	url = {https://www.nature.com/articles/s41586-025-09343-x},
	doi = {10.1038/s41586-025-09343-x},
	language = {en},
	number = {8076},
	urldate = {2026-07-08},
	journal = {Nature},
	publisher = {Nature Publishing Group},
	author = {Yang, Liuxiang and Lau, Kah Chun and Zeng, Zhidan and Zhang, Dongzhou and Tang, Hu and Yan, Bingmin and Niu, Guoliang and Gou, Huiyang and Yang, Yanping and Yang, Wenge and Luo, Duan and Mao, Ho-kwang},
	month = aug,
	year = {2025},
	pages = {370--375},
}

@article{tong_structural_2024,
	title = {Structural transition and migration of incoherent twin boundary in diamond},
	volume = {626},
	copyright = {2024 The Author(s), under exclusive licence to Springer Nature Limited},
	issn = {1476-4687},
	url = {https://www.nature.com/articles/s41586-023-06908-6},
	doi = {10.1038/s41586-023-06908-6},
	language = {en},
	number = {7997},
	urldate = {2026-07-08},
	journal = {Nature},
	publisher = {Nature Publishing Group},
	author = {Tong, Ke and Zhang, Xiang and Li, Zihe and Wang, Yanbin and Luo, Kun and Li, Chenming and Jin, Tianye and Chang, Yuqing and Zhao, Song and Wu, Yingju and Gao, Yufei and Li, Baozhong and Gao, Guoying and Zhao, Zhisheng and Wang, Lin and Nie, Anmin and Yu, Dongli and Liu, Zhongyuan and Soldatov, Alexander V. and Hu, Wentao and Xu, Bo and Tian, Yongjun},
	month = feb,
	year = {2024},
	pages = {79--85},
}

@article{huang_tracking_2023,
	title = {Tracking cubic ice at molecular resolution},
	volume = {617},
	copyright = {2023 The Author(s), under exclusive licence to Springer Nature Limited},
	issn = {1476-4687},
	url = {https://www.nature.com/articles/s41586-023-05864-5},
	doi = {10.1038/s41586-023-05864-5},
	language = {en},
	number = {7959},
	urldate = {2026-07-08},
	journal = {Nature},
	publisher = {Nature Publishing Group},
	author = {Huang, Xudan and Wang, Lifen and Liu, Keyang and Liao, Lei and Sun, Huacong and Wang, Jianlin and Tian, Xuezeng and Xu, Zhi and Wang, Wenlong and Liu, Lei and Jiang, Ying and Chen, Ji and Wang, Enge and Bai, Xuedong},
	month = may,
	year = {2023},
	pages = {86--91},
}

@article{lee_multiple_2026,
	title = {Multiple freezing–melting pathways of high-density ice through ice {XXI} phase at room temperature},
	volume = {25},
	copyright = {2025 The Author(s)},
	issn = {1476-4660},
	url = {https://www.nature.com/articles/s41563-025-02364-x},
	doi = {10.1038/s41563-025-02364-x},
	language = {en},
	number = {2},
	urldate = {2026-07-08},
	journal = {Nature Materials},
	publisher = {Nature Publishing Group},
	author = {Lee, Yun-Hee and Kim, Jin Kyun and Kim, Yong-Jae and Kim, Minju and Cho, Yong Chan and Husband, Rachel J. and Strohm, Cornelius and Ehrenreich-Petersen, Emma and Glazyrin, Konstantin and Laurus, Torsten and Graafsma, Heinz and Bauer, Robert P. C. and Lehmkühler, Felix and Appel, Karen and Konôpková, Zuzana and Tang, Minxue and Dwivedi, Anand Prashant and Sztuck-Dambietz, Jolanta and Randolph, Lisa and Buakor, Khachiwan and Humphries, Oliver and Baehtz, Carsten and Eklund, Tobias and Mohrbach, Lisa Katharina and Mondal, Anshuman and Marquardt, Hauke and O’Bannon, Earl Francis and Amann-Winkel, Katrin and Yoo, Choong-Shik and Zastrau, Ulf and Liermann, Hanns-Peter and Nada, Hiroki and Lee, Geun Woo},
	month = feb,
	year = {2026},
	pages = {302--309},
}

@article{salzmann_extent_2015,
	title = {Extent of stacking disorder in diamond},
	volume = {59},
	issn = {0925-9635},
	url = {https://www.sciencedirect.com/science/article/pii/S0925963515300388},
	doi = {10.1016/j.diamond.2015.09.007},
	urldate = {2026-07-08},
	journal = {Diamond and Related Materials},
	author = {Salzmann, Christoph G. and Murray, Benjamin J. and Shephard, Jacob J.},
	month = oct,
	year = {2015},
	pages = {69--72},
}
\bibliographystyle{sciencemag}

\section*{Acknowledgments}

We thank Susanne Kempter, Philipp Altpeter and Dmitri Efetov for laboratory, and clean room support. We greatly acknowledge Deutsches Elektronen-Synchrotron (DESY) P12 BioSAXS EMBL and P62 SAXSMAT beamlines at PETRA III (Hamburg, Germany), and the beamline scientists Clement E. Blanchet, Dmytro Soloviov and Andre Luiz Coelho Conceicao for the great support during the beamtimes. 

Funding: G.P., X.Y., and T.L. acknowledge funding from the ERC consolidator grant “DNA Funs” (project ID: 818635) and the Deutsche Forschungsgemeinschaft (DFG; German Research Foundation)  through the project MagDNA 8426405.  E.K. and B.N. acknowledges financial support from the Bavarian Collaborative Research Project Solar Technologies Go Hybrid (SolTech). T.L. further acknowledges support from the DFG through the cluster of excellence e-conversion 2.0 EXC 2089/1-390776260. 

Author contributions: X.Y, T.L. and G.P. initiated the research. X.Y. and G.P. designed the tetrapod structures and interfaces, conducted the crystallization experiments and performed SEM analysis. X.Y. and E.K. performed the SAXS measurements and analysis supervised by B.N.. Optical images were taken by T.L. and G.P.. X.Y., T.L. and G.P wrote the manuscript, with inputs from all authors. 

Competing interests: None declared. 

Data and materials availability: All data needed to evaluate the conclusions in the paper are present in the paper or the supplementary materials. 

License information: Copyright © 2026 the authors, some rights reserved; exclusive licensee American Association for the Advancement of Science. No claim to original US government works. https://www.science.org/about/science-licenses-journal- article-reuse 

\subsection*{Supplementary materials}
Materials and Methods\\
Supplementary Text\\
Figs. S1 to S33\\
Tables S1 to S4\\
References \textit{(48-\arabic{enumiv})}\\ 


\newpage


\renewcommand{\thefigure}{S\arabic{figure}}
\renewcommand{\thetable}{S\arabic{table}}
\renewcommand{\theequation}{S\arabic{equation}}
\renewcommand{\thepage}{S\arabic{page}}
\setcounter{figure}{0}
\setcounter{table}{0}
\setcounter{equation}{0}
\setcounter{page}{1} 


\begin{center}
\section*{Supplementary Materials for\\ \scititle}

    Xin Yin$^{1}$,
	Ekaterina Kostyurina$^{1}$,
    Bert Nickel$^{1}$,
	Tim Liedl$^{1\ast}$,
    Gregor Posnjak$^{1\ast}$ \\ 
\small$^\ast$Corresponding author. Email: tim.liedl@lmu.de; gregor.posnjak@lmu.de\\

\end{center}

\subsubsection*{This PDF file includes:}
Materials and Methods\\
Supplementary Text\\
Figures S1 to S33\\
Tables S1 to S4\\


\newpage


\subsection*{Materials and Methods}
\begin{figure}
	\centering
	\includegraphics[width=0.9\textwidth]{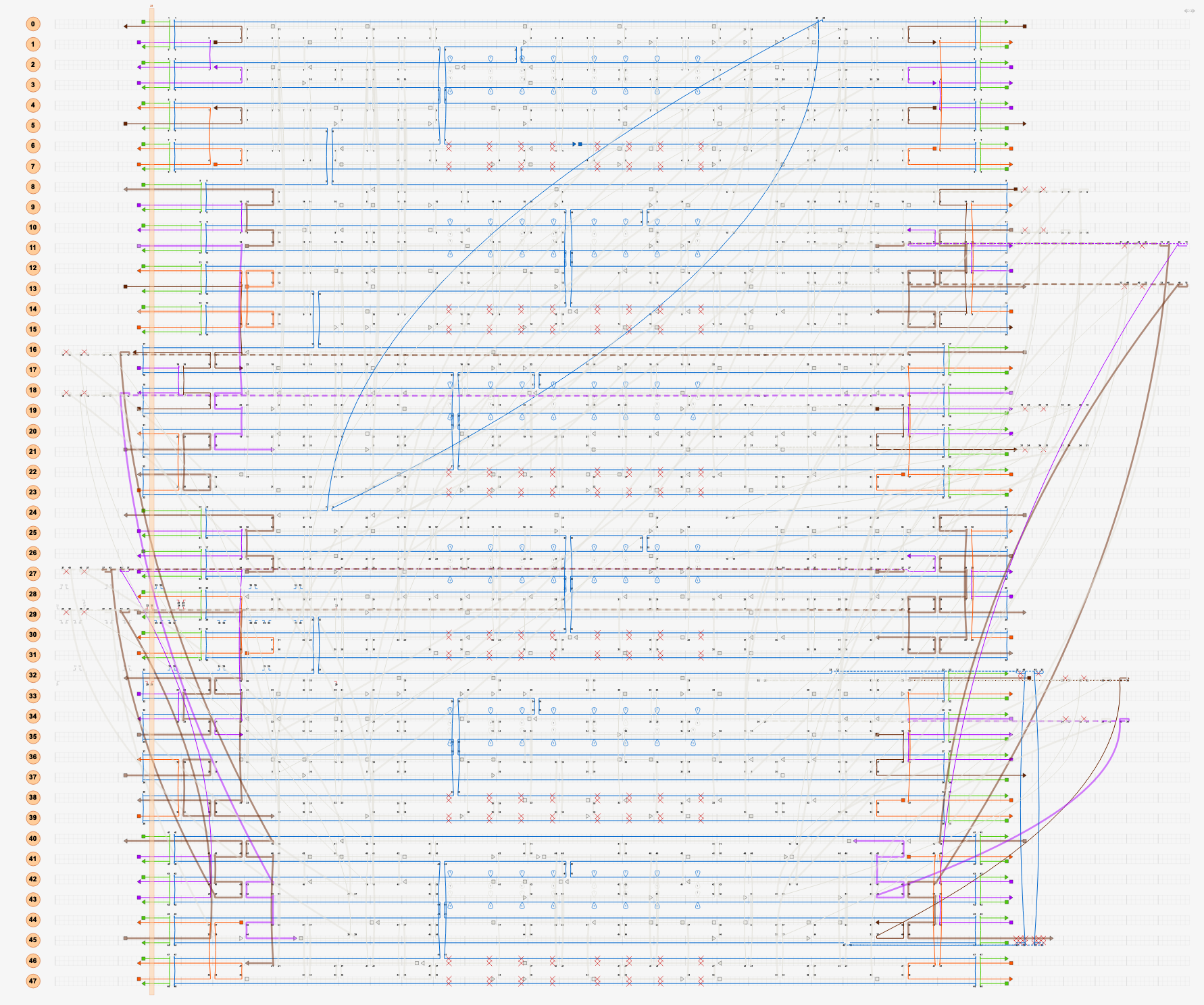} 
	\caption{\textbf{Cadnano design for monomer A.} 
    Scaffold and "mini scaffolds" are shown in blue and green. The staples with poly(C) repelling extensions shared between monomer mA and mB are shown in purple. The staples with mA-specific poly(C) extensions are colored orange and those with mA-specific binding extensions are colored brown. For clarity, the binding extensions are depicted in an extended form using 4-nt overhang.}
	\label{mA_cadnano}
\end{figure}

\begin{figure} 
	\centering
	\includegraphics[width=0.9\textwidth]{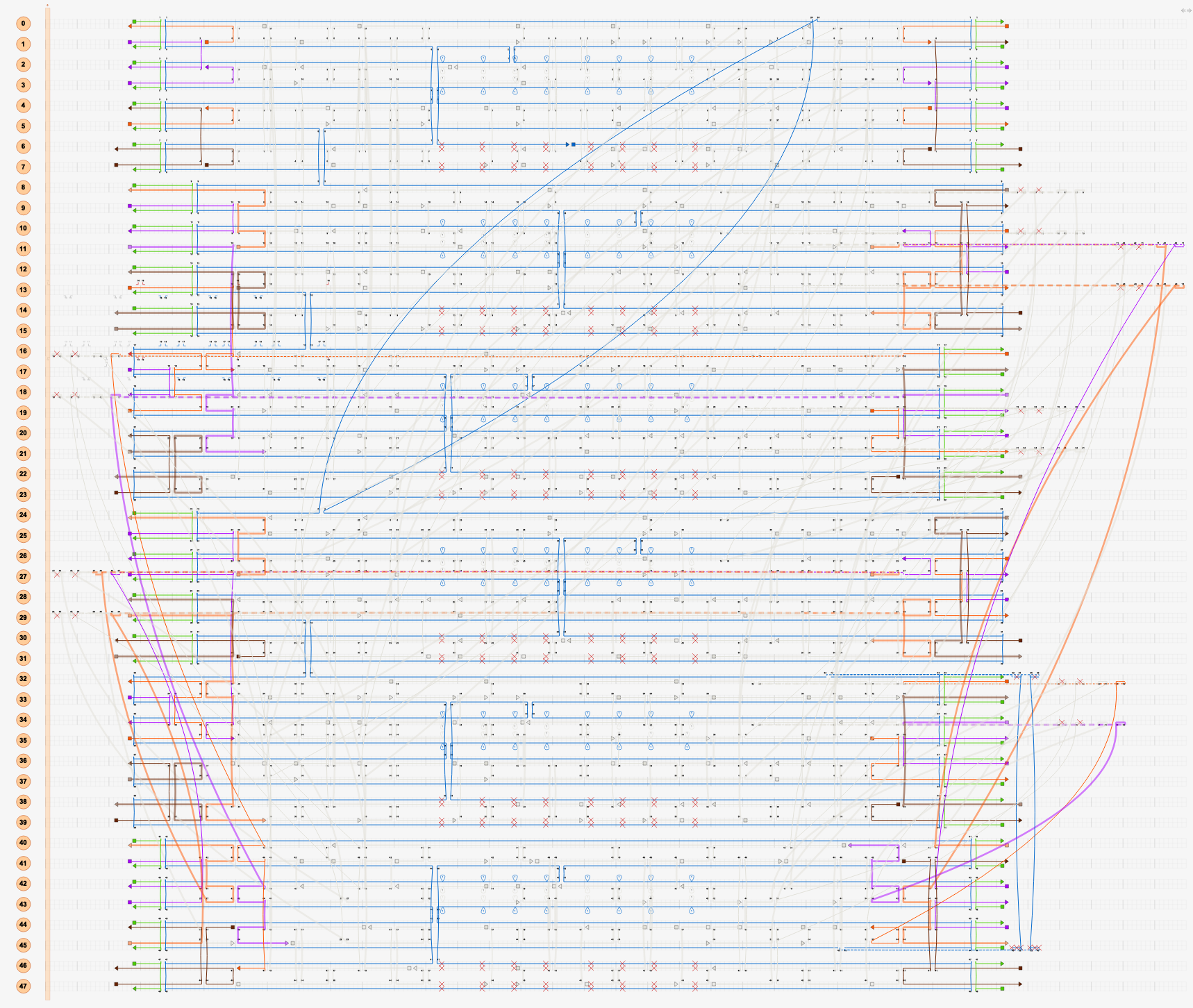} 
	\caption{\textbf{Cadnano design for monomer B.}
Scaffold and "mini scaffolds" are shown in blue and green. The staples with poly(C) repelling extensions shared between monomer mA and mB are shown in purple. The staples with mB-specific poly(C) extensions are colored orange and those with mB-specific binding extensions are colored brown. For clarity, the binding extensions are depicted in an extended form using 4-nt overhangs.}
	\label{mB_cadnano}
\end{figure}

\noindent \textbf{Tetrapod monomer and binding interface design}
The core part of the DNA origami tetrapod used in this study has been described in detail previously~\cite{posnjak_diamond-lattice_2024}. In brief, the structure has four legs, or 24-helix bundles, pointing into four directions separated by the tehtrahedral angle 109.5°. Towards the center of the tetrapod, each 24-helix bundle splits into three 8-helix bundles that bend outwards \cite{dietz_folding_2009} by 71° and then merge with the 8-helix bundles coming from two other legs. Figure~\ref{mA_cadnano} and Figure~\ref{mB_cadnano} shows the caDNAno \cite{douglas_rapid_2009} diagram of the structure, including the "mini scaffolds" (short:GCCGATGGTAGCCG , long: AGCATTAGTAGTTCCTTGATGATTACGA) at the ends of the helices, providing a non-staggered interface.

At each interface, the ends of 24 DNA helices can interact with the interfaces of the neighboring tetrapods. In our experiments, 18 helices (grey circles in Figure~\ref{clathrate_figure1}D) were extended with the sequence CCC. Three pairs of helices (pink and blue circles in Figure~\ref{clathrate_figure1}D, brown colored staples in Figure~\ref{mA_cadnano} and Figure~\ref{mB_cadnano}) were extended with one of the sequences shown in table \ref{SequenceTable} (the left end of the sequences is connected to the DNA origami body). Note that these three pairs of helices on each interface and all four legs of the tetrapod always carry the same sequence.  These sequences can be different on different monomers (mA-mB assemblies). 

\noindent \textbf{Folding and purification protocol}
The mA- and mB-type of  the DNA-origami tetrapod monomers were folded by mixing the in-house-produced \cite{douglas_dna-nanotube-induced_2007} p8634 scaffold (15 nM) with approximately 200 core staples (120 nM) in a 100 $\mu$L aliquot. The poly(C)-modified end strands (120 nM; colored purple and orange in ~Figure~\ref{mA_cadnano} and ~Figure~\ref{mB_cadnano}) for mA, or mB were added accordingly. All staples were ordered from Integrated DNA Technologies, standard desalted, 200 $\mu$M in water in each well. The folding solution contained 20 mM MgCl$_{2}$ and 1xTE buffer (10 mM Tris-HCl and 1 mM EDTA, pH adjusted to 8 with NaOH). The mixture was annealed with a folding temperature protocol (~Table~\ref{folding_table}) in a Jena Analytik Trio thermocycler.

After folding, the tetrapod monomers were purified by PEG precipitation. The folding mixture was mixed in a 1:1 ratio with a PEG purification buffer (15\% wt. PEG-8000, 500 mM NaCl and 1xTE) in a 2 mL tube and centrifuged at 20,000 RCF for 33 min at 4 °C. \cite{stahl_facile_2014} After centrifugation, the supernatant was removed and the pellet was resuspended in a buffer containing 1xTE and 5 mM MgCl$_{2}$. The mixture was then shaken at 34$^\circ$C for 1h. The concentration of tetrapod monomers was measured with a Nanodrop ND-1000 UV-Vis spectrometer, and adjusted to 50 nM.
\begin{table} 
	\centering
	\caption{\textbf{Temperature annealing protocol for monomer folding}}
	\label{folding_table} 	\begin{tabular}{lccr} 
		\\
		\hline
		Temperature($^\circ$C) & Time per $^\circ$C (min)\\
		\hline
		67 & 15 \\
		60-50 & 40\\
		25 & storage\\
		\hline
	\end{tabular}
\end{table}

\noindent \textbf{Crystallization protocol}
The two types of monomers were mixed in a 1:1 ratio (10 nM for each type) together with the "mini scaffold" strands (8 $\mu$M; colored green in ~Figure~\ref{mA_cadnano} and ~Figure~\ref{mB_cadnano}) and respective end staples for each crystallization experiment (120 nM; colored purple, orange and brown in ~Figure~\ref{mA_cadnano} and ~Figure~\ref{mB_cadnano}) in 70 $\mu$L aliquots containing 1xTE, and MgCl$_{2}$ with concentrations  ranging from 18 to 70 mM. The mixtures were annealed in PCR SingleCap 8er-SoftStrips 0.2 mL tubes (Biozym) in a Jena Analytik Trio thermocycler according to a crystallization annealing protocol (Table~\ref{crystallization_table}). During the first temperature ramp and incubation at 40$^\circ$C, all the forming networks sink to the bottom of the tube.  35 $\mu$L of supernatant was carefully removed from the top, and the sample was annealed with a second temperature ramp. As this step takes 18 days, partial evaporation of liquid could not be prevented. The resulting samples had a volume of approximately 20 $\mu$L and the resulting MgCl$_{2}$ concentration accordingly was 1.75 times higher than at the beginning.

\begin{table}
	\centering
	\caption{\textbf{Temperature annealing protocol for crystallization}}
	\label{crystallization_table}
\begin{tabular}{lccr} 
		\\
		\hline
		Temperature($^\circ$C) & Time per 0.1$^\circ$C (min)\\
		\hline
		55-33 & 15 \\
		40 & 8 (h) & remove 35 $\mu$L supernatant\\
		55-33 & 120\\
		\hline
	\end{tabular}
\end{table}

\begin{table}
	\centering
	\caption{\textbf{DNA sequence of binding extensions}}
	\label{SequenceTable}
\begin{tabular}{lccr} 
		\\
		\hline
		name & sequence for mA& sequence for mB & hybridization energy \\
		\hline
		6nt & 5' TCAGTA & 5' TACTGA & -7.94 kcal/mol \\
		5nt & 5' ACGAT & 5' ATCGT & -7.65 kcal/mol\\
		4nt & 5' GACT & 5' AGTC & -5.13 kcal/mol\\
            non-specific & GGG/CCC & CCC/GGG & -6.14 kcal/mol\\
            non-specific & GG/CCC & CCC/GG & -3.07 kcal/mol\\
		\hline
	\end{tabular}
\end{table}

\noindent \textbf{Silicification protocol}
The resulting crystals were encapsulated in silica to increase their mechanical stability\cite{posnjak_diamond-lattice_2024}, \cite{nguyen_dna-origami-templated_2019}, \cite{majewski_resilient_2021}. The crystallization solution was dilluted to 130 $\mu$L with a buffer containing 1xTE and 5 mM MgCl$_{2}$ in a 2 mL tube and shaken at 650 rpm and 4$^\circ$C. 1.6 $\mu$L of 1:3 diluted TMAPS (trimethyl[3-(trimethoxysilyl)propyl]ammonium chloride; TCI Germany, 50\% in methanol) in methanol was added to the solution while shaking. After 30 min, 1.0 $\mu$L of TEOS (tetraethyl orthosilicate, Merck) was added to the solution. The reaction temperature was gradually raised to room temperature. The temperature protocol is shown in~Table~\ref{silicification_table}. Afterwards, the reaction was stopped by filling up the tube with Milli-Q water. After 12 h of sedimentation, the supernatant (around 1.9 mL) was removed and replaced with isopropanol. After the crystals sedimented again, 1.9 mL supernatant was removed. 

\begin{table}
	\centering
	\caption{\textbf{Temperature protocol for silicification}}
	\label{silicification_table}

	\begin{tabular}{lccr} 
		\\
		\hline
		Temperature ($^\circ$C) & Time per $^\circ$C (h) \\
		\hline
		4 & 4.5  \\
		10 & 2\\
		16 & 2\\
        22 & 18\\
		\hline
	\end{tabular}
\end{table}

\noindent \textbf{Scanning electron microscopy measurements}
The remaining \~100 $\mu$L of liquid was gently shaken to redisperse the crystals, and 2 $\mu$L of the solution was drop-casted on a cleaned glass microscopy cover slip treated with Ar-plasma. After drying, it was sputtered with a 60:40 gold/palladium target (Edwards sputter coater S150B). For SEM imaging, we used 4kV acceleration voltage, 20 $\mu$m aperture, and 10 mm working distance with secondary electron, or in-lens detector of Raith eLine SEM.

\noindent \textbf{Small-angle x-ray scattering (SAXS)}

Synchrotron small-angle X-ray scattering (SAXS) measurements were performed at the P12 EMBL BioSAXS and P62 SAXSMAT beamlines of PETRA III, DESY (Hamburg, Germany). The beamline instrumentation has been described previously~\cite{blanchet_versatile_2015,haas_new_2023}. All measurements were carried out in borosilicate glass capillaries (Hilgenberg) with an inner diameter of 1.5 mm and a wall thickness of 0.05 mm. The samples were aligned so that the X-ray beam passed through the bottom of the capillaries, where the crystals had sedimented under gravity. Scattering from an empty capillary was measured separately and subtracted as background.

\noindent \textbf{Modeling of diffraction patterns}

\begin{figure}[!h] 
	\centering
	\includegraphics[width=0.8\textwidth]{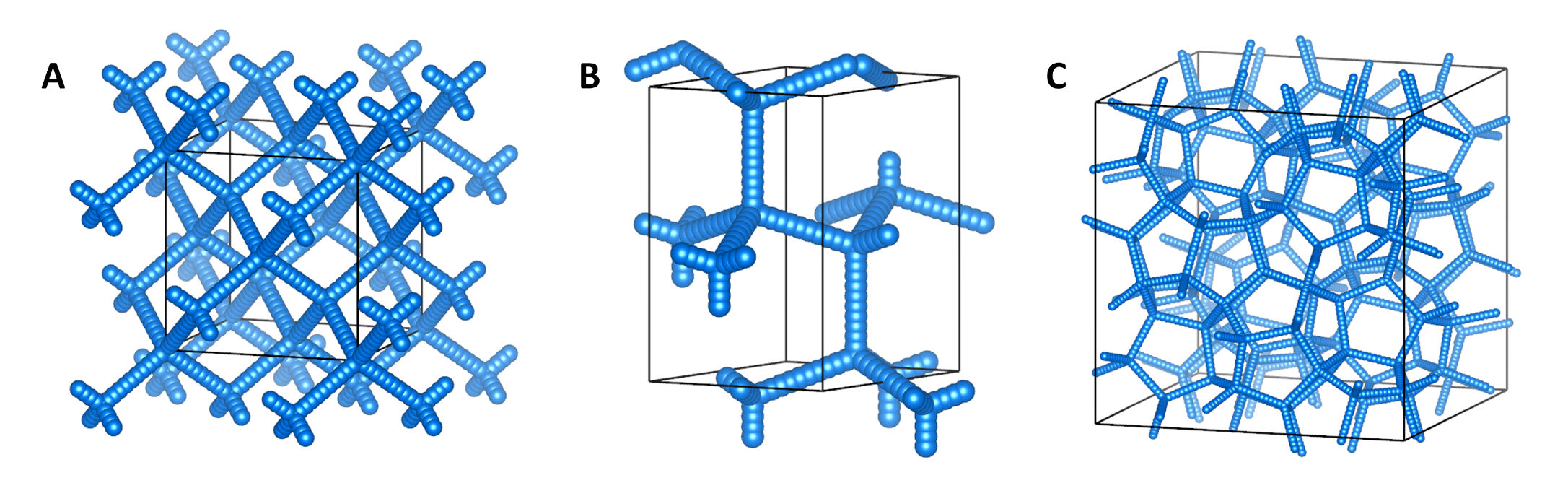} 
	\caption{\textbf{Models of the crystallographic structures for calculating x-ray diffraction patterns}. \textbf{A}: Cubic diamond, \textbf{B}: hexagonal diamond, \textbf{C}: sII clathrate. Images were created in VESTA-v.3\cite{momma_vesta_2011}.}
	\label{fig:vesta}
\end{figure}

Models of the DNA origami crystal structures were generated from the crystallographic frameworks of atomic diamond cubic, hexagonal diamond, and type-II clathrate lattices. To approximate a homogeneous electron density distribution along the DNA origami tetrapod arms, each crystallographic bond was discretized by inserting ten equally spaced intermediate points between neighboring lattice vertices. The resulting scattering-density unit cell models (Figure~\ref{fig:vesta}) were used to calculate powder diffraction patterns in VESTA-v.3~\cite{momma_vesta_2011}. For comparison with experimental SAXS data, the lattice parameter of each model was refined by uniformly scaling the unit cell to match the observed Bragg peak positions.

\noindent \textbf{Quantification of Phase Fractions from SAXS Data}

The relative fractions of the diamond cubic, hexagonal diamond, and sII clathrate phases were determined by fitting experimental SAXS profiles with theoretical scattering references for the individual phases described above. For each phase, the theoretical peak positions and relative intensities were extracted from the corresponding reference pattern within the selected \(q\)-range. Theoretical reflections were represented as a sum of Gaussian peaks with adjustable peak width and lattice scaling parameters to account for finite domain size effects and small variations in lattice spacing between samples.

The scattering intensity of a mixture was modeled as

\begin{equation}
I(q)=\sum_i A_i R_i(q)+B(q),
\end{equation}

where \(R_i(q)\) denotes the broadened reference pattern of phase \(i\), \(A_i\) is the fitted amplitude of phase \(i\), and \(B(q)\) is a smooth background contribution. The background was modeled using a spline function. Spline anchor regions were defined in \(q\)-ranges outside the diffraction peaks to guide the baseline shape. These anchor regions were assigned zero weight during the phase-fraction optimization and therefore did not contribute to the determination of the phase amplitudes.

The broadened reference pattern of phase \(i\) was represented as

\begin{equation}
R_i(q)=
\sum_j I_{ij}
\exp\!\left[
-\frac{\left(q-s_i q_{ij}\right)^2}
{2\sigma_i^2}
\right],
\end{equation}

where \(q_{ij}\) and \(I_{ij}\) are the position and relative intensity of the reflection \(j\) of phase \(i\), respectively. The parameter \(s_i\) is a phase-specific \(q\)-scaling factor that accounts for small differences in the lattice parameter between the experimental sample and the theoretical reference, and \(\sigma_i\) is the Gaussian broadening parameter describing finite crystallite size and instrumental broadening.

Model parameters were optimized by nonlinear least-squares fitting of the calculated scattering profile to the experimental SAXS data. Equal weighting was applied to all data points within the fitting range. Phase-specific scaling factors \(s_i\) were allowed to vary within \(\pm 2\%\) of their nominal values, while peak broadening parameters \(\sigma_i\) were fitted independently for each phase.

The relative phase fractions were calculated from the fitted amplitudes according to

\begin{equation}
f_i=
\frac{A_i}
{\sum_j A_j},
\end{equation}

where \(f_i\) is the fraction of phase \(i\).

Parameter uncertainties were estimated from the covariance matrix of the least-squares fit. Standard errors of the fitted amplitudes and phase fractions were obtained by propagation of the parameter uncertainties derived from the covariance matrix.


\clearpage

\subsection*{Supplementary Text}

\subsubsection*{Triple diamond structures}
\begin{figure}
	\centering
	\includegraphics[width=0.6\textwidth]{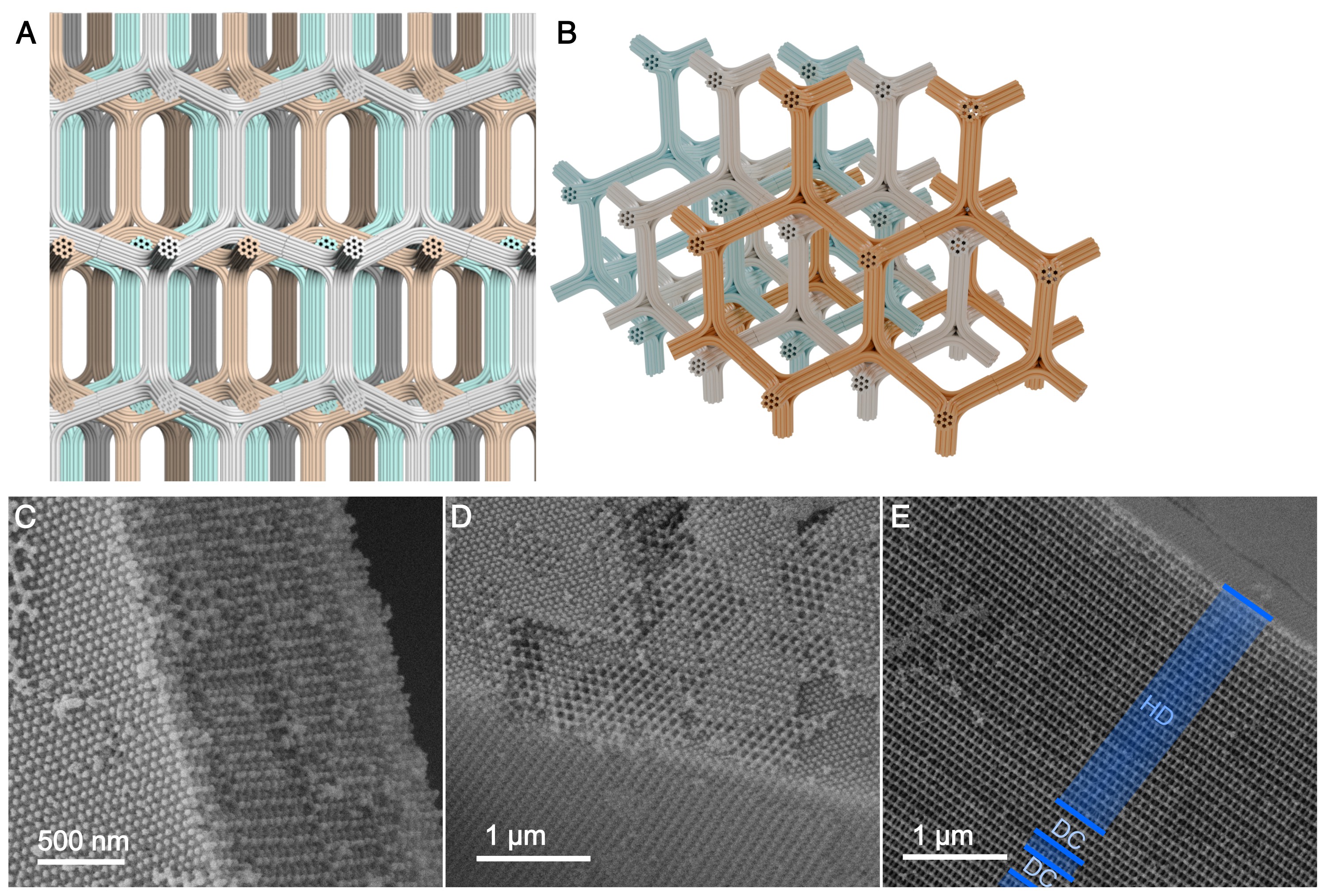} 
	\caption{\textbf{Triple hexagonal diamond and triple diamond cubic}. Models of the (\textbf{A}) {100} facet of triple hexagonal diamond, and (B) triple cubic diamond (side view). The three interpenetrating lattices are drawn in gray, brown, and turquoise. Panels (\textbf{C}) and (\textbf{D}) show SEM images corresponding to these two models. The densely populated {001}, and (111) planes, respectively, can be seen on the (C) left, and (D) upper side of the images.  (\textbf{E}) SEM image of a crystal facet showing stacking-disordered mix of triple diamond cubic (DC) and triple hexagonal diamond (HD). Triple hexagonal diamond layers are marked with blue shading.}
	\label{triple_side}
\end{figure}

With the occurrence of triple diamond, the hexagonal pattern of the $\{001\}$ and $\{111\}$ planes for hexagonal diamond, and diamond cubic, respectively, becomes more densely occupied by tetrapods with one arm pointing out of the plane, producing a distinct pattern compared to the single-lattice crystals (Figure~\ref{clathrate_figure4}). From the side view of these crystals (Figure~\ref{clathrate_figure4}B,C,F, Figures~\ref{triple_side} and \ref{tripleHD_side}), one can see a different denser pattern on the $\{100\}$ planes of triple HD, and the non-preffered $\{111\}$ planes of the triple DC. In addition to the DNA origami rods belonging to the primary lattice, we observe another network of rods corresponding to the tetrapod arms from an interpenetrated network connecting adjacent (001) or (111) planes (Figure~\ref{triple_side}C, E). However, distinguishing between double and triple lattices from the side view alone is difficult, as the rods belonging to the third lattice appear behind the primary lattice (Figure~\ref{triple_side}A,B). 

\begin{figure}
	\centering
	\includegraphics[width=0.6\textwidth]{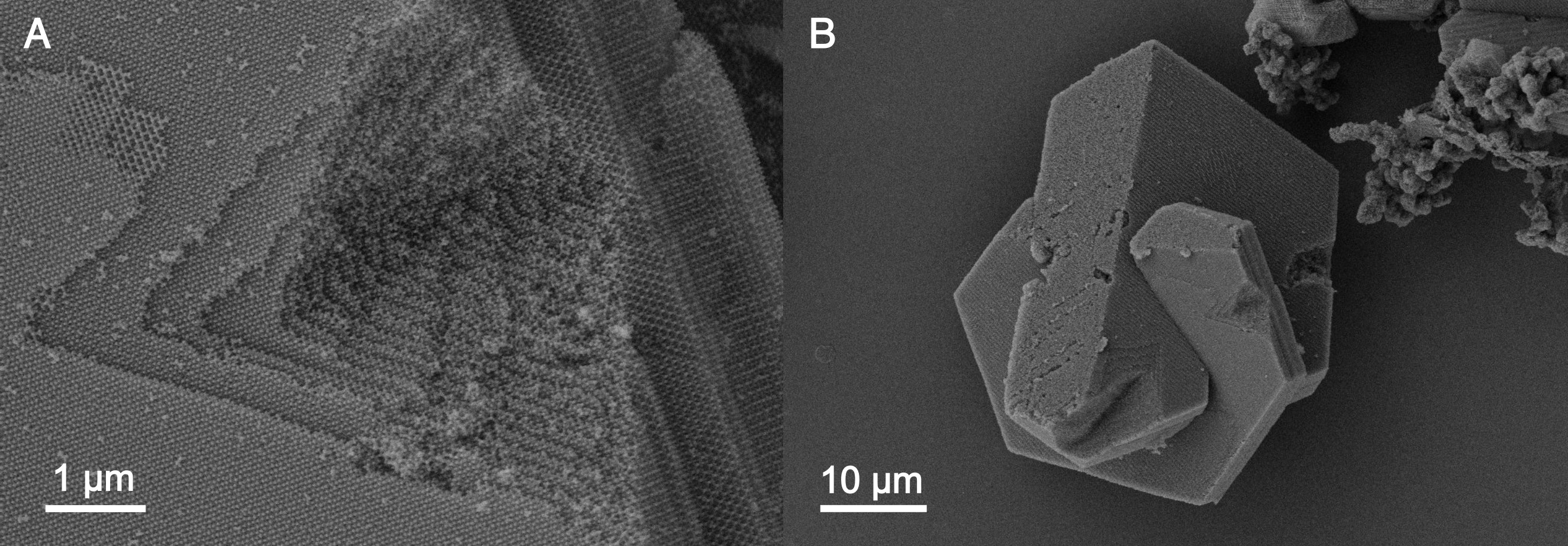} 
	\caption{\textbf{Triple diamond structure, visible inside a crystal void.} (\textbf{A}) Zoomed-in SEM image of a void area showing multiple layers of triple HD structure, with stacking disorder . (\textbf{B}) Zoomed-out image showing the position of the void. The origin of such voids is not clear, but based on the shape and position, it is most likely that an additional hexagonal diamond platelet was lodged in the crystal, preventing crystal growth and leaving a distinct imprint. We speculate that the platelet became dislodged during post-crystallization pipetting, since otherwise the gap could have been filled-in by subsequent crystal growth. The exposed inside layers of the crystal all show the distinct triple-lattice pattern, indicating the triple diamond structures are a bulk phase.}
	\label{triple_inside}
\end{figure}

The triple HD and triple DC lattices can be distinguished by the morphology of the crystals, as both triple structures feature the same characteristic angles as their single-lattice counterparts. The angle between the base (001) plane and $\{100\}$ side planes of HD is 90$^\circ$(Figure~\ref{triple_side}C), while the angle between the $\{111\}$ planes of diamond cubic is the tetrahedral angle (109.5$^\circ$; Figure~\ref{triple_side}D). In the case of stacking, the DC or triple DC fraction can be recognized by the zig-zag in the side surface of the hexagonal crystals (Figure~\ref{stacking4nt}, Figure~\ref{triple_side}). From occasional voids in the crystal, which expose the inner structure of the crystals, one can derive that the triple structure occurs as a bulk phase (Figure~\ref{triple_inside}).

\subsubsection*{Identification of crystal phases from SAXS data}
\begin{figure} 
	\centering
	\includegraphics[width=0.6\textwidth]{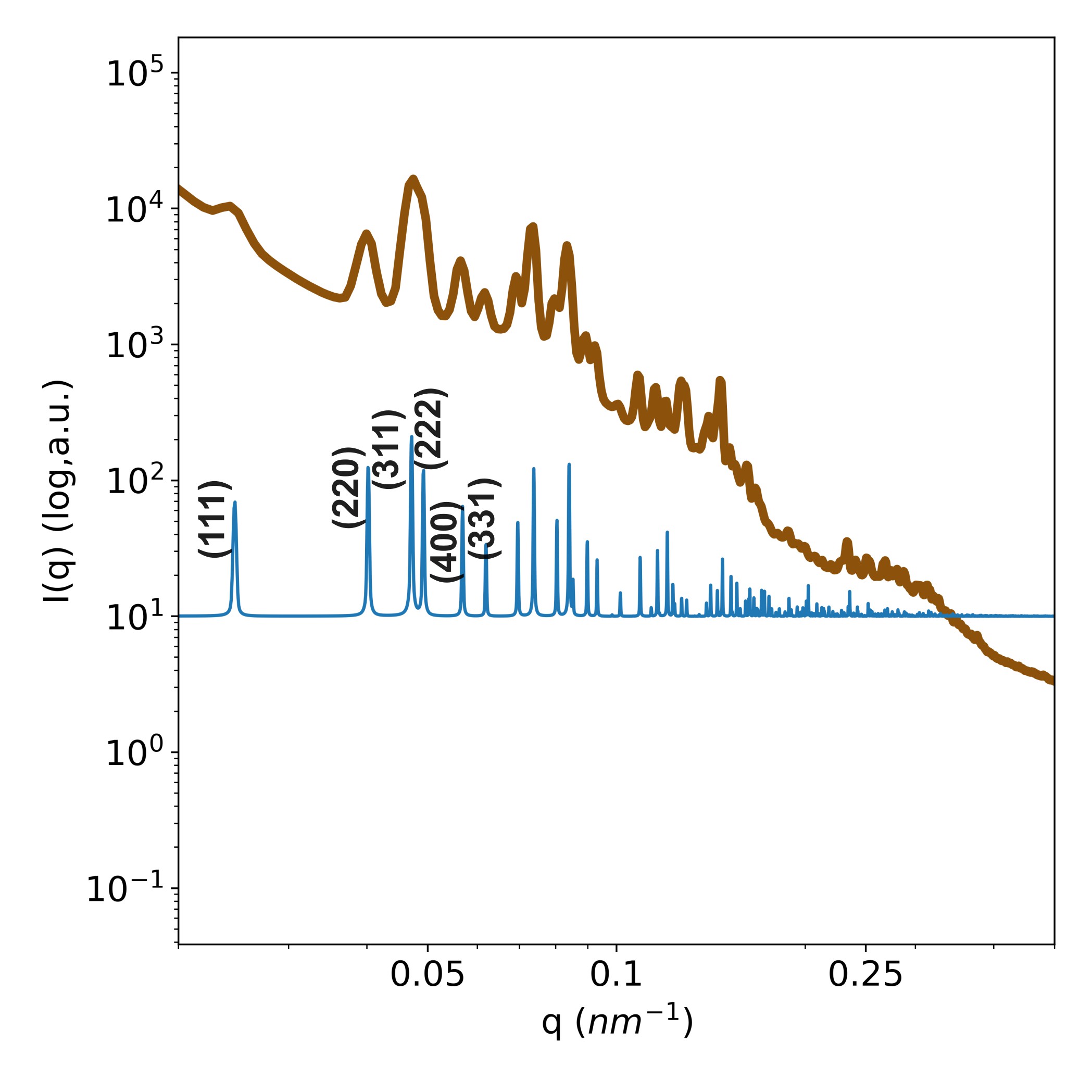} 
	\caption{\textbf{Simulated (blue) and experimental (brown) SAXS intensities for sII clathrate.} The simulated peaks agree well with the experimental data. The first six peaks are indexed.}
	\label{saxssimulation}
\end{figure}

\begin{figure} 
	\centering
	\includegraphics[width=0.6\textwidth]{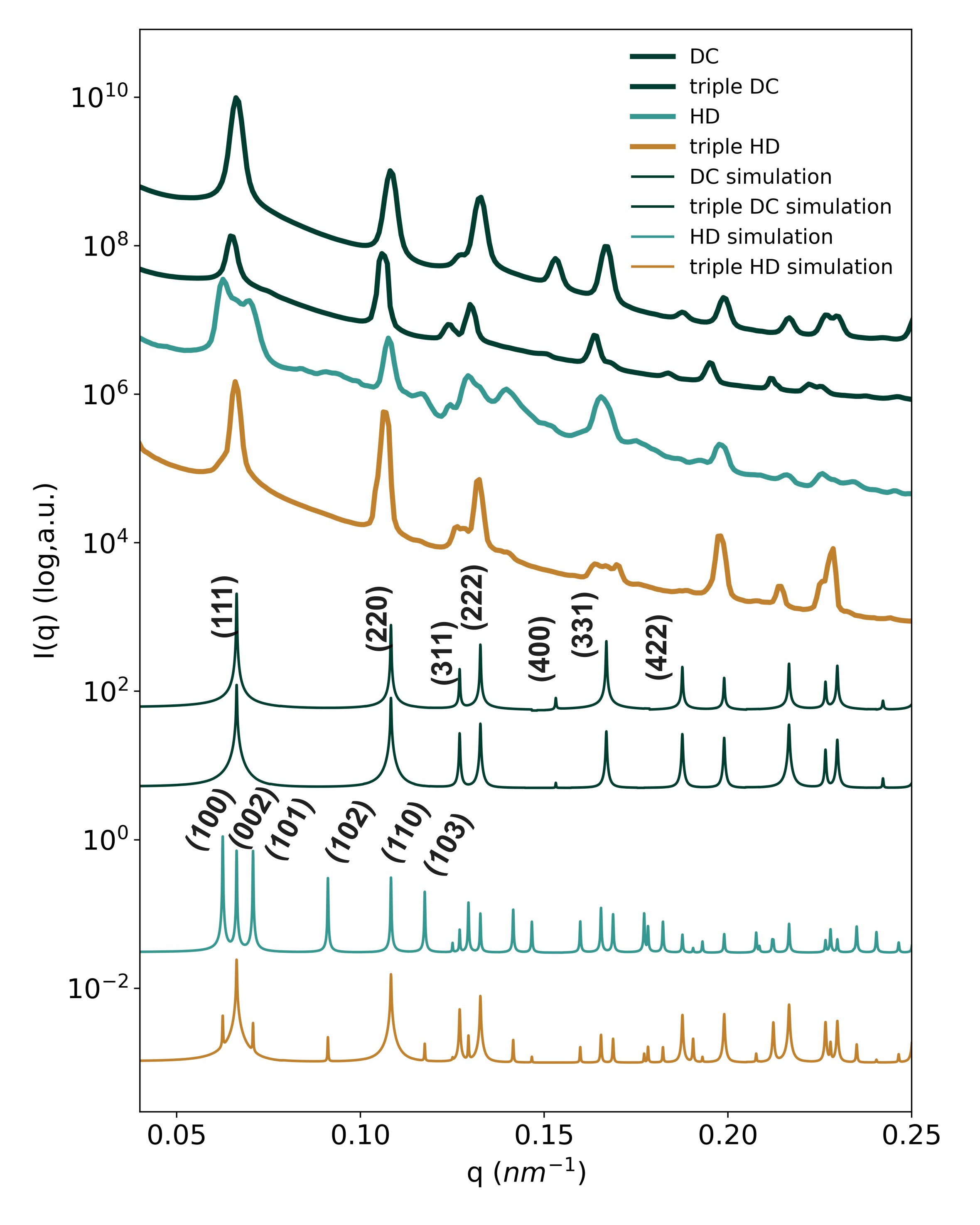} 
	\caption{\textbf{Simulated and experimental SAXS intensities for DC, triple DC, and triple HD.} The characteristic peaks are indexed in the simulated curves. The (002) peak of the HD structure and the (111) peak of the DC have exactly the same position, as they correspond to the crystal planes in both structures, which differ only by their stacking sequence. The HD structure can be identified by the additional (100), (101), and (102) peaks. Triple DC displays the same peaks as DC, however, the intensities of the (311), and the (400) peaks are slightly increased and decreased, respectively. The peak positions of triple HD are also the same as those of HD, with multiple peaks having much lower intensities, especially the (100), (101), and (102) peaks. This makes the spectrum of the triple HD hard to distinguish from single and triple DC spectra in experiments. The simulated peaks agree well with the experimental data.}
	\label{saxssimulation_1}
\end{figure}

The crystal structures were further verified by SAXS measurement as described in the methods. (Figure~\ref{saxssimulation}, Figure~\ref{saxssimulation_1}). The presence of HD is confirmed by the emergence of HD-characteristic peaks (100), (101), and (102), and the (002) peak overlaps with the (111) peak of DC. Clathrate can be identified by the $\{111\}$ peak at around 0.025 nm$^{-1}$ (Figure~\ref{saxssimulation}). 

\begin{figure} 
	\centering
	\includegraphics[width=0.9\textwidth]{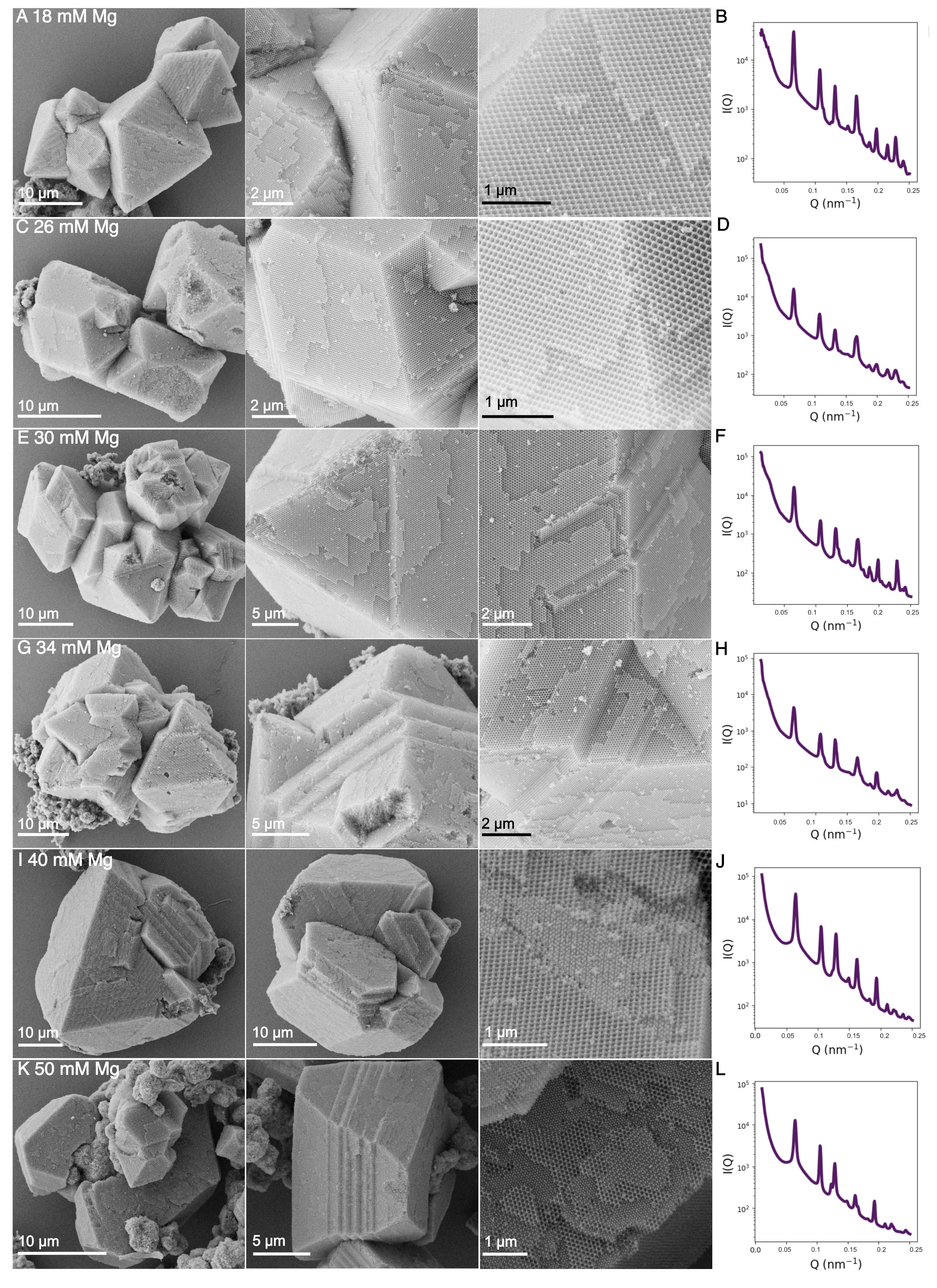} 
	\caption{\textbf{SEM images and SAXS intensities of samples where the monomers carry the binding extension 3T 6nt 3G.}
		With increasing concentrations of  MgCl$_{2}$ (top to bottom) we observe the formation of more twinning planes. At concentrations above 40 mM MgCl$_{2}$ triple lattices prevail.}
	\label{6nt3G} 
\end{figure}

\begin{figure} 
	\centering
	\includegraphics[width=0.9\textwidth]{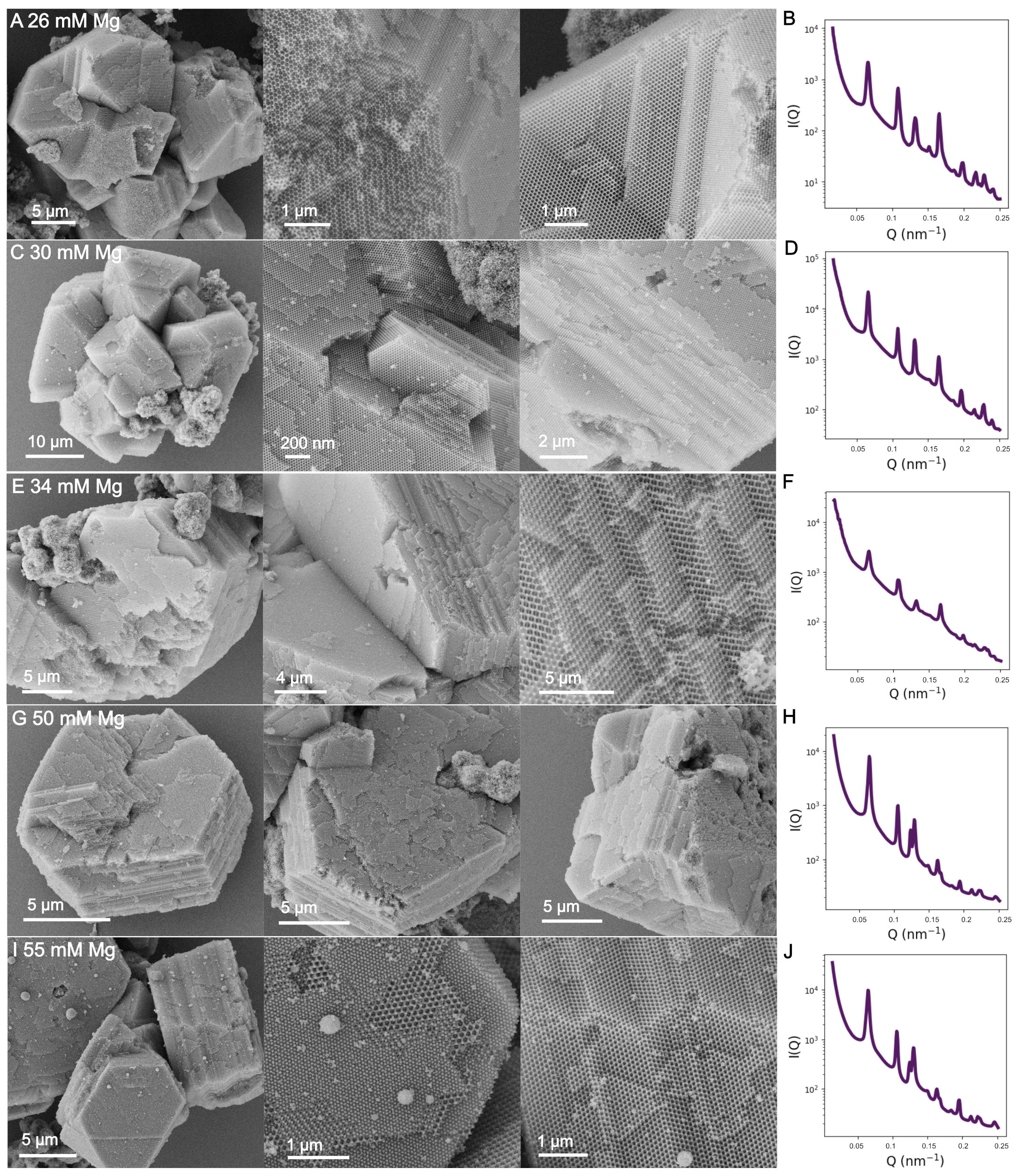}
	\caption{\textbf{SEM images and SAXS intensities of samples where the monomers carry the binding extension 3T 5nt 3G.}
		More twinning planes can be observed already at relatively low concentrations of MgCl$_{2}$. With increasing MgCl$_{2}$ concentrations we observe the formation of more twinning planes and a change of morphology towards hexagonal prisms. At concentrations above 50 mM MgCl$_{2}$ triple lattices prevail.}
	\label{5nt3G} 
\end{figure}

Simulations show that interpenetration by a second or third lattice does not alter the selection rules of either DC or HD; however, the relative scattering intensities are modulated. In particular, the intensity of the DC (311) peak increases in the presence of triple DC, which is consistent with the observations for samples carrying binding extensions "3T 6nt 3G" and "3T 5nt 3G" at elevated Mg$^{2+}$ concentrations (Figure~\ref{6nt3G}E,F, Figure~\ref{5nt3G}D, E). 

\begin{figure} 
	\centering
	\includegraphics[width=0.9\textwidth]{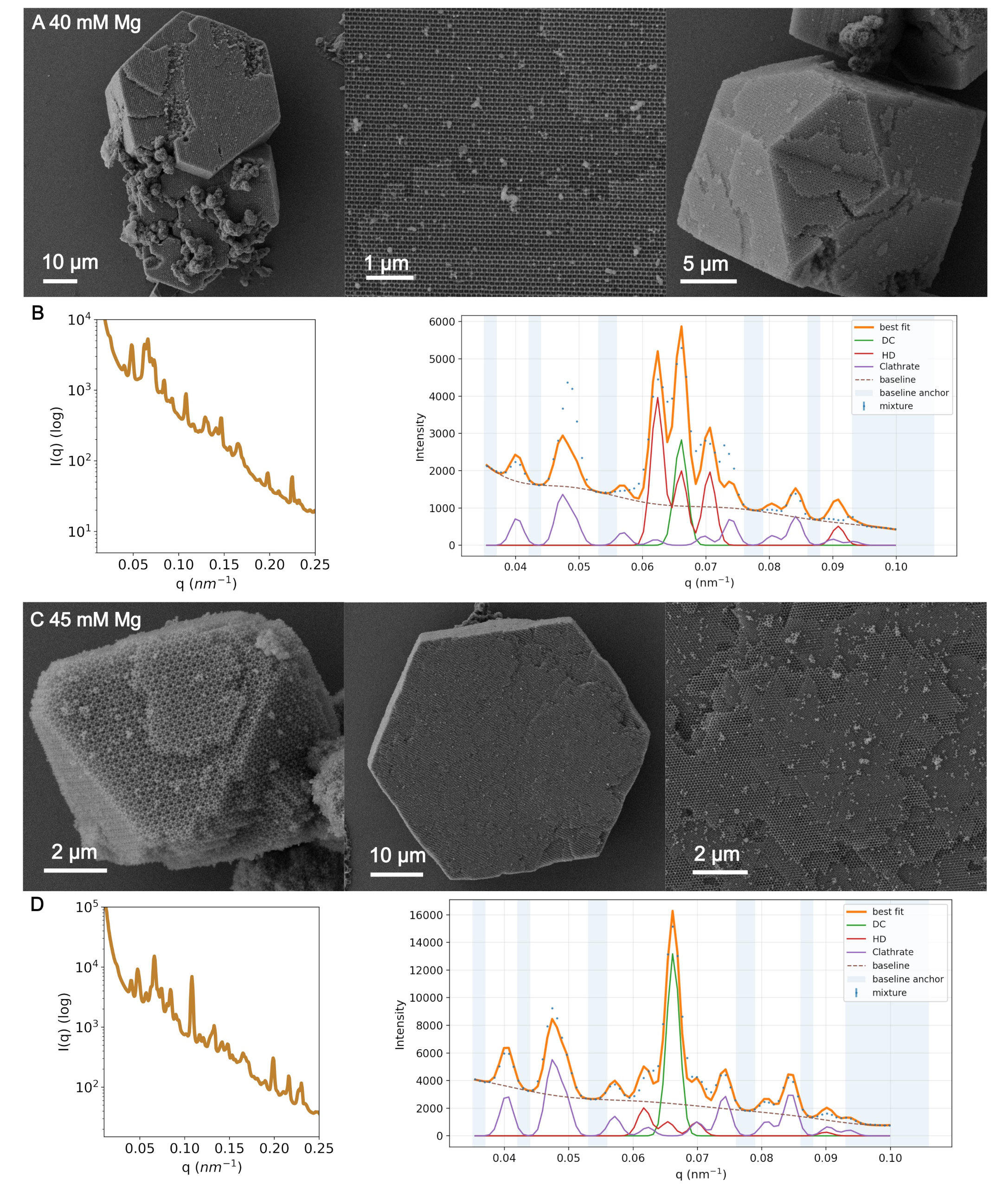} 
	\caption{\textbf{SEM images and SAXS intensities of samples where the monomers carry the binding extension 3T 4nt 3G.}
		At (\textbf{A}, \textbf{B}) 40 mM MgCl$_{2}$ concentration and (\textbf{C}, \textbf{D}) 45 mM MgCl$_{2}$ concentration we observe clathrate structures, mixed with hexagonal diamond crystals. The middle panel of (A) shows a $\{100\}$ side view of a high-purity HD structure, where in \~70 layers only 3 are DC. Rightmost panel of A, and leftmost panel of C show examples of clathrate structures. At 45 mM MgCl$_{2}$ concentration (\textbf{C}, \textbf{D}) triple lattices prevail, as seen from the dominant (111) peak, corresponding to the triple DC phase. Panel C, right, is a view of a $\{001\}$ crystal plane of a triple HD crystal.}
	\label{4nt40_45} 
\end{figure}

\begin{figure} 
	\centering
	\includegraphics[width=0.9\textwidth]{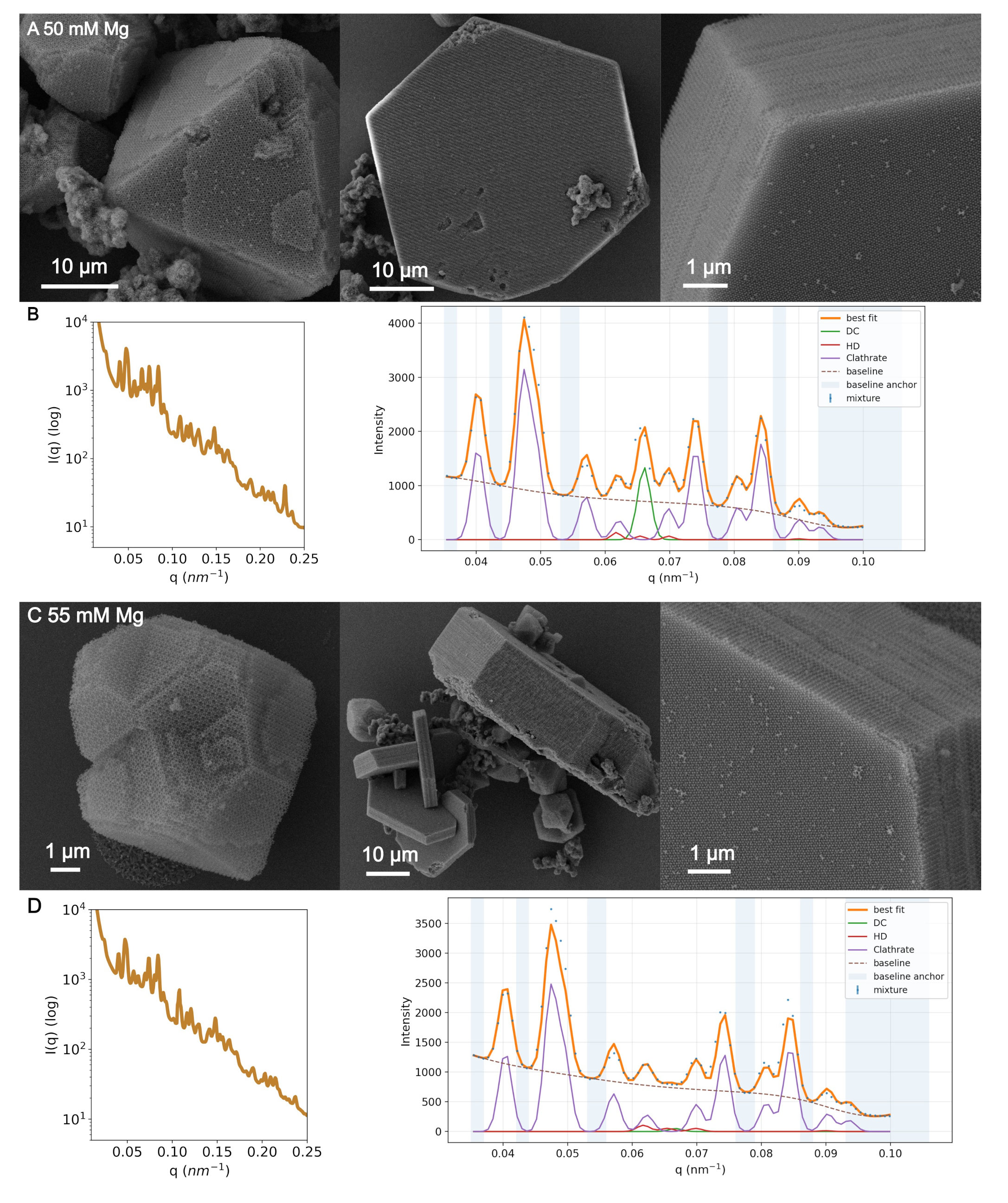} 
	\caption{\textbf{SEM images and SAXS intensities of samples with the binding extensions 3T 4nt 3G on mA and mB.}
		At (\textbf{A}, \textbf{B}) 50 mM MgCl$_{2}$ concentration and (\textbf{C}, \textbf{D}) 55 mM MgCl$_{2}$ concentration we observed an increasing fraction of clathrate in addition to the stacking disordered DC/HD. Meanwhile, the hexagonal shaped crystals (\textbf{A} middle and right panel; \textbf{C} middle and right panel) show cleaner top surface with homogeneous triple crystal pattern.}
	\label{4nt50_55} 
\end{figure}

\begin{figure} 
	\centering
	\includegraphics[width=0.9\textwidth]{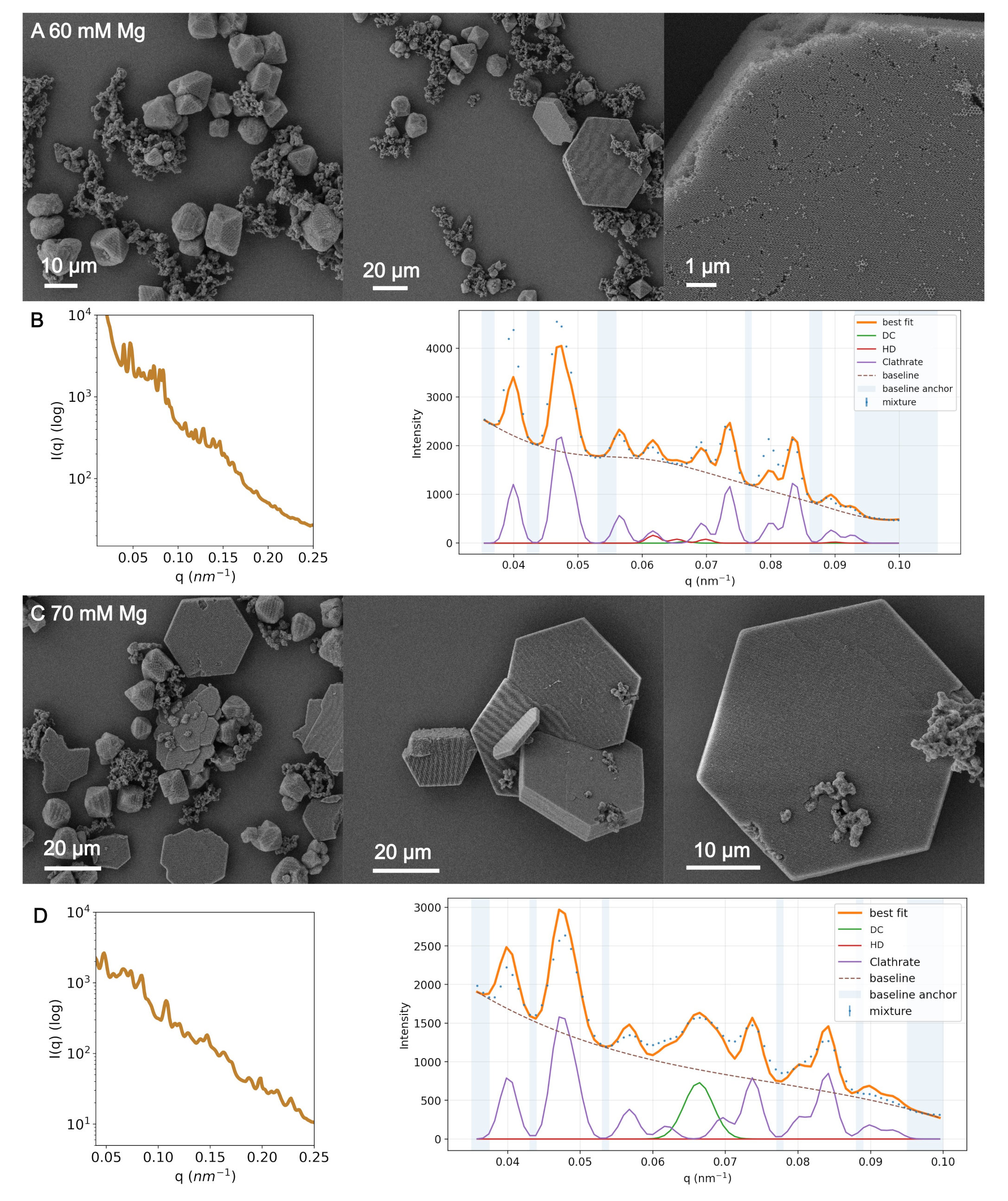} 
	\caption{\textbf{SEM images and SAXS intensities of samples with the binding extensions 3T 4nt 3G on mA and mB.}
		At (\textbf{A}, \textbf{B}) 60 mM, and (\textbf{C}, \textbf{D}) 70 mM concentration of MgCl$_{2}$ we observe an increasing fraction of clathrate in addition to the stacking disordered DC/HD. The hexagonally shaped crystals (\textbf{A} middle and right panel; \textbf{C} middle and right panel) show cleaner top surface with homogeneous triple crystal pattern.}
	\label{4nt60_70} 
\end{figure}

\begin{figure} 
	\centering
	\includegraphics[width=0.9\textwidth]{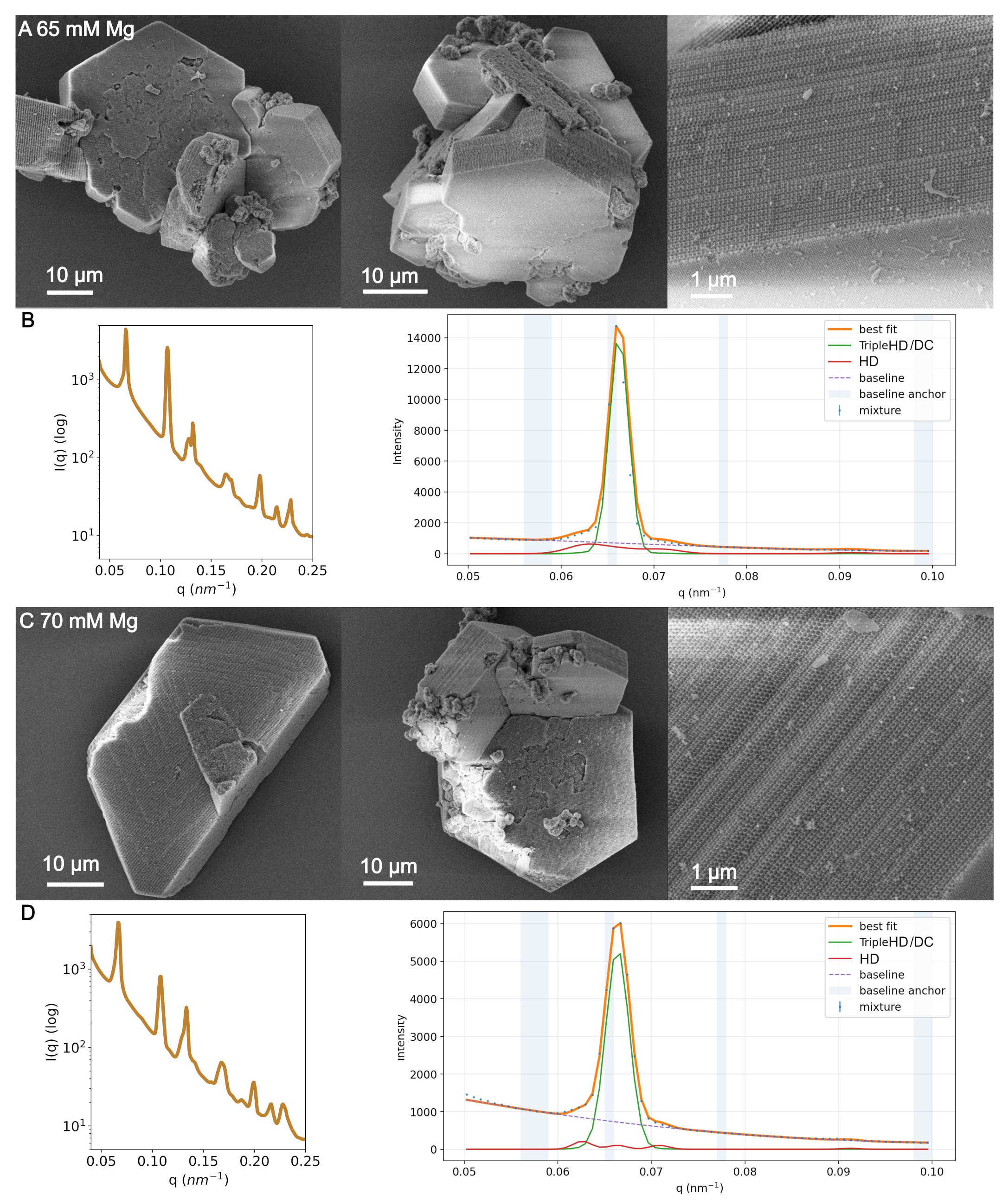} 
	\caption{\textbf{SEM images and SAXS intensities of samples with binding extension 7T 3G on mA and 3T 4nt 3G on mB.}
		At 60 mM and 70 mM MgCl$_{2}$ concentrations we observe only hexagonally shaped crystals with triple structure.}
	\label{4nt_7T65} 
\end{figure}

\begin{figure} 
	\centering
	\includegraphics[width=0.9\textwidth]{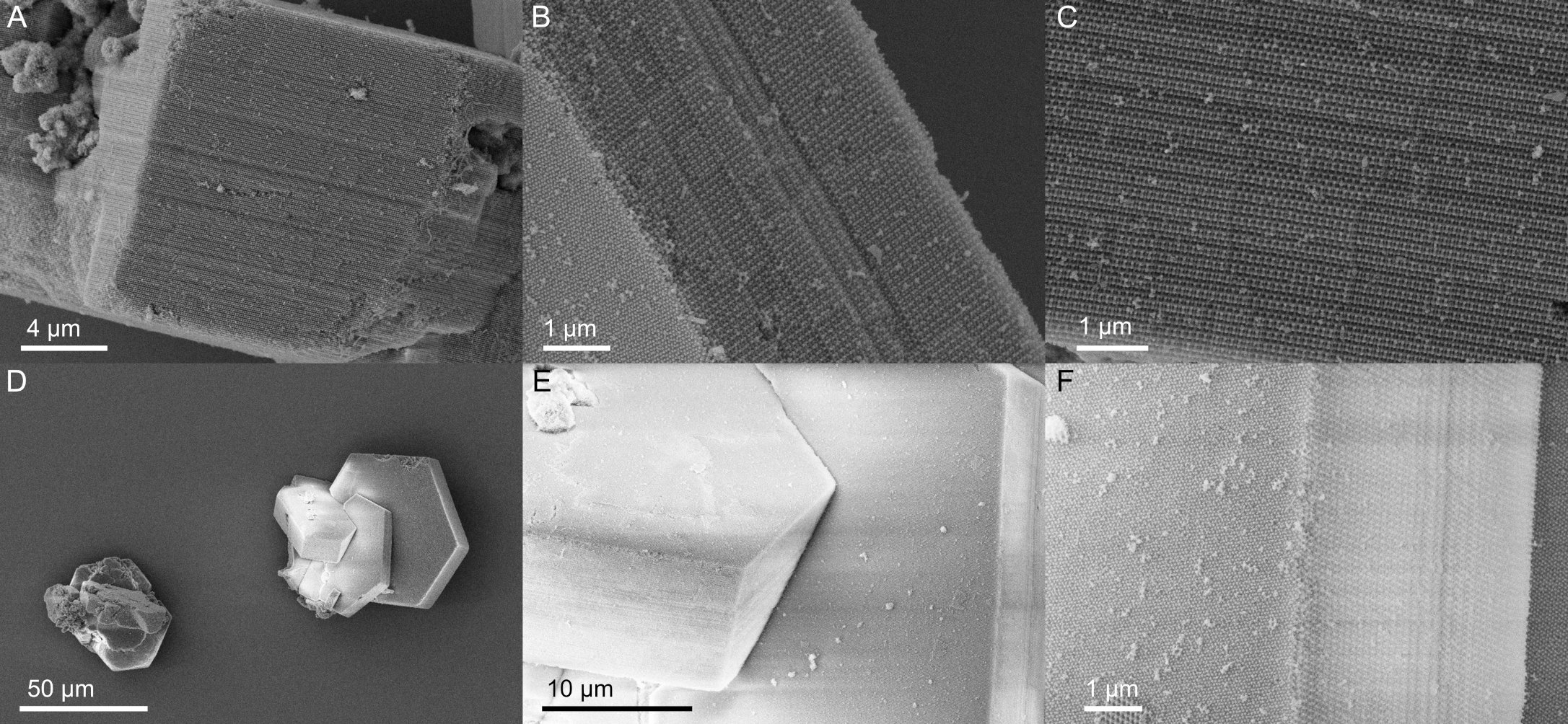} 
	\caption{\textbf{Example SEM images of side view of samples with binding extensions 7T 3G on mA and 3T 4nt 3G on mB at 65 mM MgCl$_{2}$ concentration.}
		A flat side surface with occasional wavy part together with a homogeneous triple structure on the (001) surface suggests that triple HD is the dominant phase.}
	\label{tripleHD_side} 
\end{figure}

When HD becomes interpenetrated and forms triple HD, several of its characteristic peaks are partially canceled. As a result, the SAXS profile of triple HD closely resembles those of DC and triple DC (Figure~\ref{saxssimulation_1}). Due to the large size of the triple crystals, high-intensity (111)/ (002) and (220)/ (110) peaks are typically observed in samples also containing clathrate, as shown in Figure~\ref{4nt40_45}B, Figure~\ref{4nt50_55}, and Figure~\ref{4nt60_70}. Consequently, distinguishing triple HD from DC, and triple DC based solely on SAXS fitting is challenging. In samples where side-view images are available, the flat $\{100\}$ side surfaces of the hexagonal-shaped crystals show that triple HD is the dominant phase (Figure~\ref{4nt_7T65}, Figure~\ref{tripleHD_side}). 

\begin{figure}
	\centering
	\includegraphics[width=0.9\textwidth]{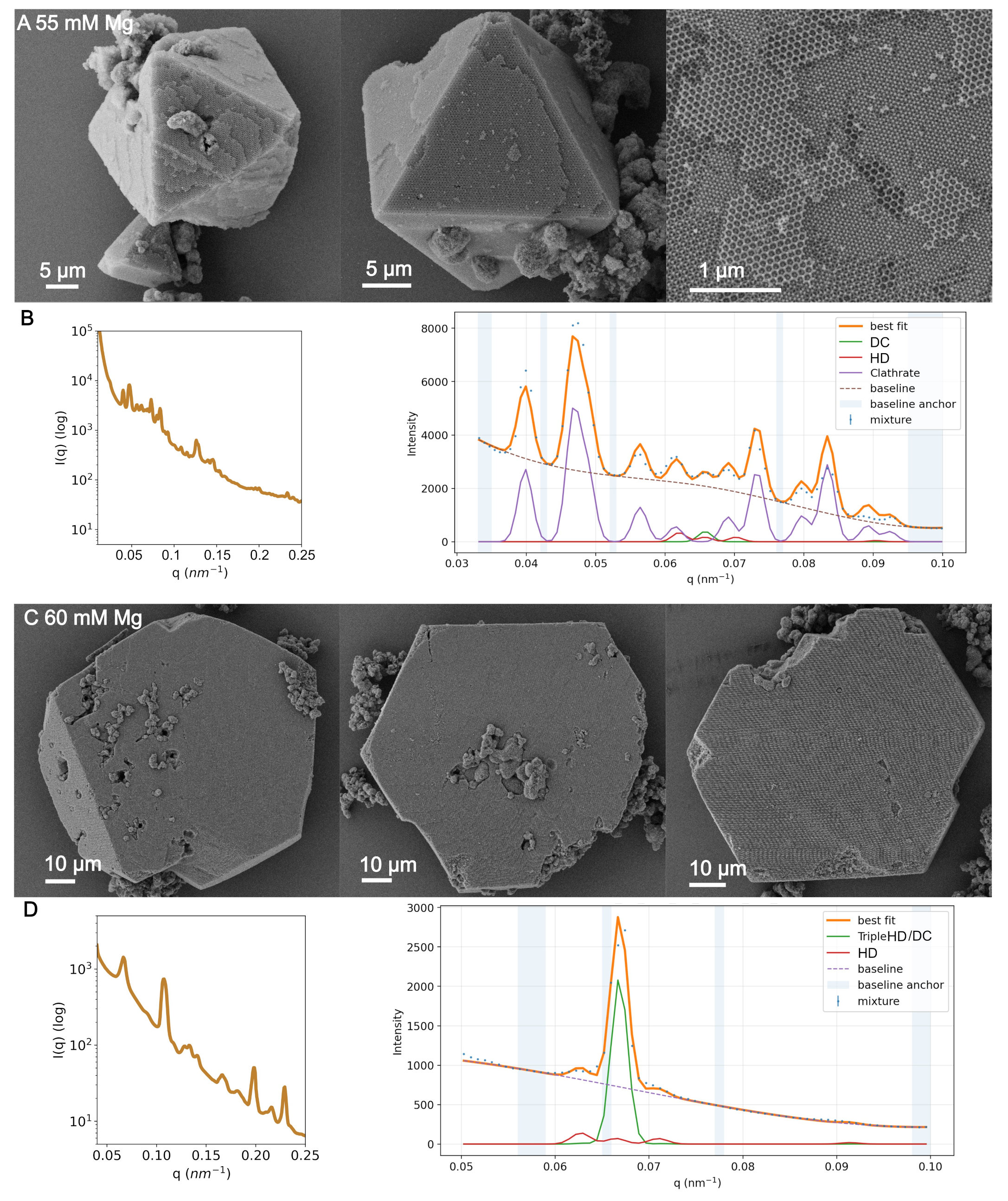} 
	\caption{\textbf{SEM images and SAXS intensities of samples with binding extensions 7T 3G on mA and 3T 4nt 3G on mB.}
		At (\textbf{A}, \textbf{B}) 55 mM MgCl$_{2}$ concentration we observed a large fraction of clathrate and occasionally triple stacking-disordered crystals. At (\textbf{C}, \textbf{D}) 60 mM MgCl$_{2}$ concentration clathrate crystals are not present.}
	\label{4nt_7T55} 
\end{figure}

\begin{figure} 
	\centering
	\includegraphics[width=0.6\textwidth]{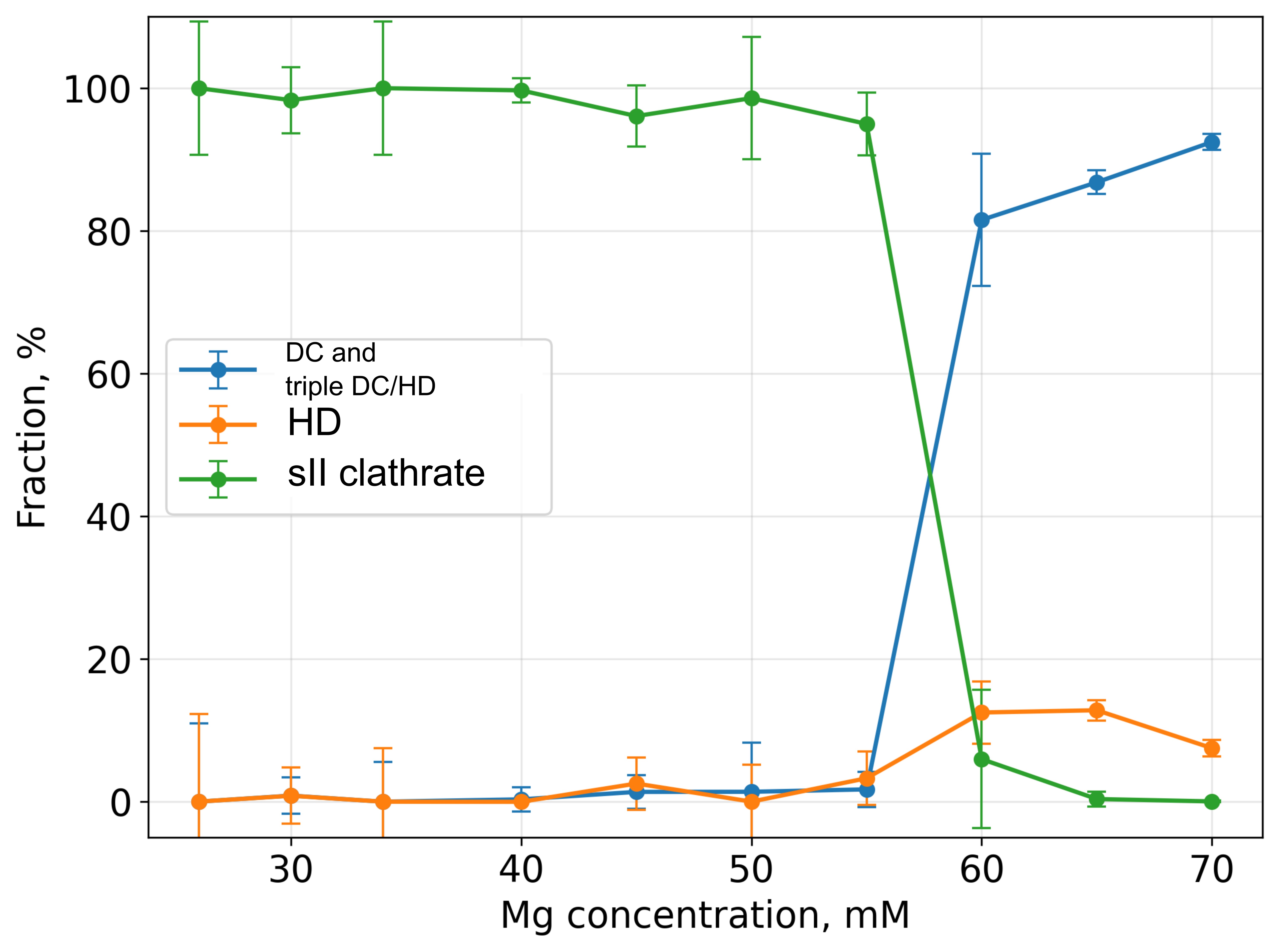} 
	\caption{\textbf{Phase fractions fitted from SAXS intensities of samples with binding extension 7T 3G on mA and 3T 4nt 3G on mB.}
		At lower MgCl$_{2}$ concentrations we observe pure sII clathrate phase. At concentrations above 55 mM MgCl$_{2}$, triple DC/HD lattices prevail. Error bars denote the fitting residuals.}
	\label{4nt7_fraction} 
\end{figure}

For the samples shown in Figure~\ref{4nt_7T55}B and Figure~\ref{4nt60_70}, some crystals reach sizes of approximately 100 $\mu$m and preferentially align with the (001) plane parallel to the substrate during drop-casting. As a result, side-view images are difficult to obtain, limiting direct morphological verification of the crystal structure. Nevertheless, these crystals exhibit well-defined hexagonal morphologies with sharp edges and a high degree of crystallinity in the exposed (001) plane. Additionally, the low fraction of non-interpenetrated HD obtained from phase-fraction analysis suggests that the HD phase is predominantly present as triple HD (Figure~\ref{4nt7_fraction}). Together, these observations support the assignment of triple HD as the dominant phase at these conditions.

\newpage
\begin{figure} 
	\centering
	\includegraphics[width=0.6\textwidth]{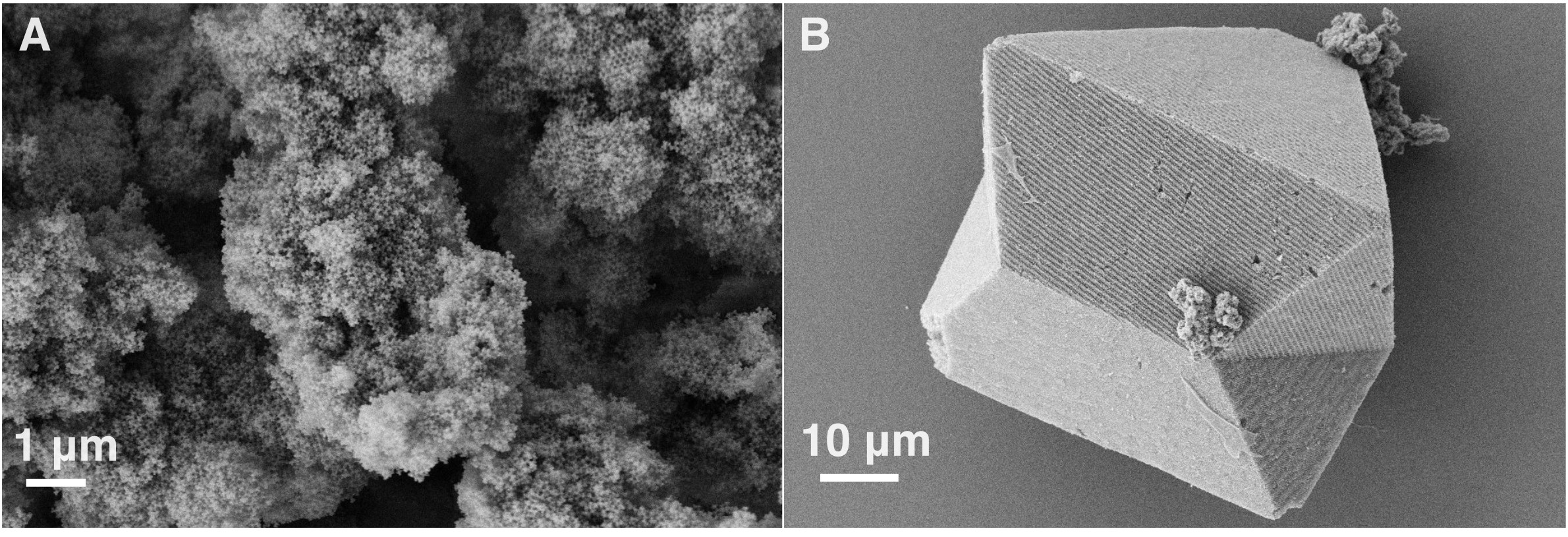} 
	\caption{\textbf{Samples with binding extensions without and with poly(G).} SEM images of a sample with binding extensions (\textbf{A}) 3T 6nt, and (\textbf{B}) 3T 6nt 2G. No crystallites can be found in the sample without the extension 2G.
		 }
	\label{6ntcompare}
\end{figure}

\begin{figure} 
	\centering
	\includegraphics[width=0.9\textwidth]{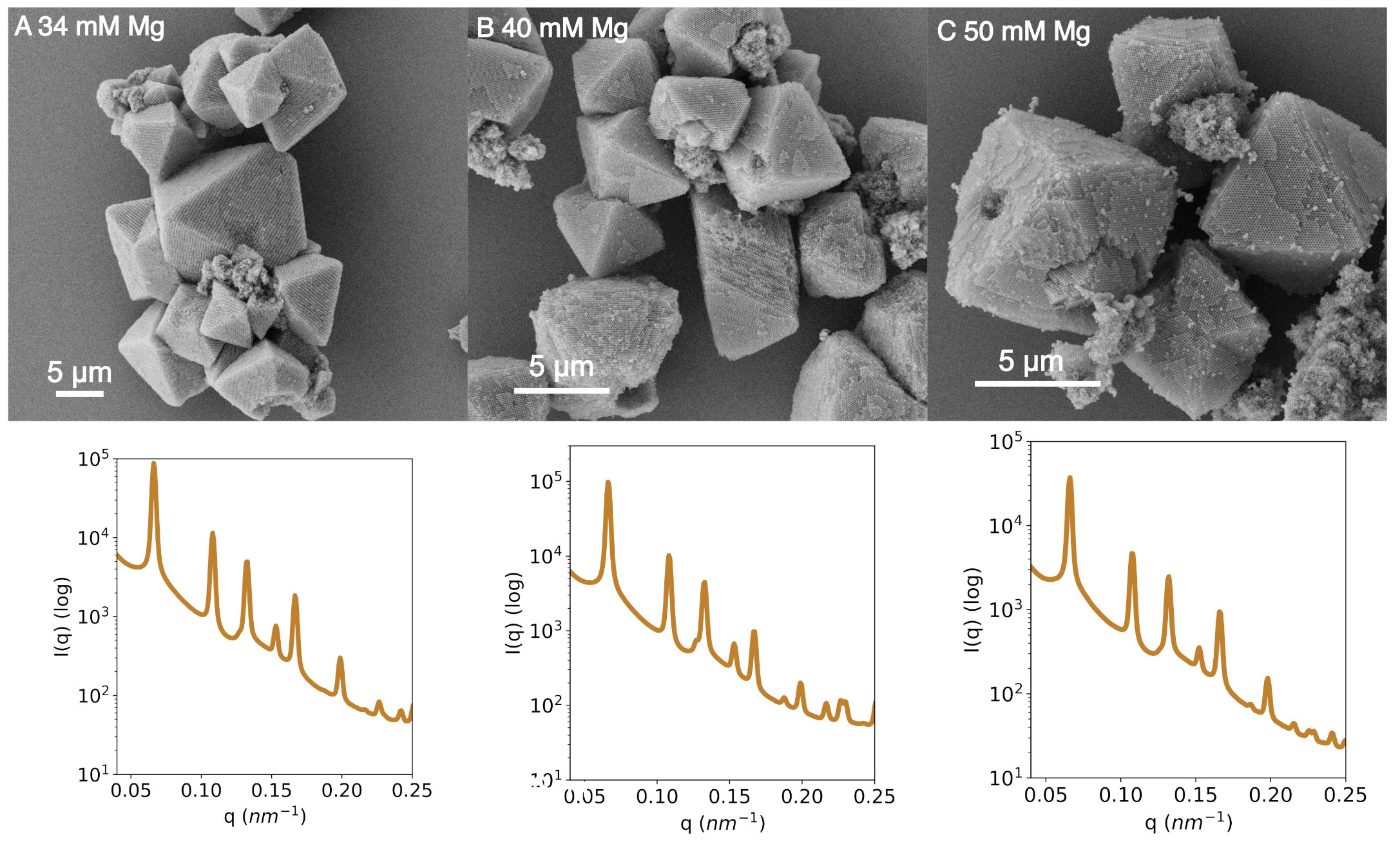} 
	\caption{\textbf{SEM images and corresponding SAXS intensities of samples assembled from a single mA monomer with extension 3T 3G.}
		Crystals grown from a monomer carrying the binding extension 3T 3G form DC for a wide range of Mg$^{2+}$ concentrations. This observation suggests that a torsional potential strongly favoring the staggered configuration prevents the formation of alternative non-specific configurations.}
	\label{3T3G} 
\end{figure}

\begin{figure} 
	\centering
	\includegraphics[width=0.9\textwidth]{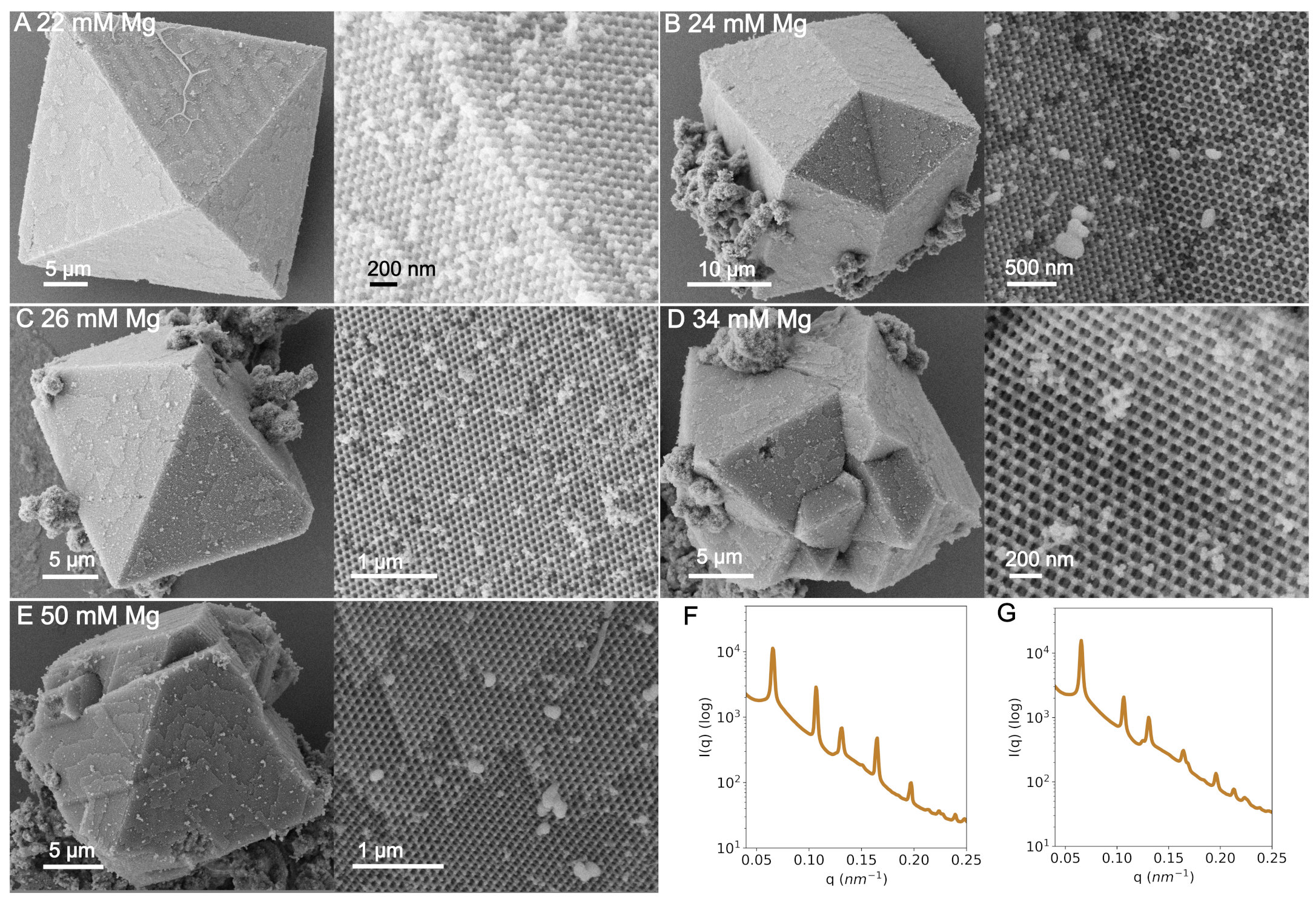} 
	\caption{\textbf{SEM images and SAXS intensities of samples with mA and mB monomers carrying the binding extension 3T 6nt 2G.}
		For this binding extension, DC form reliably under all tested conditions. The twinned crystal shown in panel B (24 mM MgCl$_{2}$) is an example of occasional twinning. With increasing concentrations of MgCl$_{2}$ more twinning planes can be observed. Triple DC occurs at 50 mM MgCl$_{2}$ concentration. Despite these morphology changes, the SAXS intensities for samples with (F) 34 mM MgCl$_{2}$, and (G) 50 mM MgCl$_{2}$ are unchanged, indicating that the bulk crystal structure is preserved as DC.}
	\label{6nt2G} 
\end{figure}

\begin{figure} 
	\centering
	\includegraphics[width=0.9\textwidth]{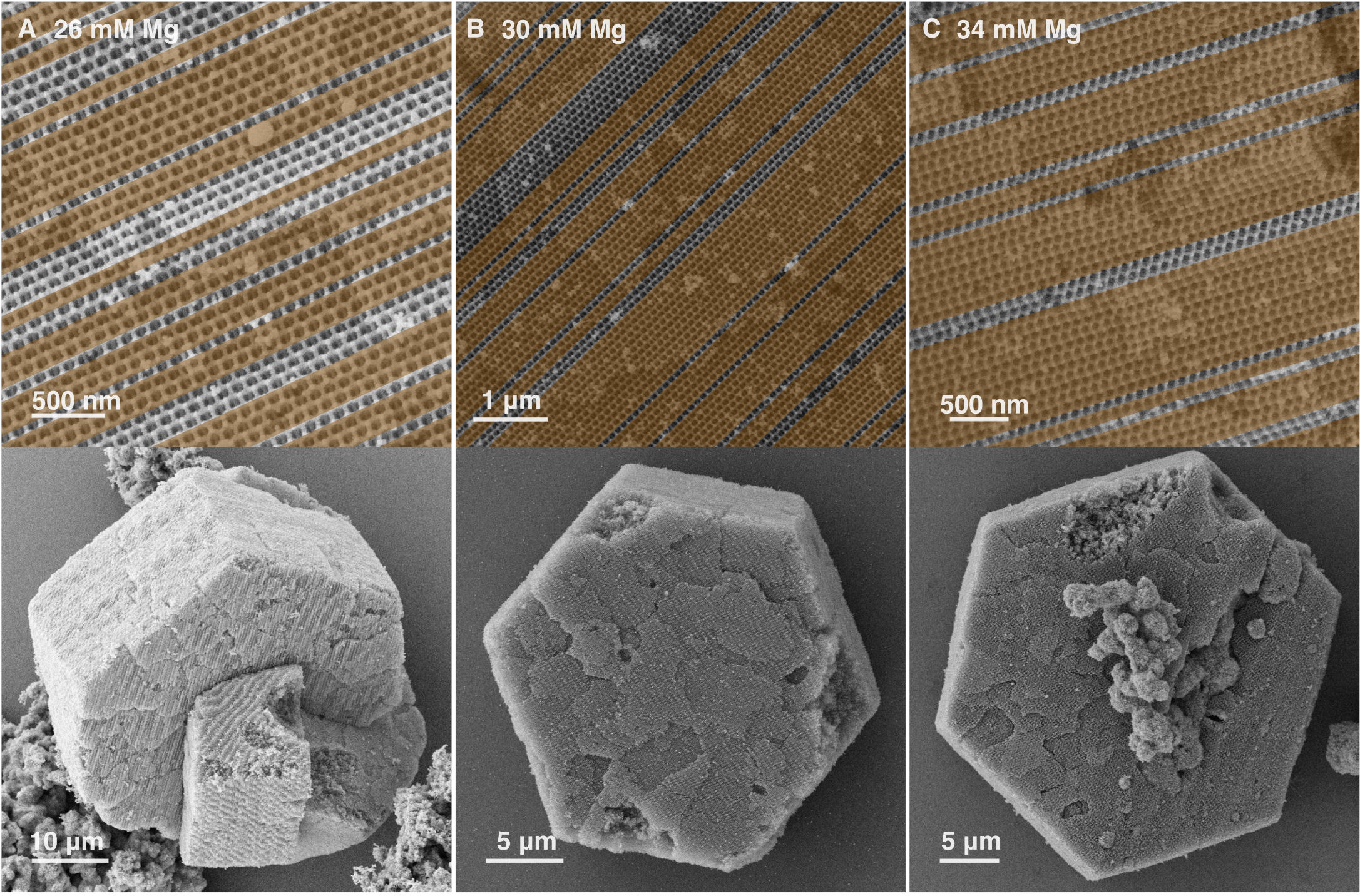} 
	\caption{\textbf{Examples of stacking disorder.} SEM images of samples with 3T 4nt 3G binding sequences and (\textbf{A}) 26 mM, (\textbf{B}) 30 mM, and (\textbf{C}) 34 mM MgCl$_{2}$, with stacking disorder along (001)HD/(111)DC direction. The stacking of the (111) plane of DC and (001) plane of HD follows an irregular sequence. The HD layers are marked with brown shading.}
	\label{stacking4nt}
\end{figure}

\begin{figure} 
	\centering
	\includegraphics[width=0.9\textwidth]{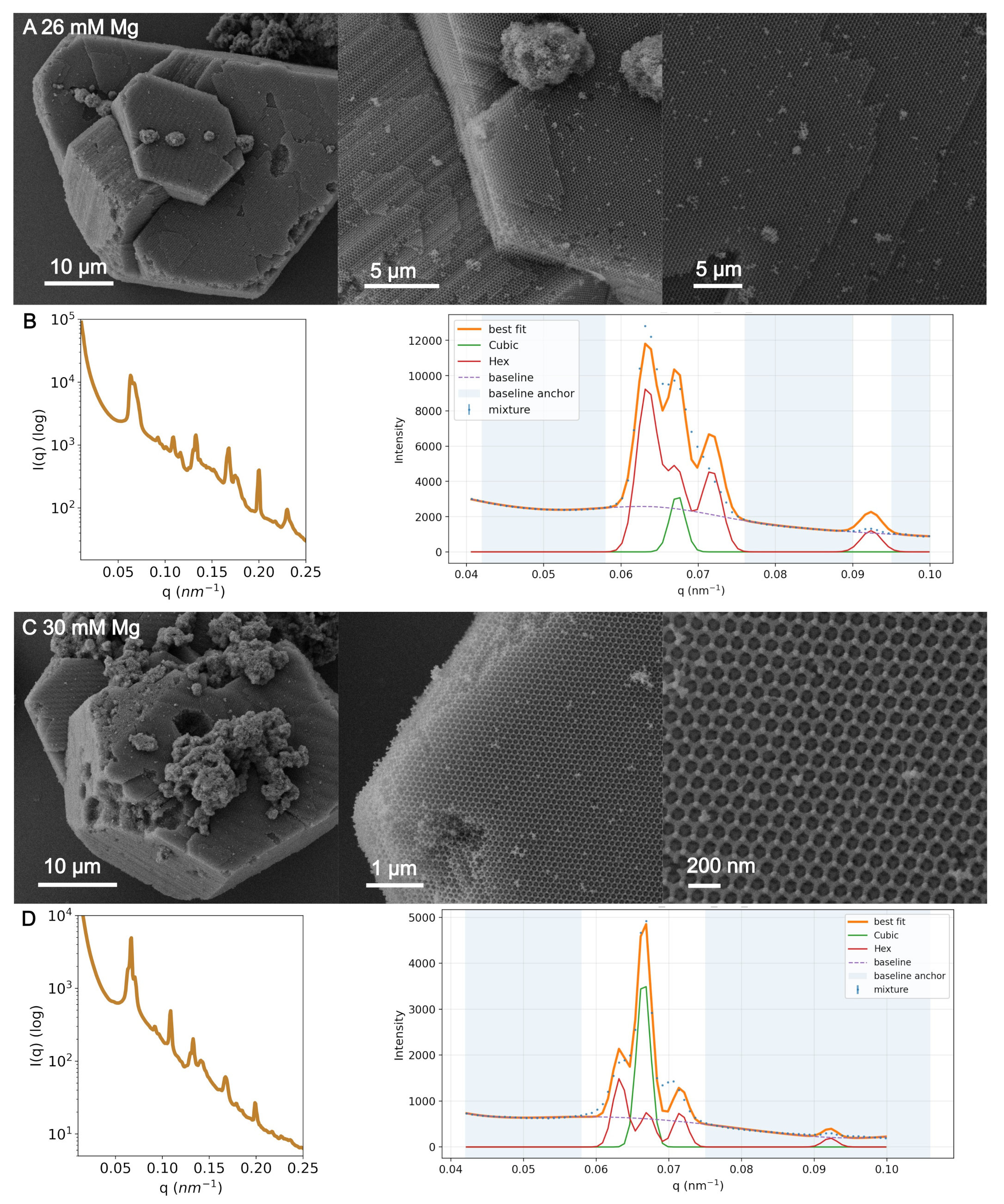} 
	\caption{\textbf{SEM images and SAXS intensities of samples where the monomers carry the binding extension 3T4nt3G.}
		At (\textbf{A}, \textbf{B}) 26 mM, and (\textbf{C}, \textbf{D}) 30 mM concentration of MgCl$_{2}$ we observe the emergence of HD. SAXS fitting (\textbf{B}, and \textbf{D}, right panels) indicates a high fraction of HD (90 \% and 50 \%, respectively).}
	\label{4nt26_30} 
\end{figure}

\begin{figure} 
	\centering
	\includegraphics[width=0.9\textwidth]{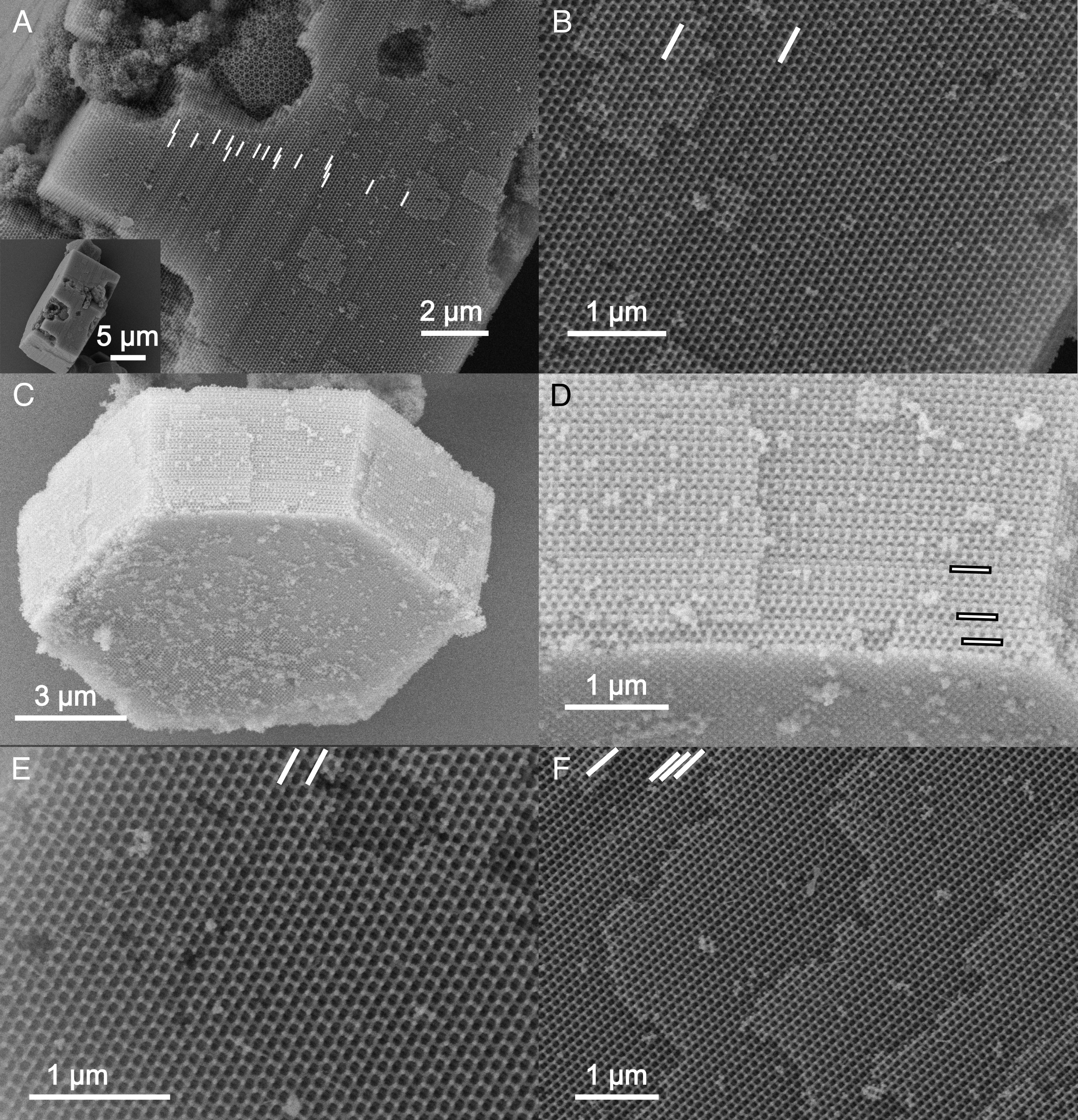} 
	\caption{\textbf{SEM images of $\{100\}$ facets of high quality HD crystals.}
		(\textbf{A, B}) A hexagonal diamond crystal, showing 17 isolated staggered layers among 152 total layers, including 28 (panel A, left), and 43 consecutive HD layers (panel A, right, and panel B); (\textbf{C, D})A tilted view of the HD crystal in  Figure~\ref{clathrate_figure2}G, showing three staggered among 45 total crystal layers; (\textbf{E}) A HD crystal showing two staggered layers among 52 total crystal layers; (\textbf{F}) Sample showing four staggered layers among 85 total crystal layers, with 70 consecutive HD layers. White bars mark the positions of the staggered layers. The staggered connections can be recognised by the Y-like pattern between the free tetrapod arms. The eclipsed connections, corresponding to the HD structure, can be recognised by the I-like pattern between the free tetrapod arms.}
	\label{Hex} 
\end{figure}

\begin{figure} 
	\centering
	\includegraphics[width=0.9\textwidth]{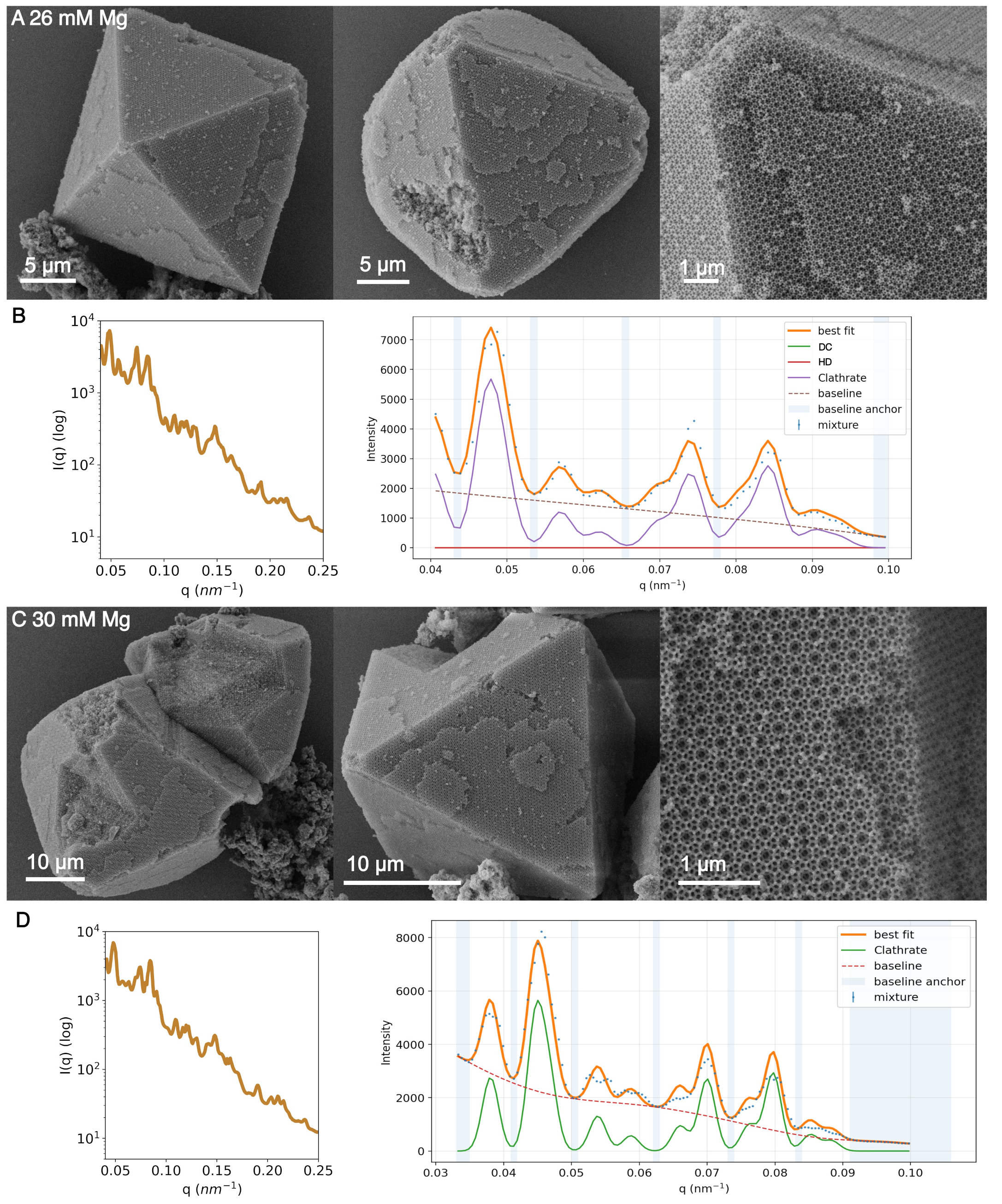} 
	\caption{\textbf{SEM images and SAXS intensities of samples with binding extension 7T 3G on mA and 3T 4nt 3G on mB.}
		At (\textbf{A}, \textbf{B}) 26 mM MgCl$_{2}$ concentration and (\textbf{C}, \textbf{D}) 30 mM MgCl$_{2}$ concentration we observe pure sII clathrate.}
	\label{4nt_7T26} 
\end{figure}

\begin{figure} 
	\centering
	\includegraphics[width=0.9\textwidth]{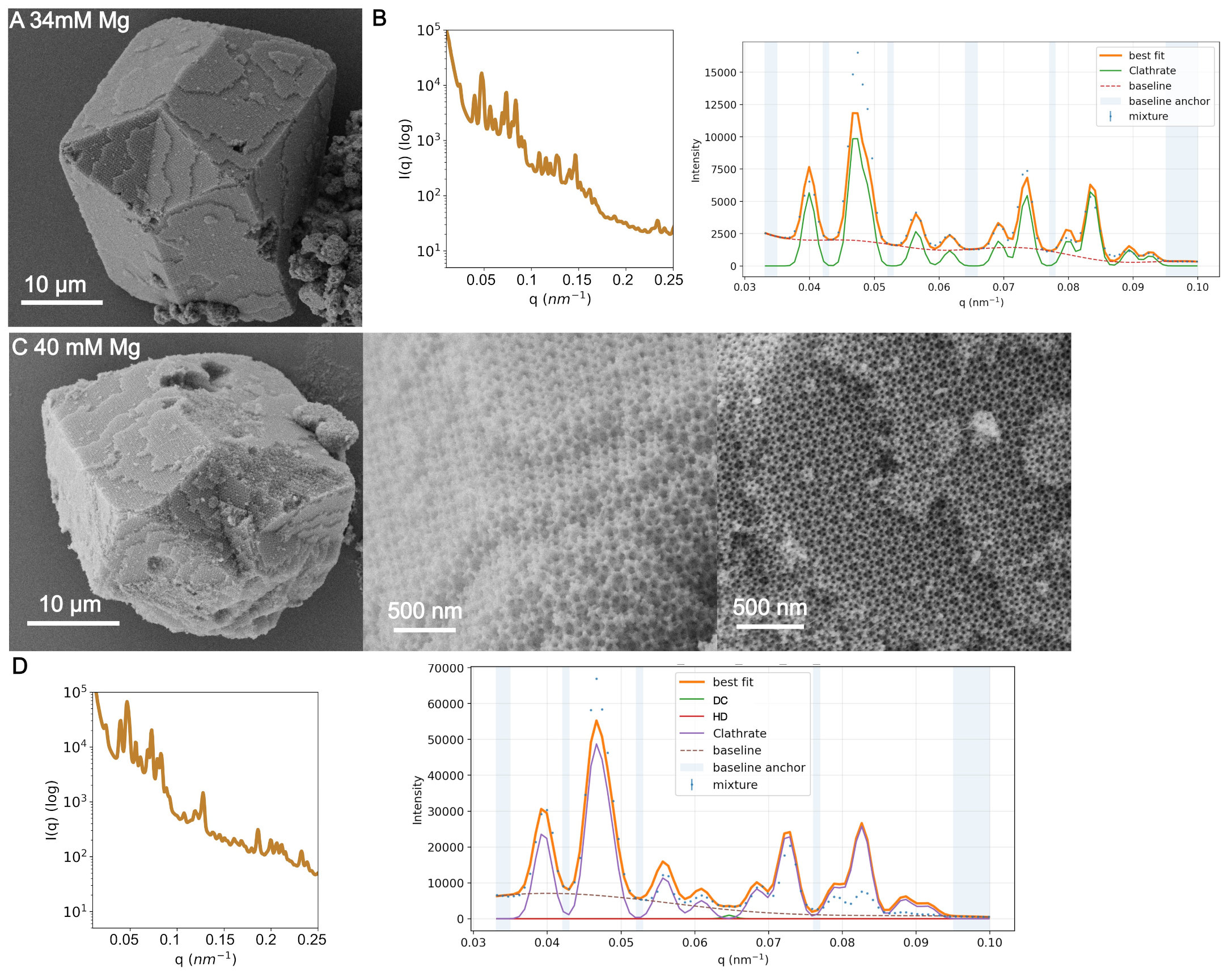} 
	\caption{\textbf{SEM images and SAXS intensities of samples with binding extension 7T 3G on mA and 3T 4nt 3G on mB.}
		At (\textbf{A}, \textbf{B}) 34 mM MgCl$_{2}$ concentration we observed pure clathrate. At (\textbf{C}, \textbf{D}) 40 mM MgCl$_{2}$ concentration we seldom observed the DC/HD stacking structures in addition to the clathrate.}
	\label{4nt_7T34} 
\end{figure}

\begin{figure} 
	\centering
	\includegraphics[width=0.6\textwidth]{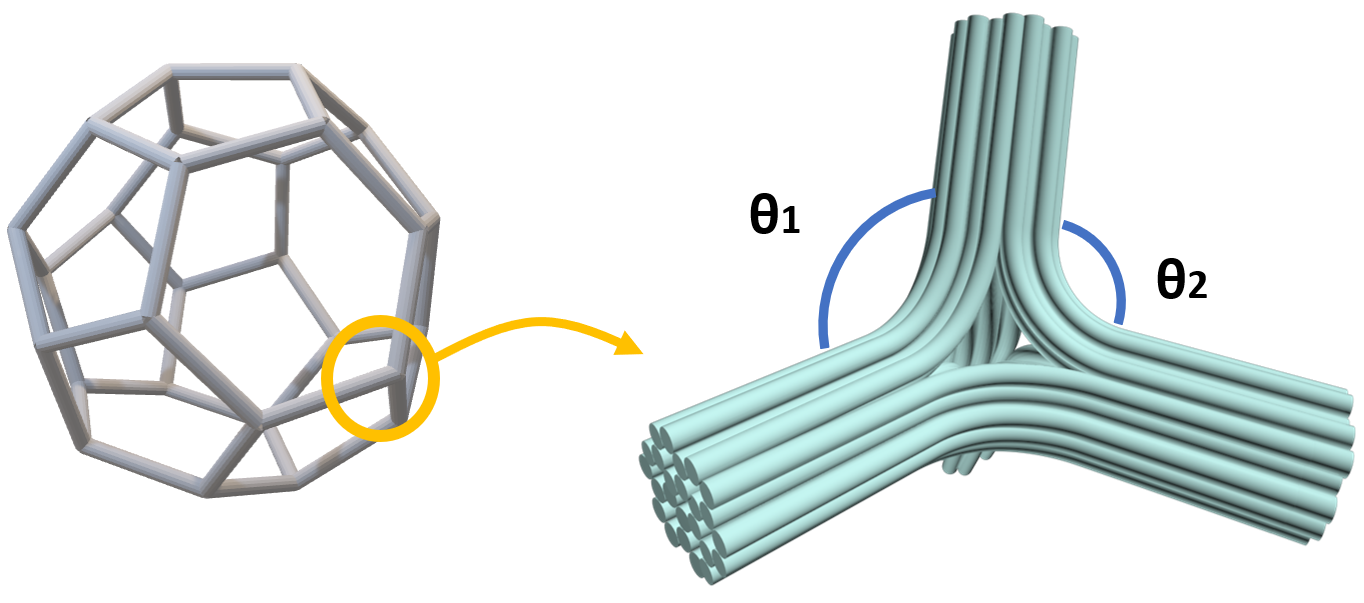} 
	\caption{\textbf{Deformation of tetrapod monomers in hexakaidecahedral cages.}
		To form the hexagonal (with 120° internal angles) and pentagonal (108°) faces of the hexakaidecahedral cages in the sII clathrate, the angles between the arms of the tetrapods ($\theta_1$ and $\theta_2$) deviate from the ideal tetrahedral angle of 109.5°.}
	\label{deformed_tetrapod} 
\end{figure}

\begin{figure} 
	\centering
	\includegraphics[width=0.9\textwidth]{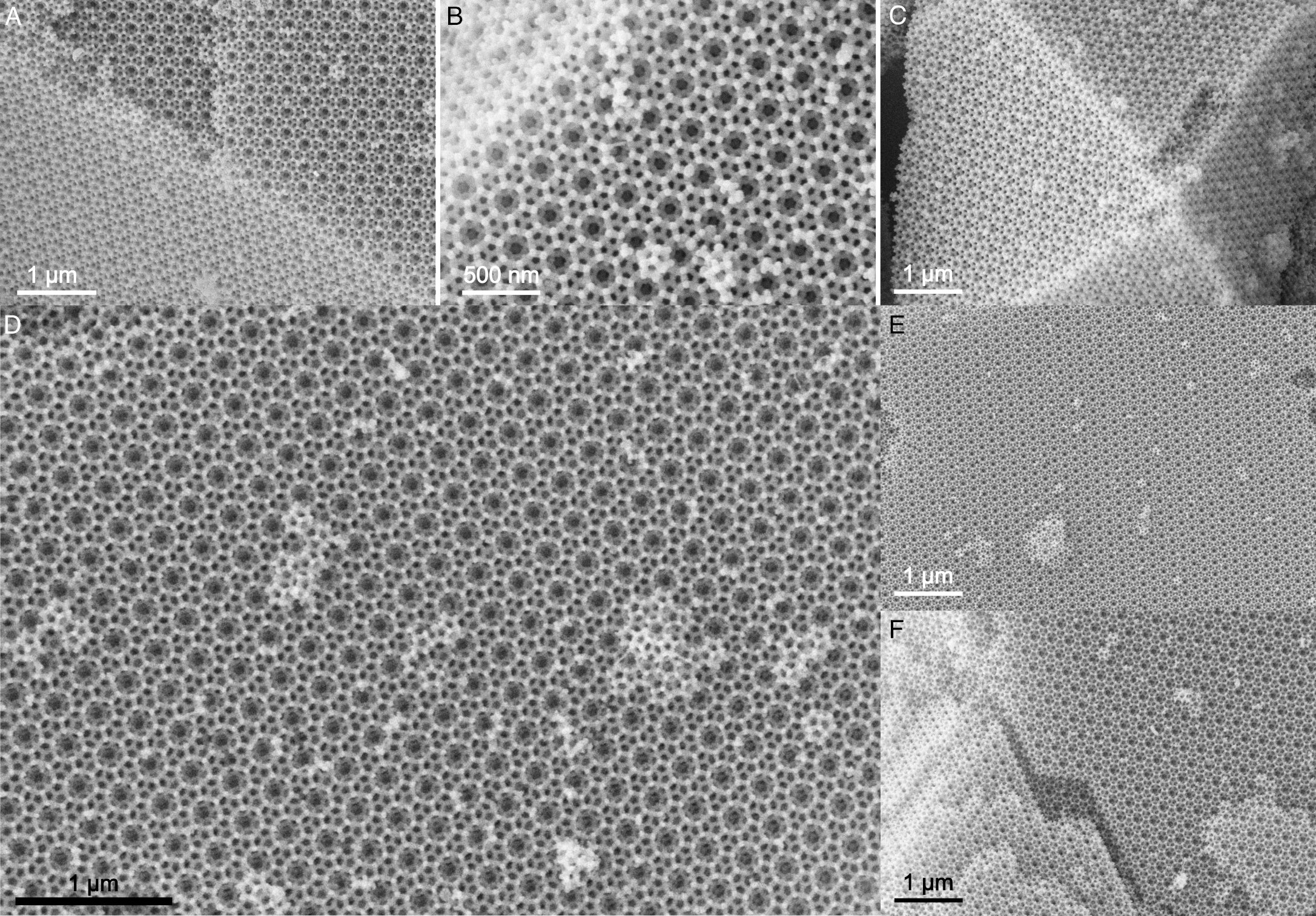} 
	\caption{\textbf{Zoomed-in SEM images of sII clathrate crystals.}(\textbf{A}, \textbf{B}) Examples of sII clathrate crystals with sharp edges. (\textbf{C}) Four {111} planes meet and form a sharp tip. (\textbf{D}) An image of a clathrate {111} plane with islands of dodecahedral cages, indicating the early formation of the next layer. (\textbf{E}) Example of large area, high quality clathrate {111} surface. (\textbf{F}) Clathrate (111) surface with a grain boundary.}
	\label{clathrate_detail}
\end{figure}

\begin{figure} 
	\centering
	\includegraphics[width=0.9\textwidth]{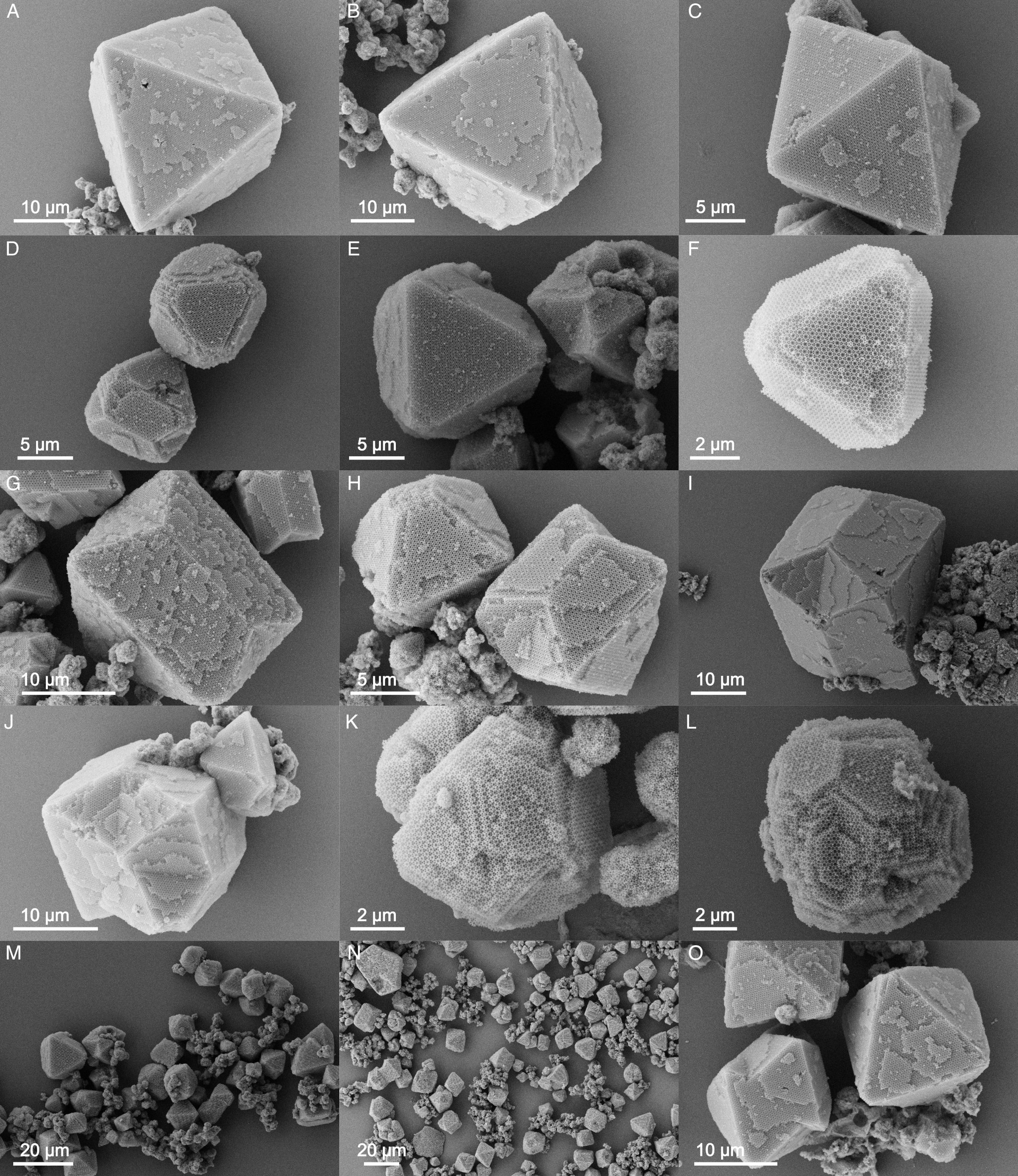} 
	\caption{\textbf{SEM images of sII clathrate crystals.}(\textbf{A-C}) Examples of clathrate with octahedral morphology with binding extension 7T 3G on mA and 3T 4nt 3G on mB. (\textbf{D-F}) Examples of clathrate with truncated octahedral morphology with binding extension 3T 4nt 3G on mA and mB. A rectangular shaped $\{100\}$ plane occurs at the tips. (\textbf{G-J}) Examples of twinned clathrate crystals with binding extension 7T 3G on mA and 3T 4nt 3G on mB. (\textbf{K-L}) Examples of twinned clathrate crystals with binding extension 3T 4nt 3G. (\textbf{M-O}) Zoomed out SEM images of sII clathrate samples.}
	\label{clathrate}
\end{figure}

\begin{figure}
	\centering
	\includegraphics[width=0.9\textwidth]{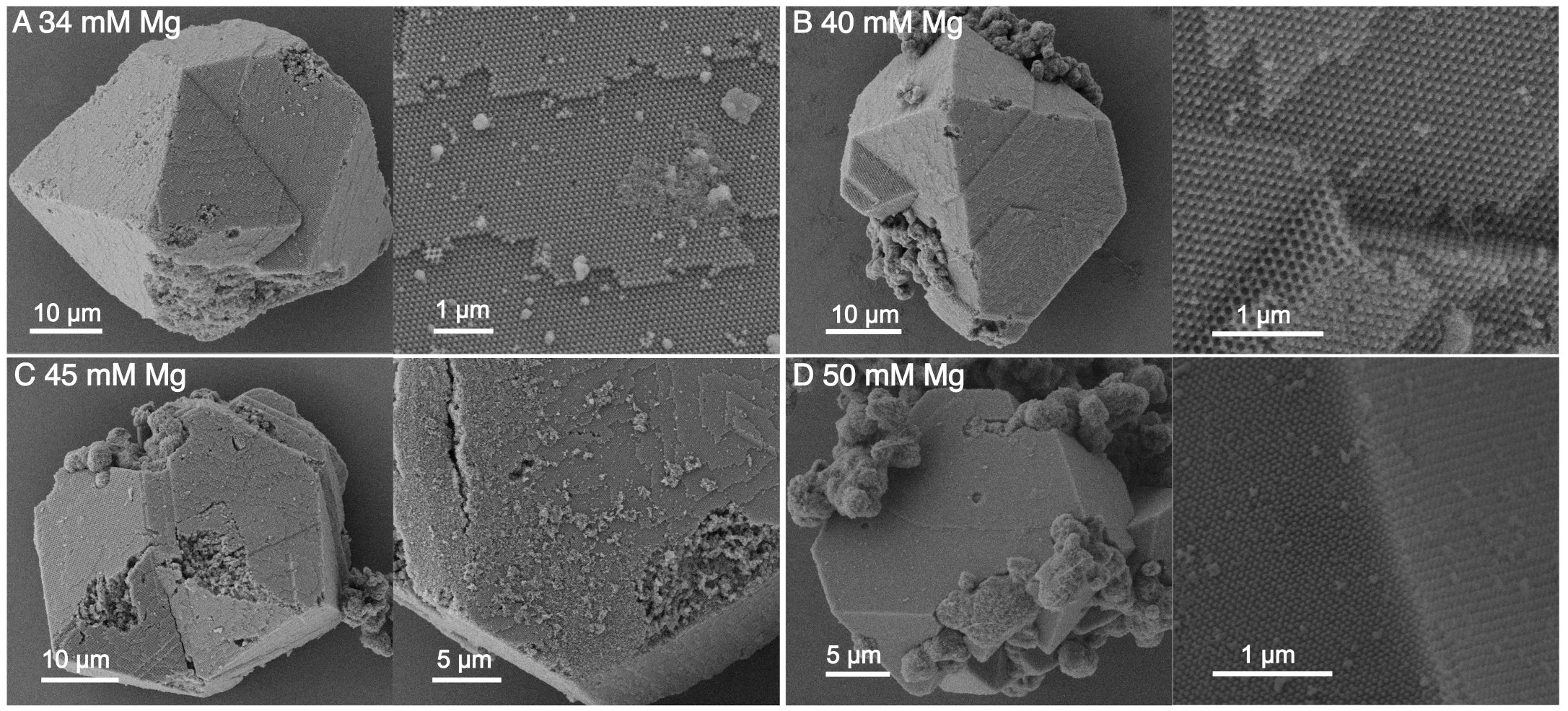} 
	\caption{\textbf{SEM images of samples assembled from a single mB monomer with extension 3T 4nt 3G.}
		At MgCl$_{2}$ concentrations up to 40 mM we observe the formation of single DC. At concentrations above 40 mM MgCl$_{2}$ triple lattices prevail.}
	\label{DB4nt3G} 
\end{figure}

\begin{figure} 
	\centering
	\includegraphics[width=0.9\textwidth]{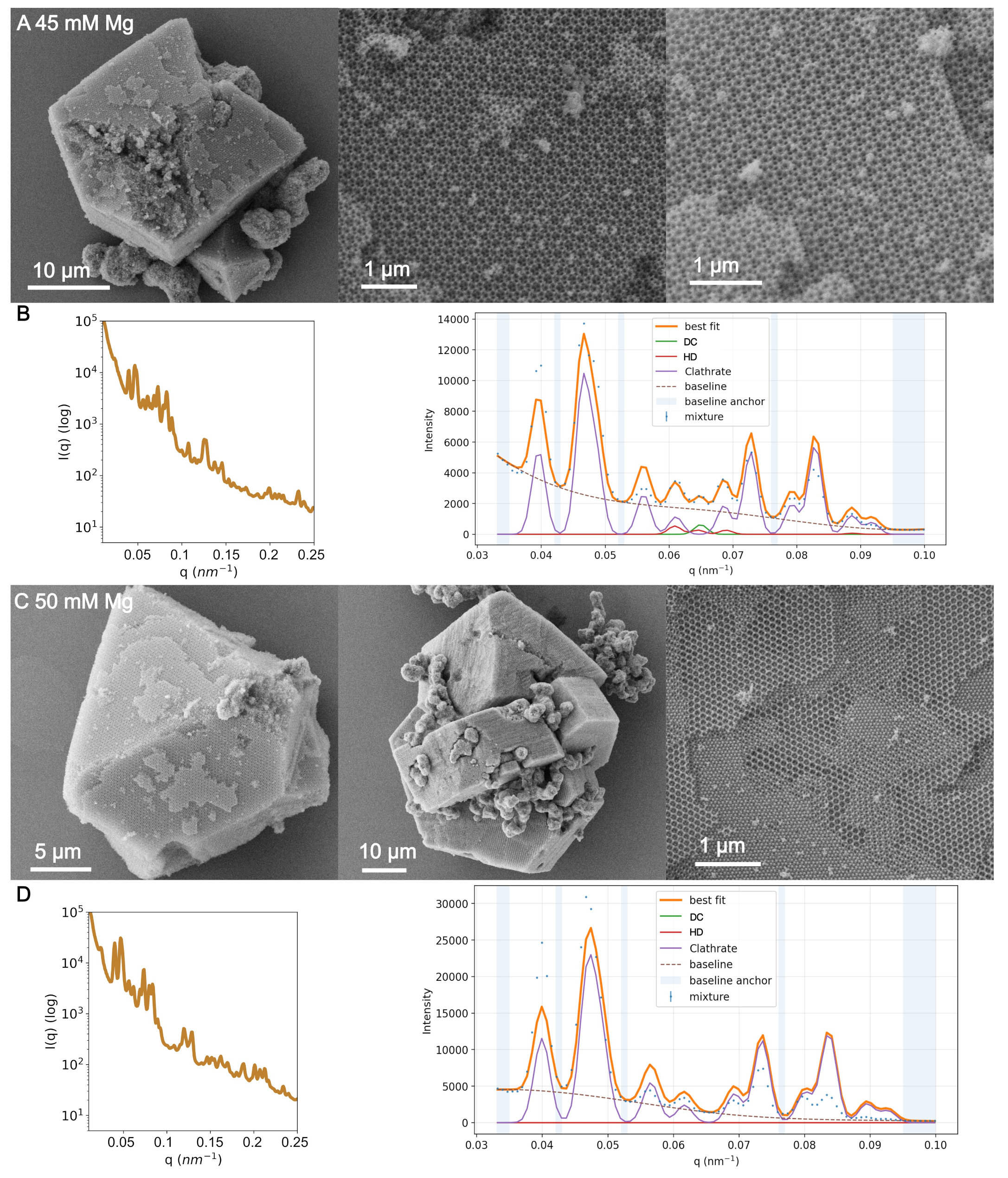} 
	\caption{\textbf{SEM images and SAXS intensities of samples with binding extensions 7T 3G on mA and 3T 4nt 3G on mB.}
		At (\textbf{A}, \textbf{B}) 45 mM MgCl$_{2}$ concentration we observed a large fraction of clathrate and some DC/HD stacking. At (\textbf{C}, \textbf{D}) 50 mM MgCl$_{2}$ concentration we seldom observed the clathrate crystals. The most common phase is DC/HD stacking-disordered structures with the emergence of triple lattice.}
	\label{4nt_7T45} 
\end{figure}

\clearpage

\subsubsection*{Bond tuning}
The staggered binding configuration is favored in samples with longer specific regions: "3T 6nt 2G", "3T 6nt 3G" and "3T 5nt 3G" making 6 connections per pair of monomers. With these sequences, the dominant phase in the system is diamond cubic, with different frequencies of twinning depending on the staggered to eclipsed configuration strength ratio, and high Mg$^{2+}$ concentration.. 

In contrast, binding between poly(C) and poly(G) regions occurs between 12 bonds per patch. The "3T 4nt 3G" sequence on both monomers with an even shorter specific region disfavors staggered, and prefers eclipsed binding.  Likewise, increasing Mg$^{2+}$ concentrations weaken staggered binding and strengthen the eclipsed configurations. Consequently, in samples carrying the binding extension "3T 4nt 3G" we observed hexagonally shaped crystals with stacking disorder at lower Mg$^{2+}$ concentrations, while at higher Mg2+ concentration clathrate and triple diamond structures start to form. The combination of "7T 3G" (mA) and "3T 4nt 3G" (mB) even more strongly favors the eclipsed configuration, as the A1 to B1 staggered binding mode is no longer possible. In this case, we obtained a pure clathrate phase at Mg$^{2+}$ concentrations below 34 mM, and their coexistence with triple HD and CD form at higher Mg$^{2+}$ concentrations. At Mg$^{2+}$ concentrations above 55 mM the sII clathrate stops to form, leading to triple HD crystals with some stacking disorder.

\begin{figure} 
	\centering
	\includegraphics[width=0.6\textwidth]{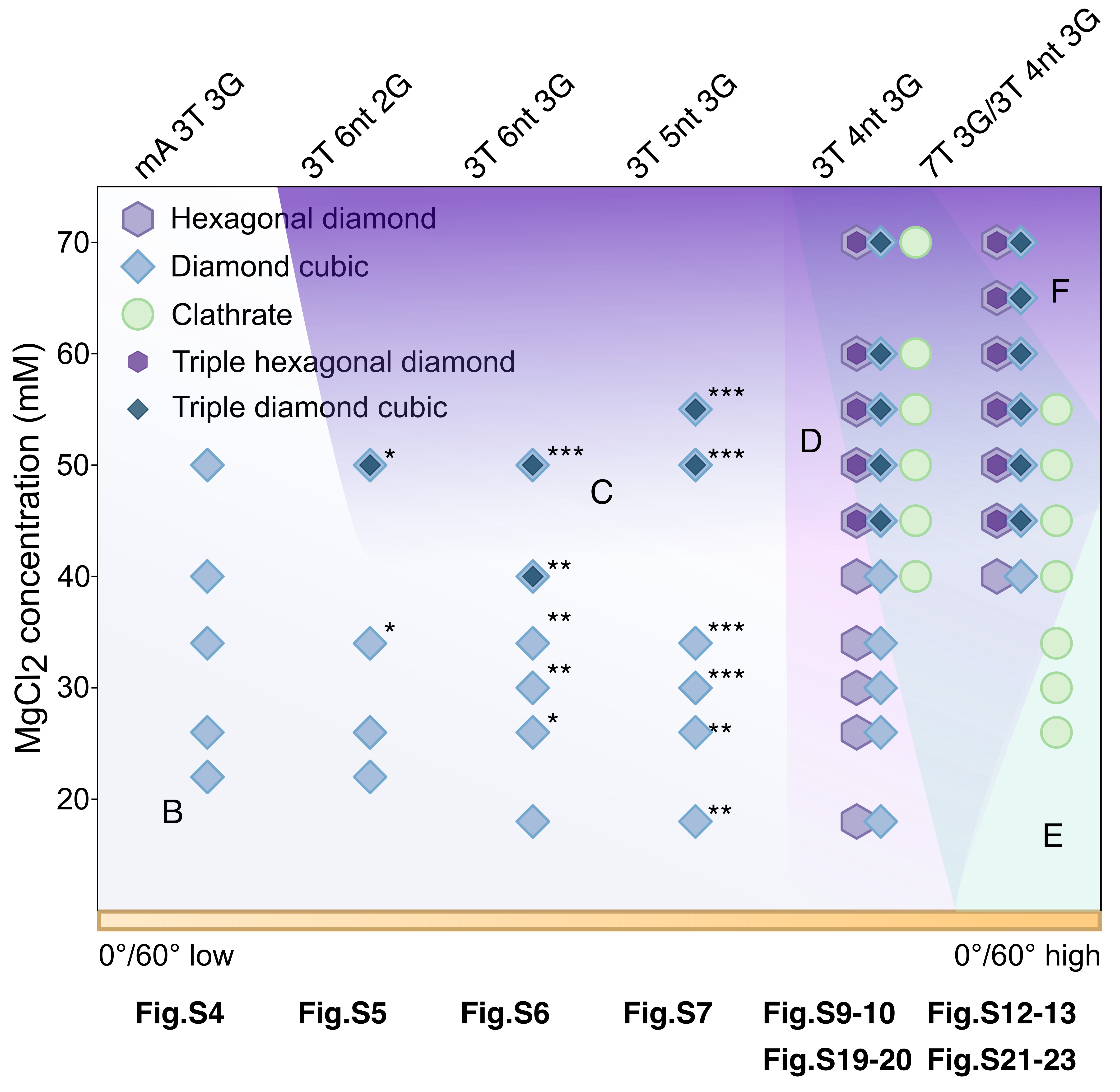} 
	\caption{\textbf{Phase diagram with sample names listed at the top and the corresponding figure numbers at the bottom.}
 The occurrence of structures, plotted over the configuration bias (eclipsed/staggered, x-axis), and the MgCl$_2$ concentration (y-axis). Each blue diamond, purple hexagon or green circle denotes the observation of diamond cubic, hexagonal diamond, or sII clathrates, respectively, at the indicated conditions. Touching diamond and hexagonal symbols correspond to stacking-disordered mixture of DC and HD. Dark filling of the diamonds and hexagons points to the occurrence of triple lattices. Asterisks denote the amount of twinning in the DC crystals: * for most crystals having a single twinning plane, ** for multiple twinning planes in most crystals, and *** for multiple twinning planes in the same direction, leading to crystals with hexagonal morphology.}
	\label{phasediagram_samples} 
\end{figure}

\begin{figure} 
	\centering
	\includegraphics[width=0.6\textwidth]{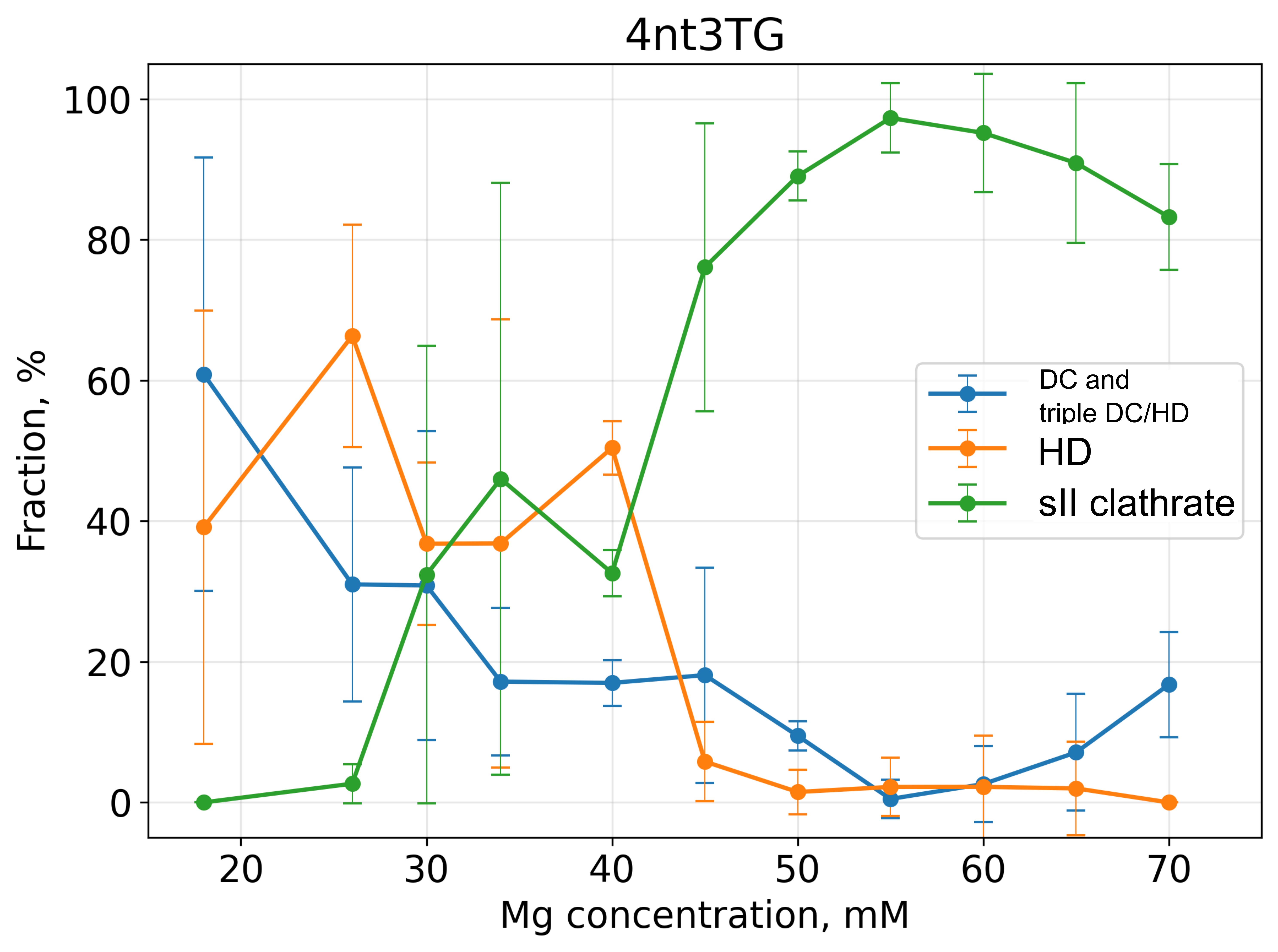} 
	\caption{\textbf{Phase fractions fitted from SAXS intensities of samples with binding extension 3T 4nt 3G on both monomers.}
		With increasing MgCl$_{2}$ concentrations (up to 40 mM) we observe an increase of HD relative to DC, corresponding to higher fractions of HD in the stacking disordered DC/HD structures. At the same time, the fraction of the sII clathrate phase increases with increasing MgCl$_{2}$ concentration, becoming the dominant phase above 45 mM MgCl$_{2}$ concentration. At the highest concentrations, the fraction of clathrate is reduced in favor of triple structures, which are fitted with the DC model, as discussed in the supplementary text. The fitting data was averaged from independent measurements of 2 or 3 samples at each concentration, the error bars denote the standard deviation. }
	\label{4nt3G_fraction} 
\end{figure}

\begin{figure} 
	\centering
	\includegraphics[width=0.6\textwidth]{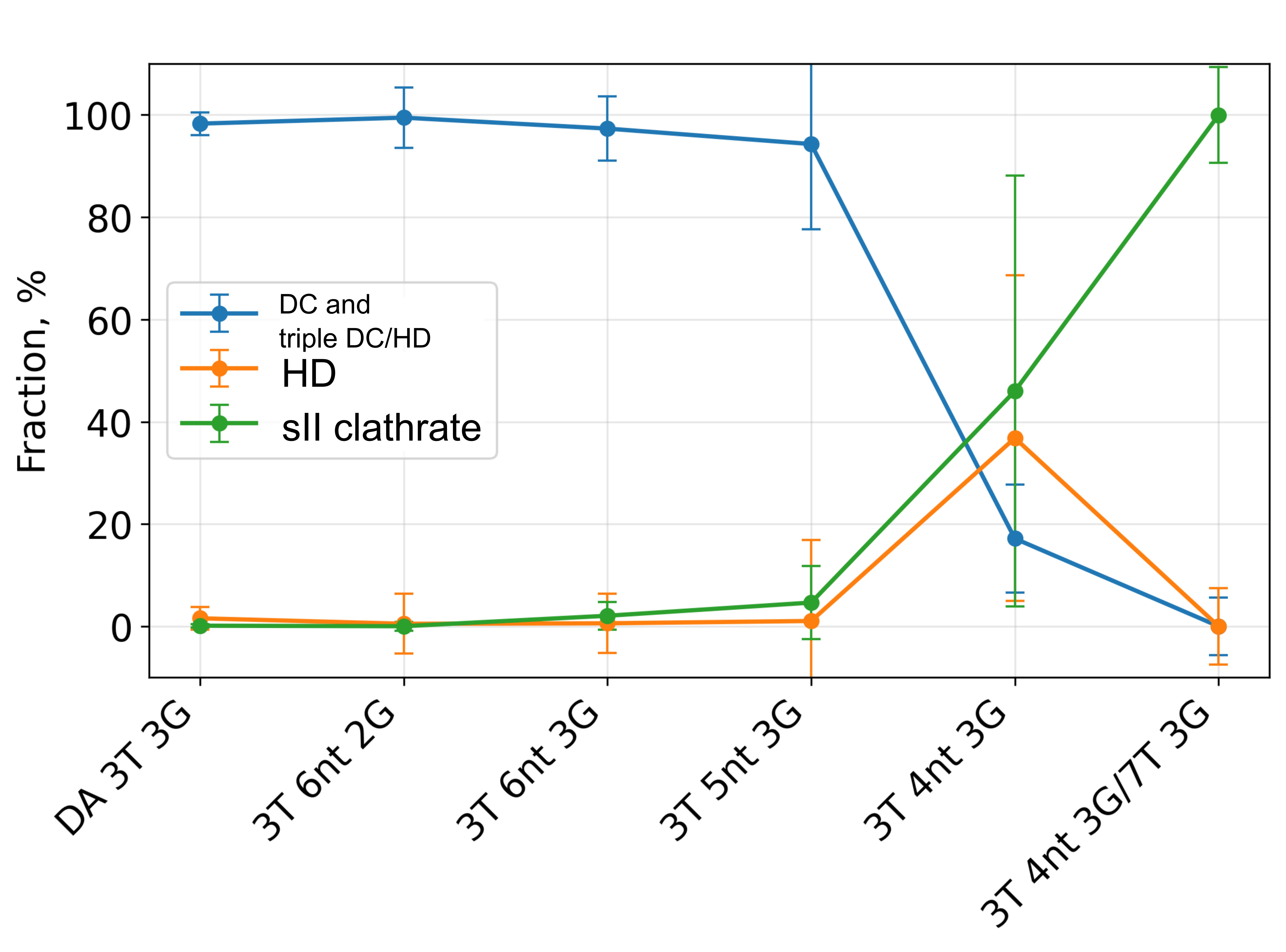} 
	\caption{\textbf{Phase fractions fitted from SAXS intensities of samples with different binding extensions at 34 mM MgCl$_{2}$ concentration.}
		Phase fractions plotted against the configuration bias (eclipsed/staggered, x-axis). With increasing eclipsed configuration, We observe a sharp change from pure DC to pure sII clathrate phase. Error bars denote the fitting residuals except for the 3T 4nt 3G samples, where they represent the standard deviation from measurements of 3 individual samples. }
	\label{34mM_fraction} 
\end{figure}

\begin{figure} 
	\centering
	\includegraphics[width=0.6\textwidth]{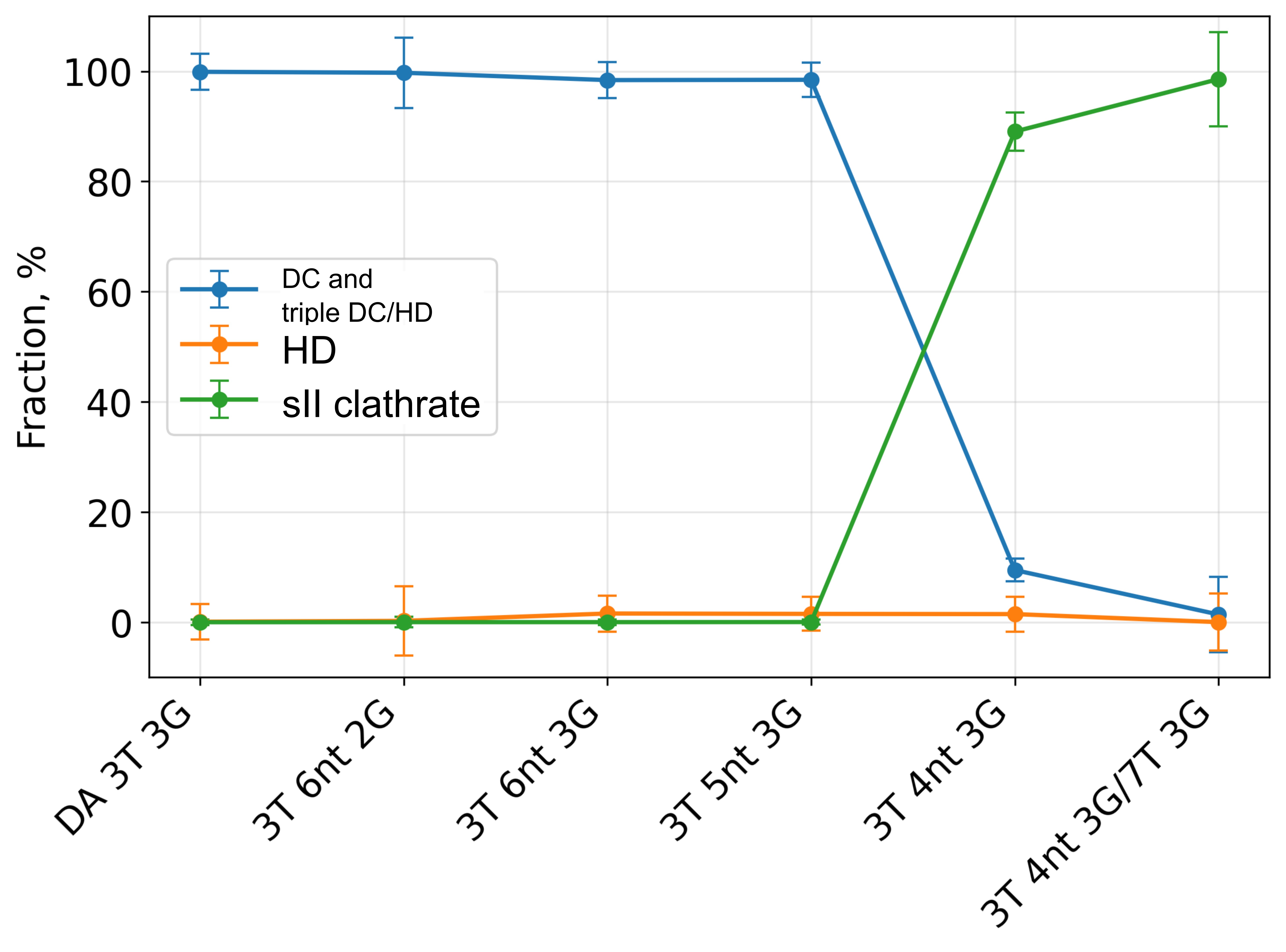} 
	\caption{\textbf{Phase fractions fitted from SAXS intensities of samples with different binding extensions at 50 mM MgCl$_{2}$ concentration.}
		Phase fractions plotted against the configuration bias (eclipsed/staggered, x-axis). With increasing eclipsed configuration, We observe a change from pure DC to triple DC and to pure sII clathrate phase. Error bars denote the fitting residuals except for the 3T 4nt 3G samples, where they represent the standard deviation from measurements of 2 or 3 individual samples. }
	\label{50mM_fraction} 
\end{figure}

\clearpage 


\end{document}